\documentstyle[12pt]{article}

\expandafter\ifx\csname amssym12.def\endcsname\relax \else\endinput\fi
\expandafter\edef\csname amssym12.def\endcsname{%
	 \catcode`\noexpand\@=\the\catcode`\@\space}
\catcode`\@=11
\def\undefine#1{\let#1\undefined}
\def\newsymbol#1#2#3#4#5{\let\next@\relax
 \ifnum#2=\@ne\let\next@\msafam@\else
 \ifnum#2=\tw@\let\next@\msbfam@\fi\fi
 \mathchardef#1="#3\next@#4#5}
\def\mathhexbox@#1#2#3{\relax
 \ifmmode\mathpalette{}{\m@th\mathchar"#1#2#3}%
 \else\leavevmode\hbox{$\m@th\mathchar"#1#2#3$}\fi}
\def\hexnumber@#1{\ifcase#1 0\or 1\or 2\or 3\or 4\or 5\or 6\or 7\or 8\or
 9\or A\or B\or C\or D\or E\or F\fi}

\font\tenmsa=msam10 scaled\magstep1
\font\sevenmsa=msam7 scaled\magstep1
\font\fivemsa=msam5 scaled\magstep1
\newfam\msafam
\textfont\msafam=\tenmsa
\scriptfont\msafam=\sevenmsa
\scriptscriptfont\msafam=\fivemsa
\edef\msafam@{\hexnumber@\msafam}
\mathchardef\dabar@"0\msafam@39
\def\dashrightarrow{\mathrel{\dabar@\dabar@\mathchar"0\msafam@4B}}
\def\dashleftarrow{\mathrel{\mathchar"0\msafam@4C\dabar@\dabar@}}

\def\ulcorner{\delimiter"4\msafam@70\msafam@70 }
\def\urcorner{\delimiter"5\msafam@71\msafam@71 }
\def\llcorner{\delimiter"4\msafam@78\msafam@78 }
\def\lrcorner{\delimiter"5\msafam@79\msafam@79 }
\def\yen{{\mathhexbox@\msafam@55 }}
\def\checkmark{{\mathhexbox@\msafam@58 }}
\def\circledR{{\mathhexbox@\msafam@72 }}
\def\maltese{{\mathhexbox@\msafam@7A }}

\font\tenmsb=msbm10 scaled\magstep1
\font\sevenmsb=msbm7 scaled\magstep1
\font\fivemsb=msbm5 scaled\magstep1
\newfam\msbfam
\textfont\msbfam=\tenmsb
\scriptfont\msbfam=\sevenmsb
\scriptscriptfont\msbfam=\fivemsb
\edef\msbfam@{\hexnumber@\msbfam}
\def\Bbb#1{{\fam\msbfam\relax#1}}
\def\widehat#1{\setbox\z@\hbox{$\m@th#1$}%
 \ifdim\wd\z@>\tw@ em\mathaccent"0\msbfam@5B{#1}%
 \else\mathaccent"0362{#1}\fi}
\def\widetilde#1{\setbox\z@\hbox{$\m@th#1$}%
 \ifdim\wd\z@>\tw@ em\mathaccent"0\msbfam@5D{#1}%
 \else\mathaccent"0365{#1}\fi}
\font\teneufm=eufm10 scaled\magstep1
\font\seveneufm=eufm7 scaled\magstep1
\font\fiveeufm=eufm5 scaled\magstep1
\newfam\eufmfam
\textfont\eufmfam=\teneufm
\scriptfont\eufmfam=\seveneufm
\scriptscriptfont\eufmfam=\fiveeufm
\def\frak#1{{\fam\eufmfam\relax#1}}

\csname amssym.def\endcsname

\expandafter\ifx\csname pre amssym.tex at\endcsname\relax \else \endinput\fi
\expandafter\chardef\csname pre amssym.tex at\endcsname=\the\catcode`\@
\catcode`\@=11
\begingroup\ifx\undefined\newsymbol \else\def\input#1 {\endgroup}\fi
 \input amssym.def \relax
\newsymbol\boxdot 1200
\newsymbol\boxplus 1201
\newsymbol\boxtimes 1202
\newsymbol\square 1003
\newsymbol\blacksquare 1004
\newsymbol\centerdot 1205
\newsymbol\lozenge 1006
\newsymbol\blacklozenge 1007
\newsymbol\circlearrowright 1308
\newsymbol\circlearrowleft 1309
\undefine\rightleftharpoons
\newsymbol\rightleftharpoons 130A
\newsymbol\leftrightharpoons 130B
\newsymbol\boxminus 120C
\newsymbol\Vdash 130D
\newsymbol\Vvdash 130E
\newsymbol\vDash 130F
\newsymbol\twoheadrightarrow 1310
\newsymbol\twoheadleftarrow 1311
\newsymbol\leftleftarrows 1312
\newsymbol\rightrightarrows 1313
\newsymbol\upuparrows 1314
\newsymbol\downdownarrows 1315
\newsymbol\upharpoonright 1316
 \let\restriction\upharpoonright
\newsymbol\downharpoonright 1317
\newsymbol\upharpoonleft 1318
\newsymbol\downharpoonleft 1319
\newsymbol\rightarrowtail 131A
\newsymbol\leftarrowtail 131B
\newsymbol\leftrightarrows 131C
\newsymbol\rightleftarrows 131D
\newsymbol\Lsh 131E
\newsymbol\Rsh 131F
\newsymbol\rightsquigarrow 1320
\newsymbol\leftrightsquigarrow 1321
\newsymbol\looparrowleft 1322
\newsymbol\looparrowright 1323
\newsymbol\circeq 1324
\newsymbol\succsim 1325
\newsymbol\gtrsim 1326
\newsymbol\gtrapprox 1327
\newsymbol\multimap 1328
\newsymbol\therefore 1329
\newsymbol\because 132A
\newsymbol\doteqdot 132B
 
\newsymbol\triangleq 132C
\newsymbol\precsim 132D
\newsymbol\lesssim 132E
\newsymbol\lessapprox 132F
\newsymbol\eqslantless 1330
\newsymbol\eqslantgtr 1331
\newsymbol\curlyeqprec 1332
\newsymbol\curlyeqsucc 1333
\newsymbol\preccurlyeq 1334
\newsymbol\leqq 1335
\newsymbol\leqslant 1336
\newsymbol\lessgtr 1337
\newsymbol\backprime 1038
\newsymbol\risingdotseq 133A
\newsymbol\fallingdotseq 133B
\newsymbol\succcurlyeq 133C
\newsymbol\geqq 133D
\newsymbol\geqslant 133E
\newsymbol\gtrless 133F
\newsymbol\sqsubset 1340
\newsymbol\sqsupset 1341
\newsymbol\vartriangleright 1342
\newsymbol\vartriangleleft 1343
\newsymbol\trianglerighteq 1344
\newsymbol\trianglelefteq 1345
\newsymbol\bigstar 1046
\newsymbol\between 1347
\newsymbol\blacktriangledown 1048
\newsymbol\blacktriangleright 1349
\newsymbol\blacktriangleleft 134A
\newsymbol\vartriangle 134D
\newsymbol\blacktriangle 104E
\newsymbol\triangledown 104F
\newsymbol\eqcirc 1350
\newsymbol\lesseqgtr 1351
\newsymbol\gtreqless 1352
\newsymbol\lesseqqgtr 1353
\newsymbol\gtreqqless 1354
\newsymbol\Rrightarrow 1356
\newsymbol\Lleftarrow 1357
\newsymbol\veebar 1259
\newsymbol\barwedge 125A
\newsymbol\doublebarwedge 125B
\undefine\angle
\newsymbol\angle 105C
\newsymbol\measuredangle 105D
\newsymbol\sphericalangle 105E
\newsymbol\varpropto 135F
\newsymbol\smallsmile 1360
\newsymbol\smallfrown 1361
\newsymbol\Subset 1362
\newsymbol\Supset 1363
\newsymbol\Cup 1264
 
\newsymbol\Cap 1265
 
\newsymbol\curlywedge 1266
\newsymbol\curlyvee 1267
\newsymbol\leftthreetimes 1268
\newsymbol\rightthreetimes 1269
\newsymbol\subseteqq 136A
\newsymbol\supseteqq 136B
\newsymbol\bumpeq 136C
\newsymbol\Bumpeq 136D
\newsymbol\lll 136E
 
\newsymbol\ggg 136F
 
\newsymbol\circledS 1073
\newsymbol\pitchfork 1374
\newsymbol\dotplus 1275
\newsymbol\backsim 1376
\newsymbol\backsimeq 1377
\newsymbol\complement 107B
\newsymbol\intercal 127C
\newsymbol\circledcirc 127D
\newsymbol\circledast 127E
\newsymbol\circleddash 127F
\newsymbol\lvertneqq 2300
\newsymbol\gvertneqq 2301
\newsymbol\nleq 2302
\newsymbol\ngeq 2303
\newsymbol\nless 2304
\newsymbol\ngtr 2305
\newsymbol\nprec 2306
\newsymbol\nsucc 2307
\newsymbol\lneqq 2308
\newsymbol\gneqq 2309
\newsymbol\nleqslant 230A
\newsymbol\ngeqslant 230B
\newsymbol\lneq 230C
\newsymbol\gneq 230D
\newsymbol\npreceq 230E
\newsymbol\nsucceq 230F
\newsymbol\precnsim 2310
\newsymbol\succnsim 2311
\newsymbol\lnsim 2312
\newsymbol\gnsim 2313
\newsymbol\nleqq 2314
\newsymbol\ngeqq 2315
\newsymbol\precneqq 2316
\newsymbol\succneqq 2317
\newsymbol\precnapprox 2318
\newsymbol\succnapprox 2319
\newsymbol\lnapprox 231A
\newsymbol\gnapprox 231B
\newsymbol\nsim 231C
\newsymbol\ncong 231D
\newsymbol\diagup 201E
\newsymbol\diagdown 201F
\newsymbol\varsubsetneq 2320
\newsymbol\varsupsetneq 2321
\newsymbol\nsubseteqq 2322
\newsymbol\nsupseteqq 2323
\newsymbol\subsetneqq 2324
\newsymbol\supsetneqq 2325
\newsymbol\varsubsetneqq 2326
\newsymbol\varsupsetneqq 2327
\newsymbol\subsetneq 2328
\newsymbol\supsetneq 2329
\newsymbol\nsubseteq 232A
\newsymbol\nsupseteq 232B
\newsymbol\nparallel 232C
\newsymbol\nmid 232D
\newsymbol\nshortmid 232E
\newsymbol\nshortparallel 232F
\newsymbol\nvdash 2330
\newsymbol\nVdash 2331
\newsymbol\nvDash 2332
\newsymbol\nVDash 2333
\newsymbol\ntrianglerighteq 2334
\newsymbol\ntrianglelefteq 2335
\newsymbol\ntriangleleft 2336
\newsymbol\ntriangleright 2337
\newsymbol\nleftarrow 2338
\newsymbol\nrightarrow 2339
\newsymbol\nLeftarrow 233A
\newsymbol\nRightarrow 233B
\newsymbol\nLeftrightarrow 233C
\newsymbol\nleftrightarrow 233D
\newsymbol\divideontimes 223E
\newsymbol\varnothing 203F
\newsymbol\nexists 2040
\newsymbol\Finv 2060
\newsymbol\Game 2061
\newsymbol\mho 2066
\newsymbol\eth 2067
\newsymbol\eqsim 2368
\newsymbol\beth 2069
\newsymbol\gimel 206A
\newsymbol\daleth 206B
\newsymbol\lessdot 236C
\newsymbol\gtrdot 236D
\newsymbol\ltimes 226E
\newsymbol\rtimes 226F
\newsymbol\shortmid 2370
\newsymbol\shortparallel 2371
\newsymbol\smallsetminus 2272
\newsymbol\thicksim 2373
\newsymbol\thickapprox 2374
\newsymbol\approxeq 2375
\newsymbol\succapprox 2376
\newsymbol\precapprox 2377
\newsymbol\curvearrowleft 2378
\newsymbol\curvearrowright 2379
\newsymbol\digamma 207A
\newsymbol\varkappa 207B
\newsymbol\Bbbk 207C
\newsymbol\hslash 207D
\undefine\hbar
\newsymbol\hbar 207E
\newsymbol\backepsilon 237F
\catcode`\@=\csname pre amssym.tex at\endcsname

\newif{\ifcomentarios}
\comentariosfalse
\renewcommand{\theequation}{\thesection.\arabic{equation}}
\newtheorem{theorem}{Theorem}
\newtheorem{lemma}[theorem]{Lemma}
\newtheorem{definition}[theorem]{Definition}
\newtheorem{proposition}[theorem]{Proposition}
\newtheorem{corollary}[theorem]{Corollary}

\newtheorem{conjecture}[theorem]{Conjecture}

\newcommand{\zerarcounters}
{
\setcounter{equation}{0}
\setcounter{theorem}{0}
}

\newcommand{\Fullbox}{\hfill{\rule{2.5mm}{2.5mm}}}
\newcommand{\EndofStatement}{\hfill\Box}

\newcommand{{\fa}}{(\phi_1\times\cdots\times\phi_n )_{in}}
\newcommand{{\fb}}{(\phi_2\times\cdots\times\phi_n )_{in}}
\newcommand{{\fc}}{(\phi_{m+1}\times\cdots\times\phi_n )_{in}}
\newcommand{{\fd}}{(\phi_{n-k+1}\times\cdots\times\phi_n )_{in}}
\newcommand{{\fe}}{(\phi_{n-k}\times\cdots\times\phi_n )_{in}}
\newcommand{{\ff}}{(\phi_{n-k+1}\times\cdots\times\phi_n )_{in}}
\newcommand{{\fg}}{(\phi_{n-k}\times\cdots\times\phi_n )_{in}}
\newcommand{{\fh}}{(\phi_{1}\times\cdots\times\phi_k )_{out}}
\newcommand{{\fz}}{(\phi_{1}\times\cdots\times\phi_m )_{out}}
\newcommand{{\fl}}{(\phi_2\times\cdots\times\phi_n )_{out}}
\newcommand{{\fm}}{\phi_{1,\ldots , n}}
\newcommand{{\fn}}{\phi_{2,\ldots , n}}

\newcommand{\undx}{\underline{x}}

\newcommand{\undV}{\underline{V}}

\newcommand{\A}{{\cal A}}

\newcommand{\calA}{{\cal A}}
\newcommand{\calB}{{\cal B}}
\newcommand{\calC}{{\cal C}}
\newcommand{\calD}{{\cal D}}
\newcommand{\calE}{{\cal E}}
\newcommand{\calF}{{\cal F}}
\newcommand{\calG}{{\cal G}}
\newcommand{\calH}{{\cal H}}

\newcommand{\calL}{{\cal L}}
\newcommand{\calM}{{\cal M}}

\newcommand{\calQ}{{\cal Q}}

\newcommand{\calS}{{\cal S}}

\newcommand{\calU}{{\cal U}}

\newcommand{\be}{\begin{equation}}
\newcommand{\ee}{\end{equation}}
\newcommand{\bma}{\begin{displaymath}}
\newcommand{\ema}{\end{displaymath}}
\newcommand{\bc}{\begin{center}}
\newcommand{\ec}{\end{center}}

\newcommand{\om}{{\omega }}

\newcommand{\Om}{\Omega}

\newcommand{\al}{{\alpha}}

\newcommand{\la}{{\langle }}
\newcommand{\ra}{{\rangle }}

\newcommand{\bege}{\beta_{g}}
\newcommand{\beha}{\beta_{h}}
\newcommand{\gage}{\gamma_{g}}
\newcommand{\gaha}{\gamma_{h}}
\newcommand{\bear}{\begin{eqnarray}}
\newcommand{\eear}{\end{eqnarray}}

\newcommand{\supp}{\, \mbox{\small{supp}} \, }
\newcommand{\uflex}
{{\scriptstyle {\raise 9pt\hbox{$\backslash$}\,\!\!\!\!\!\Bigg\vert}}}

\newcommand{\Z}{\Bbb Z}
\newcommand{\N}{\Bbb N}
\newcommand{\R}{\Bbb R}
\newcommand{\C}{\Bbb C}
\newcommand{\UM}{1\! \rm{l}}

\newcommand{\ncm}{\newcommand}
\ncm{\rncm}{\renewcommand}
\ncm{\id}{{\bf 1}}
\ncm{\beq}{\begin{equation}}
\ncm{\eeq}{\end{equation}}
\ncm{\bea}{\begin{eqnarray}}
\ncm{\beanon}{\begin{eqnarray*}}
\ncm{\eea}{\end{eqnarray}}
\ncm{\eeanon}{\end{eqnarray*}}
\ncm{\fns}{\footnotesize}
\ncm{\setc}[1]{\setcounter{equation}{#1}}
\newcounter{eqnr}
\newenvironment{eqnarrayabc}{\stepcounter{equation}
  \setcounter{eqnr}{\value{equation}}\setc{0}
  \rncm{\theequation}{\thesection.\arabic{eqnr}\alph{equation}}
  \begin{eqnarray}}{\end{eqnarray}\setc{\value{eqnr}}}
\ncm{\bealph}{\begin{eqnarrayabc}}
\ncm{\eealph}{\end{eqnarrayabc}}
\ncm{\eqboxabc}[3]{\newline\parbox[t]{1.5cm}{#1}\hfill
  \parbox[b]{12cm}{\begin{eqnarray*} #3\end{eqnarray*}}\hfill
   \parbox[b]{1.5cm}{\vspace{-0.0cm}\begin{eqnarrayabc}#2\end{eqnarrayabc}}\newline}
\ncm{\eqbox}[2]{\newline\parbox{1.5cm}{#1}\hfill
  \parbox{12cm}{\beanon #2\eeanon}\hfill
  \parbox{1cm}{\bea\eea}\newline}
\ncm{\nr}[1]{\parbox{1cm}{\begin{eqnarrayabc}#1\end{eqnarrayabc}}\\}

\ncm{\kal}[1]{\mbox{$\cal #1 $}}
\ncm{\mrk}[1]{\!\!\! #1 \!\!\!}
\ncm{\qed}{\hspace*{0.4cm}\rule{0.24cm}{0.24cm}}
\ncm{\mbold}[1]{\mbox{\boldmath $ #1 $}}
\ncm{\bm}{\mbold}
\ncm{\str}{\stackrel}
\ncm{\sub}{\subset}
\ncm{\e}{\varepsilon}
\ncm{\ka}{\kappa}
\ncm{\r}{\rho}
\ncm{\inputc}[1]{\begin{center}\input{#1}\end{center}}
\ncm{\lto}{\longrightarrow}
\ncm{\x}{\times}
\ncm{\bmm}{\bm{\cal M}}
\ncm{\cp}{{\bf P}}
\ncm{\bfp}{{\bf P}}
\ncm{\bmi}{\bm{i}}
\ncm{\bmom}{\bm{\om}}
\ncm{\bmOm}{\bm{\Om}}
\ncm{\res}{\restriction}
\ncm{\bmL}{\bm{\cal L}}
\ncm{\bmell}{\bm{\ell}}
\ncm{\bmE}{\bm{\cal E}}
\ncm{\bme}{\bm{e}}
\ncm{\bmpi}{\bm{\pi}}
\ncm{\bmr}{\bm{r}}
\ncm{\bmsigma}{\bm{\sigma}}
\ncm{\wt}{\widetilde}

\newcounter{label}[equation]

\newcounter{nroflabels}

\ncm{\Nr}[1]{\label{#1}\setcounter{label}{#1}\stepcounter{nroflabels}}

\ncm{\Ende}{\fbox{number of labelled equations:\ \arabic{nroflabels}}
\end{document}}

\ncm{\bra}{\langle}

\ncm{\ket}{\rangle}

\ncm{\ba}{\begin{array}}

\ncm{\ea}{\end{array}}

\ncm{\ox}{\otimes}

\ncm{\vfi}{\varphi}

\ncm{\Ad}{\mbox{Ad\,}}

\ncm{\End}{\mbox{End\,}}

\ncm{\Aut}{\mbox{Aut\,}}

\ncm{\bsn}{\bigskip\noindent}

\ncm{\1}{{\Bbb 1}}

\rncm{\AA}{{\frak A}}

\ncm{\BB}{{\frak B}}

\ncm{\FF}{{\frak F}}

\ncm{\GG}{{\frak G}}

\ncm{\JJ}{{\frak J}}

\ncm{\del}{\delta}

\ncm{\ga}{\gamma}

\ncm{\ep}{\epsilon}

\ncm{\La}{\Lambda}

\ncm{\cc}{\calC}

\ncm{\ccp}{\cc^p}

\ncm{\lam}{\lambda}

\ncm{\ve}{\varepsilon}

\ncm{\ttt}{{\bf t}}

\ncm{\AAA}{\hat\AA}

\begin{document}



\begin{center}
  \begin{Large} {\bf
	  Dyonic Sectors and Intertwiner Connections \\[.3cm]
	in 2+1-dimensional Lattice
	${{\sf Z}\kern-0.35em {\sf Z}}_{N} $-Higgs Models.}
  \end{Large}
\end{center}

\vspace{0.5cm}

\begin{center}

{\sc Jo\~ao C. A. Barata}\footnote{
E-mail: jbarata{@}fma.if.usp.br
}
\\
Instituto de
F\'{\i}sica da Universidade de S\~ao Paulo.
\\
P.O. Box 66318,  S\~ao Paulo 05315 970, SP, Brasil.
\\
and
\\
{\sc Florian Nill}\footnote{
 E-mail: nill{@}omega.physik.fu-berlin.de }
\\
Institut f\"ur Theoretische Physik der Freien Universit\"at Berlin.
\\
Arnimallee 14, Berlin 14195, Germany.

\end{center}


\vspace{.5cm}

{\bf Abstract:}  We construct dyonic states $\om_\r$ in
2+1-dimensional lattice $\Z_N$-Higgs models, i.e. states which
are both, electrically and magnetically charged. These states
are parametrized by $\r=(\varepsilon,\,\mu)$, where
$\varepsilon$ and $\mu$ are $\Z_N$-valued electric and magnetic
charge distributions, respectively, living on the spatial
lattice $\Z^2$. The associated Hilbert spaces $\calH_\r$ carry
charged representations $\pi_\r$ of the observable algebra
$\AA$, the global transfer matrix ${\bf t}$ and a unitary
implementation of the
group $\Z^2$ of spatial lattice translations. We prove that for
coinciding total charges
$\displaystyle q_\r = \left(\sum_x \varepsilon(x),\,
\sum_p\mu(p)\right)\in \Z_N\x\Z_N $ these representations are
dynamically equivalent and we construct a local intertwiner
connection $U(\Gamma): \calH_\r\to\calH_{\r'}$, where
$\Gamma:\r\to\r'$ is a path in the space of charge
distributions $\calD_q=\{\r:\, q_\r=q\}$.  The
holonomy of this connection is given by $\Z_N$-valued phases.
This will be the starting point for a construction of
scattering states with anyon statistics in a subsequent paper.


\newpage

\tableofcontents

\newpage


\section{Introduction}
\zerarcounters

In this paper we continue our project
initiated in \cite{FlorianBarataI} of a constructive analysis
of states with anyonic statistics in a 2+1 dimensional lattice
gauge theory. We investigate a
general class of models with the discrete Abelian gauge group
$\Z_N $
for arbitrary $N \in \N$, $N \geq 2$, and with discrete Higgs
fields.
The vacuum expectations in this theory are represented by  classical
expectations of an (euclidean) statistical mechanics model
given by the thermodynamic limit of finite volume
expectations of the form
\be
  \la B_{cl } \ra_\Lambda :=
\frac{
\int d\al d\varphi \, \, B_{cl} (\al, \varphi ) \, e^{-S_\Lambda}
}{
\int d\al d\varphi  \, \, e^{-S_\Lambda}
} ,
\label{EuCliDFoRm}
\ee
with a generalized Wilson action:
\be
S_\Lambda(\varphi , \, \al)  :=
\sum_p S_g (d\al(p)) + \sum_b S_h (d\varphi (b) - \al (b))\ .
\label{ACAOum}
\ee
Here $\varphi $ and $ \al$ are $ \Z_N$-valued Higgs and gauge fields,
respectively, on a euclidean space-time lattice $ \Z^3$. Hence $ \varphi$
lives on sites, $ \al$ lives on bonds, $ d$ denotes the lattice
exterior derivative and the above sums go over all elementary
positively oriented bonds $ b$ and plaquettes $ p$ in a finite
space-time volume $\Lambda$ of our lattice.
$B_{cl}$ is some classical observable, i.e.
a gauge invariant function of the gauge field $\al$ and of the Higgs field
$ \varphi$ with finite
support. Here, gauge invariance means invariance under
the simultaneous transformations  $ \varphi \to \varphi + \lambda$ and
$\al \to \al + d\lambda$ for arbitrary
$ \lambda: \, \Z^3 \mapsto \Z_N$, with finite support.
Above, the integrations over $ \al$ and
$ \varphi$ are actually finite sums, since these variables are
discrete (to be precise, the discrete Haar measure on $\Z_N$ is employed).

In order to have charge conjugation symmetry and reflection positivity
(i.e. a positive transfer matrix)
the actions $ S_g$ and $ S_h$ will be chosen as even functions on
$ \Z_N \equiv \{ 0, \ldots , N - 1\}$
taking their minimal value at $ 0 \in \Z_N$. Thus we have a general
Fourier expansion
\be
    S_{g/h} (n) = - \frac{1}{\sqrt{N}}
\sum_{m=0}^{N-1} \beta_{g/h} (m)
\cos
\left(
  \frac{2 \pi m n}{N}
\right) ,
\label{Essesebetas}
\ee
$ n \in \Z_N$,
where we call $ \beta_g (m)$ and $ \beta_h (m)$, $ m \in \Z_N$, the
gauge and Higgs coupling constants, respectively\footnote{
The values of $\bege (0)$ and $\beha (0)$
only determine additive constants to the
action and will be fixed by convenient normalization conditions.
}.
We also require $ \beta_{g/h}(m) = \beta_{g/h}(N-m)$, i.e.
$ \beta_{g/h}$ and $ - S_{g/h}$ are in fact Fourier transforms of each
other. This model describes a $\Z_N$-Higgs model where the radial degree of
freedom of the Higgs field is frozen, i.e. $|\phi |=1$ and
$\phi (x) = e^{\frac{2\pi i }{N} \varphi (x)}$.

We will be interested in the so-called ``free charge phase'' of
this model, which, roughly speaking, is obtained whenever, for all
$0 \neq n \in \Z_N $,
\bear
 S_g (n) - S_g (0)  & \geq c_g      & \gg 0 ,
\\
 S_h (n) - S_h (0)  & \leq c_h^{-1} & \ll 1 ,
\eear
for large enough positive constants $ c_g$ and $ c_h$. In this region
convergent polymer and cluster expansions are available and have been
analyzed in detail in \cite{FlorianBarataI}, see also Appendix
\ref{SegundoApendicehlkjhglkdgdg}
for a short review.

The first analysis of the structure of the charged states in this
phase had been performed for the case of the group $\Z_2$ by
Fredenhagen and Marcu in \cite{FredMarcu}.  In that work it had been
shown that electrically charged states exist in $d+1$ dimensions, $d
\geq 2$ in the ``free charge phase'' of the model. The ideas employed
by the authors involved a wide combination of methods from Algebraic
Quantum Field Theory and Classical Statistical Mechanics. Later on the
existence of electrically charged particles in the same model had been
shown in \cite{BarFredenhagen} and the existence of multi-particle
scattering states of these particles had been proven in
\cite{BarataII}, combining methods and results of \cite{FredMarcu} and
of \cite{teilchen}.

In \cite{FlorianBarataI} we extended some of these results to the
$\Z_N$-Higgs model mentioned above and showed, after previous results
of Gaebler on the $\Z_2$ case \cite{Gaebler}, the existence of
magnetically charged states in $2+1$ dimensions. In
\cite{FlorianBarataI} we also proved the existence of electrically and
of magnetically charged particles in this model.

Our intention here is to show the existence of dyonic states
$\omega_\rho$ in the ``free charge phase'' of our $\Z_N$-Higgs model,
i.e. states carrying simultaneously electric and magnetic charges,
$\rho = (\varepsilon,\mu)$. This had been performed in the $\Z_2$ case
in \cite{Gaebler}.  We construct the associated charged
representations of the observable algebra ${\frak A}$ as the
GNS-triples $(\pi_\r, \calH_\r, \Om_\r)$ obtained from $\om_\r$.
These representations fall into equivalence classes labeled by the
total charges $\displaystyle q_\r = \left(
  \sum_x\varepsilon(x),\sum_p\mu(p)\right)\in\Z_N\x\Z_N$.  We show
that for each choice of $q\in\Z_N\x\Z_N$ the {\em state bundle}
$$
\calB_q \; := \; \bigcup_{\r\in\calD_q} (\r,\calH_\r)
$$
over the discrete base space $\calD_q :=\{\r:\,q_\r=q\}$ is
equipped with a non-flat connection, i.e. a collection of unitary
parallel transporters $U_{\r',\r}:\calH_\r\to\calH_{\r'}$ depending on
paths $\Gamma$ from $\r$ to $\r'$ in $\calD_q$ such that $\pi_{\r'} =
\Ad U_{\r',\r}\circ\pi_\r$.  The holonomy of this connection is given
by $\Z_N$-valued phases which appear as winding numbers between
``electric'' Wilson loops and ``magnetic'' vortex loops in the
euclidean functional integral picture.

In an upcoming paper \cite{FBIII} this construction will be the
starting point for an analysis of the anyonic statistics of scattering
states in these models.

At this point we should also mention the previous work of J.
Fr\"ohlich and P. A. Marchetti on the construction of dyonic and
anyonic states in the framework of euclidean lattice field theories
(\cite{FrMarch1}, \cite{FrMarch2} and \cite{FrMarch3}).  In their
approach the analogue of our Hilbert spaces $\calH_\rho$ are obtained
by Osterwalder-Schrader reconstruction methods which makes it
difficult to discuss the representation theoretic background of the
observable algebra.

In difference with their approach we work with the Hamiltonian
description. In particular, we construct our states as functionals on
the quasi-local observable algebra generated by the time-zero fields.
As usual, in this approach the euclidean description reappears in the
form of local transfer matrices $T_V$, whose ground state expectation
values are given by functional integrals of the type
(\ref{EuCliDFoRm}). Still, we interpret and motivate our constructions
mostly algebraically and consider the functional integral techniques
only as a technical tool.

We now describe the plan of this paper in some more detail.

In Section 2 we introduce our basic setting.  We follow the standard
canonical quantization prescription of gauge theories in $A_0=0$ gauge
to define the local algebras generated by the time-zero fields.  We
then define an ``euclidean dynamics'' in terms of local transfer
matrices whose ground states give rise to finite volume euclidean
functional integral expectations with actions
(\ref{ACAOum})-(\ref{Essesebetas}).  We also review the notion of {\em
  external} charges (related to a violation of Gauss' law) in order to
distinguish them from the {\em dynamical} charges (``superselection
sectors'') that we are interested in this work.

Section 3 is devoted to the construction of dyonic sectors.  We start
with generalizing the Fredenhagen-Marcu prescription to obtain dyonic
states $\om_\r$ on our observable algebra $\AA$. We then construct the
associated GNS-representations $(\pi_\r,\calH_\r,\Om_\r)$ of $\AA$ and
implement the euclidean dynamics by a global transfer matrix on
$\calH_\r$. We show that as representations of the ``dynamic closure''
$\hat\AA\supset\AA$ (i.e. the $*$-algebra generated by $\AA$ and the
global transfer matrix) these representations are irreducible and
pairwise equivalent provided their total charges $q_\r\in\Z_N\x\Z_N$
coincide. (We also conjecture that they are dynamically inequivalent,
if their total charges disagree).  Finally we show that the infimum of
the energy spectrum in the dyonic sectors is uniquely fixed by
requiring charge conjugation symmetry and cluster properties of
correlation functions for infinite space-like separation.

In Section 4 we construct an intertwiner algebra on $\Bbb H_q =
\oplus_{\r\in\calD_q}\calH_\r$ by defining electric and magnetic
``charge transporters'' $\calE_q(b),\,\calM_q(b)\in\calB(\Bbb H_q)$
living on bonds $b$ in $\Z^2$ and fulfilling local Weyl commutation
relations. In terms of these intertwiners we obtain a unitary
connection $U(\Gamma):\calH_\r\to\calH_{\r'}$ intertwining $\pi_\r$
and $\pi_{\r'}$ for any path $\Gamma:\r\to\r'$ in $\calD_q$. The
holonomy of this connection is given by $\Z_N$-valued phases.  We
conclude by applying our connection to construct a unitary
implementation of the translation group in the dyonic sectors.

We remark at this point that we do not touch the question of the
existence of dyonic {\em particles} in these models, i.e.  of
particles in the dyonic sectors carrying simultaneously electric and
magnetic charges.  To study the existence of such particles requires
adaptation of the known Bethe-Salpeter kernels methods for situations
involving charged particles in lattice models. This will be performed
elsewhere. Clearly, if these particles exist they should be expected
to show anyonic statistics among themselves.

A good part of the methods and results used here has been extracted
from \cite{FlorianBarataI} and we will often refer to this paper when
necessary. In particular, we will not repeat the proof of the
convergence of the polymer expansion we are going to use since this
point has been discussed in detail in \cite{FlorianBarataI}, see
however Appendix A for a short review.  In fact, although polymer
expansions are the main technical tool of this work (as well as of all
other works on $\Z_N$-Higgs models cited above), our aim here is to
formulate theorems and present results in a way which can be followed
without any detailed knowledge on cluster expansions.  Following this
strategy, we abandon all statistical mechanics aspects of our proofs
to Appendix B and reserve the main body of this work to exploit
algebraic and quantum field theoretical concepts.

{\em Remarks on the notation}.  Due to a different focus our notation
will differ in many points from that of \cite{FlorianBarataI}. We will
change our notation according to our needs of emphasis and clarity.
The symbol $\Box$ indicates ``end of statement'' and the symbol
$\rule{2.5mm}{2.5mm}$ indicates ``end of proof''.  Products of
operators run from the left to the right, i.e., $\prod_{a=1}^{n}A_a$
means $A_1 \cdots A_n$.  For an invertible operator $B$, $\mbox{Ad}\,
B$ denotes the automorphism $B \cdot B^{-1}$. If $\frak A \subset
\calB (\calH)$ is an algebra acting on a Hilbert space $\calH$ then we
denote by ${\frak A} '$ the {\em commutant } of $\frak A$, i.e. the
set of all operators of $ \calB (\calH )$ which commute with all
elements of $\frak A$.  Here $\calB (\calH)$ is the algebra of all
bounded operators acting on $\calH$.

{\em Acknowledgments. } We would like to thank K. Fredenhagen
for stimulating interest and several useful discussions.

\section{The Basic Setting}
\zerarcounters

We will always consider the lattice $\Z^d$, $d =2$, $3$ as a
chain complex and denote by $(\Z^d)_p$ the elementary
positively oriented p-cells in $\Z^d$. We also use the standard terminology
``sites'', ``bonds'' and ``plaquettes'' for 0-, 1- and 2-cells,
respectively.
By a (finite) volume $V\subset\Z^d$ we mean the closed chain sub-complex
generated by a (finite) union of
elementary d-cells in $(\Z^d)_d$. We denote by $V_p$
the set of elementary oriented p-cells in $V$ where, by
definition, a p-cell is contained in $V_p\subset(\Z^d)_p$ if and
only if it lies in the boundary of some (p+1)-cell contained in
$V_{p+1}$. We denote by $\cc_p(V)\equiv\Z V_p$
the set of p-chains in V and by
$\ccp(V)\equiv\ccp(V,\Z_N)$ the set of $\Z_N$-valued
cochains with support in $V$ (i.e. group homomorphisms
$\al:\,\cc_p(V)\to\Z_N$). As usual we identify $\ccp(V)$ with the
group of $\Z_N$-valued functions on $(\Z^d)_p$ with support in
$V_p$. Hence, for $V\subset W$ we have the natural inclusion
$\ccp(V)\subset\ccp(W)$. We also denote $\ccp := \ccp(\Z^d)$ and define
$\ccp_{loc}\subset\ccp$ as the set of p-cochains with finite
support. Often we will identify an elementary p-cell
$c\in(\Z^d)_p$ with its characteristic p-cochain (i.e. taking the
value $1\in\Z_N$ on $c$ and $0\in\Z_N$ else).

Considered as a finite Abelian group $\ccp(V)\cong\Z_N^{|V_p|}$ is
self-dual for all finite $V$, the pairing $\ccp\x\ccp\to U(1)$
being given by the homomorphism
\be
(\al,\,\beta)\mapsto e^{i\bra\al,\,\beta\ket}
:= \exp \left( \frac{2\pi i}{N}\sum_{c\in V_p}\al(c)\beta(c) \right)
\label{pair}
\ee
We denote by $d:\,\ccp\to\cc^{p+1}$ and
$d^*:\,\ccp\to\cc^{p-1}$ the exterior derivative and its
adjoint, such that
$$
e^{i\la\al,\,d\beta\ra} = e^{i\la d^*\al,\,\beta\ra}
$$
for all $\al\in\ccp_{loc}$ and all $\beta\in\cc_{loc}^{p-1}$.

In the main body of this paper we will be working with
``time-zero'' fields, i.e. cochains defined on the {\em spatial}
lattice $\Z^2$.
The translation to the euclidean functional integral formalism
will bring us to a {\em space-time} lattice $\Z^3$.
To unload the notation we will use the same symbols for both
pictures as long as the meaning becomes obvious from the
context.
Hence, in both pictures $\vfi\in\cc^0$ will denote the Higgs
field, $\al\in\cc^1$ will denote the gauge field and a gauge
transformation consists of a mapping
$(\vfi,\,\al)\mapsto(\vfi+\lambda,\,\al+d\lambda)$ with
$\lambda\in\cc^0_{loc}$.

\setcounter{theorem}{0}
\subsection{The Local Algebras}

As is well known, a $d+1$-dimensional lattice system described
by (\ref{EuCliDFoRm},\ref{ACAOum}) can  typically also
be described as a {\em quantum spin system},
using the transfer matrix formalism.
By this we mean an operator algebra living on a $ d$-dimensional
spatial lattice, together with discrete ``euclidean time'' translations
given by $ e^{-t H}$, $ t \in \Z$, where $ T \equiv e^{-H}$ is the transfer
matrix.
In this formulation expectations like
(\ref{EuCliDFoRm}) represent the vacuum or ground state of the
``euclidean'' dynamics defined by the transfer matrix.

We have described in detail the quantum spin system of our model in
\cite{FlorianBarataI} (see also \cite{FredMarcu}).
It corresponds to the Weyl form of the usual canonical
quantization prescription in
temporal ($\alpha_0=0$) gauge.
Let us recall here its main ingredients.

On the spatial lattice $\Z^2$ we
introduce the local algebra of time-zero Higgs and gauge fields in the
following way.
To each $x\in{(\Z^2)_0}$ we associate the {unitary} $\Z_N$-fields
$P_H(x )$ and $Q_H(x )$ and to each $b \in {(\Z^2)_1}$
we associate the {unitary} $\Z_N$-fields
$P_G(b )$ and $Q_G(b )$
(the subscripts $G$ and $H$ stand for ``gauge'' and ``Higgs'', respectively)
satisfying the relations:
\be
P_H (x )^{N} =
Q_H (x )^{N} =
P_G (b )^{N} =
Q_G (b )^{N} = \UM ,
\ee
and the $\Z_N$-Weyl algebra relations
\be
P_G(\alpha )Q_G(\beta )=
e^{- i\la\alpha , \, \beta   \ra}
Q_G(\beta )P_G(\alpha ) ,
\ee
\be
P_H(\gamma )Q_H(\delta )=
e^{-i\la\gamma , \, \delta   \ra}
Q_H(\delta )P_H(\gamma )  ,
\ee
where $\alpha $, $\beta\in\cc_{loc}^1$ and
$\gamma $, $\delta\in\cc_{loc}^0$ play the r\^ole of test
functions, i.e.
$
\displaystyle
P_H (\gamma):=
\prod_{x\in {(\Z^2)_0}}P_H (x)^{\gamma(x)}
$ etc.
Operators localized at different sites and bonds commute
and the $G$-operators commute with
the $H$-operators.

We denote $[\delta Q_H](\alpha):= Q_H (d^*\alpha)$,
$[\delta^* P_G] (\beta ):= P_G (d \beta )$ etc.,
\footnote{There was a misprint in these definitions in \cite{FlorianBarataI}.}
where $d$ is the exterior derivative on cochains and $d^*$ is its adjoint.

We will realize these operators by attaching to each lattice point $ x$
a Hilbert space $ \calH_{x}$ and to each
lattice bond $ b$
a Hilbert space $ \calH_{b}$, where
${\cal H}_{x} \cong {\cal H}_{b} \cong {\Bbb C}^N$.
The operators
$Q_H (x )$,
$P_H (x )$,
$Q_G (b )$ and
$P_G (b )$
are given on
$
{\cal H}_{x}
$, and
$
{\cal H}_{b}
$,
respectively, as matrices with matrix elements:
\bear
P_H (x )_{a,\, b}  = P_G (b )_{a,\, b} & =  &
\delta_{a, \, b+1  (\mbox{\footnotesize{mod}}\, N) } \; \; \mbox{ and }
\\
Q_H (x )_{a,\, b}  = Q_G (b )_{a,\, b} & =  &
\delta_{a,\, b} e^{\frac{2\pi i}{N} a}  ,
\eear
for $a$, $b\in \{0 , \ldots , N-1 \}$.

The operators $Q_H$ and $Q_G$ have to be interpreted as the $\Z_N$
versions of the Higgs field and gauge field, respectively:
$Q_H (x )=e^{\frac{2\pi i}{N}\varphi (x )}$,
$Q_G (x )=e^{\frac{2\pi i}{N} \al (b )}$, with
$\varphi$ and $\al$ taking values in $\Z_N$.
The operators $P_H$ and $P_G$ are their respective
canonically conjugated ``exponentiated momenta'', i.e. shift operators
by one $\Z_N$-unit. Hence these operators indeed provide the
Weyl form of the canonical quantization in $\alpha_0=0$ gauge,
see also equation (\ref{canquant}) below.

We denote by ${\frak F}_{loc}$ the $*$-algebra generated by these operators.
Denoting by ${\frak F}(V)$ the $\mbox{C}^{*}$-sub-algebra
generated by
$Q_H (x )$,
$P_H (x )$,
$Q_G (b )$ and
$P_G (b )$  for $x\in V_0$, $b \in V_1$, $V\subset \Z^2$ finite,
one has ${\frak F}_{loc}=\cup_{|V | < \infty}{\frak F}(V)$.
The algebra ${\frak F}(V)$ acts on
$
{\cal H}_{V}:=
\otimes_{x \in V_0  }
{\cal H}_{x}
\otimes_{b \in V_1}
{\cal H}_{b}
$.
We will denote by ${\frak F}$ the unique $\mbox{C}^{*}$-algebra generated by
${\frak F}_{loc}$. Without mentioning explicitly we will
frequently use that by continuity states $\om$ on $\FF$ or $*$-automorphisms
$\gamma$ of $\FF$ are uniquely determined by their definition
on $\FF_{loc}$.

Let $ \calS$ denote the group of spatial lattice translations by
$ a\in \Z^2$ and rotations by multiples of $ \pi /2 $. $ \calS$ acts
naturally on $\frak F$ as a group of $ *$-automorphisms $ \tau_g$,
$ g\in \calS$, given by $ \tau_g (Q_H (x ))= Q_H (gx )$ etc.

We also have the group of local (time independent) gauge
transformations
$\calG\equiv\cc_{loc}^0$
acting as $*$-automorphisms on $\frak F$ by
\be
Q_H (\alpha ) \mapsto e^{-i \la \alpha , \, \lambda \ra } Q_H
(\alpha )\ ,
\qquad
Q_G (\gamma ) \mapsto e^{-i \la \gamma , \, d\lambda  \ra} Q_G (\gamma ) ,
\ee
and leaving all operators $ P_H (x )$ and $P_G (b )$
invariant.
These gauge transformations are implemented by the unitaries
$ G (\lambda ) := \prod_{x } G (x )^{\lambda (x )}$,
where
\footnote{Here we replace our notation ${\calQ}(x)$ of
\cite{FlorianBarataI} by the more
common one $G(x)$ and also correct a misprint in equation. (2.9) of
\cite{FlorianBarataI}.}
\be
   G(x):=P_H(x )\left[\delta^{*}P_G \right] (x ) .
\ee
Note that
$G(x )^{*} = G(x )^{-1} = G(x )^{N-1}$.
The operator $G(x )$ is the generator
of a $\Z_N$ gauge transformation at the point $x$, as one can
easily checks. It can be interpreted as the lattice analog of
$\exp \left( - 2 \pi i  (\mbox{div } {\Bbb E}  - \rho) / N \right)$.

We denote the gauge group algebra $\GG\subset\FF$ as the Abelian
$C^*$-sub-algebra generated by $\{G(x)\,|\,x\in\Z^2\}$ and put
$\GG_{loc} := \GG\cap\FF_{loc}$ and $\GG(V) := \GG\cap\FF(V)$.

The algebra of observables ${\frak A}$ is defined as the gauge invariant
sub-algebra of ${\frak F}$ :
\be
   {\frak A}:=
\{A \in {\frak F}: \; G(x )AG(x )^*=A\;
\mbox{for all } x \in \Z^2  \} ,
\ee
i.e. the commutant of $\GG$ in $\FF$.
We call ${\frak A}(V ) := {\frak F}(V)\cap {\frak A}$
and $ {\frak A}_{loc} := \cup_{V } {\frak A} (V )$.
Then $ {\frak A}_{loc}$ is
the norm dense sub-algebra generated by $P_G (b )$,
$P_H (x)$ and
\be
Q_{GH}(b):=Q_G(b)\left[\delta Q_H\right](b)^*
\label{Q_GH}
\ee
for all $x\in(\Z^2)_0$ and $b\in(\Z^2)_1$.

We now provide a convenient ``ket vector'' notation for the local
Hilbert spaces $\calH_V$. Identifying
$\calH_x\cong\calH_b\cong\C^N\cong\ell^2(\Z_N)$
we may naturally denote ON-basis elements of $\calH_{x/b}$ by
$|a\ket,\ a\in\Z_N$, i.e. the characteristic functions on
$\Z_N$. Correspondingly, ON-basis elements of $\calH_V$ are
labeled by
$$
|\varphi,\al\ket = \bigotimes_{ x\in V_0, b\in V_1}\,
\left( |\varphi(x)\ket\ox|\al(b)\ket\right) ,
$$
where $(\varphi,\al)$ run through all ``classical configurations'',
i.e. $\Z_N$-valued 0- and 1-cochains, respectively, with
support in V.

With this notation the representation 
of our local field algebras $\calF(V)$ is immediately recognized as
the Weyl form of the canonical quantization in $\al_0=0$ gauge, i.e.
\be
\ba{rcl}
Q_H(\chi) \,|\vfi,\al\ra &=& e^{i\la\vfi,\,\chi\ra} \,|\vfi,\al\ra \\
Q_G(\beta) \,|\vfi,\al\ra &=& e^{i\la \al,\,\beta\ra} \,|\vfi,\al\ra \\
P_H(\chi) \,|\vfi,\al\ra &=& |\vfi +\chi,\al\ra \\
P_G(\beta) \,|\vfi,\al\ra &=& |\vfi ,\,\al+\beta\ra
\ea
\label{canquant}
\ee
where $\chi$ and $\beta$ denote $\Z_N$-valued 0- and 1-cochains,
respectively, with support in V. The action of the gauge transformations on
$\calH_V$ now takes the usual form
$$
G(\lam)\,|\vfi,\al\ket = |\vfi+\lam, \al+d\lam\ket .
$$

It is well known that for a single site $x\in\Z^2$ the operators $Q_H(x)$ and
$P_H(x)$ generate $\End\calH_x$ and similarly for $Q_G(b)$ and $P_G(b)$.
Hence ${\frak F}(V)\cong\End (\calH_V)$ for all $V$ implying ${\frak F}_{loc}$
and ${\frak F}$ to be simple \cite{Mu}. On the other hand, the observable
algebra
${\frak A}$ is not simple since it contains ${\frak G}$ in its center.
However, since eventually we are only interested in representations
$\pi$ of ${\frak A}$ satisfying $\pi(G(x))=\UM$ for all $x\in\Z^2$
(i.e. representations without external charges, see Section
\ref{2.4} below)
we might as well consider
$$
{\frak B} := {\frak A}/{\frak J} .
$$
as our ``essential'' observable algebra, where ${\frak J}\subset{\frak A}$ is the
two-sided closed ideal generated by $\{G(x)-\UM\}_{x\in\Z^2}$.
In the obvious way we also define ${\frak B}(V) = {\frak A}(V)/{\frak J}(V)$
and ${\frak B}_{loc}=\cup_{V\subset\Z^2}{\frak B}(V)$.
Then ${\frak B}$ is the $C^*$-closure of ${\frak B}_{loc}$ and by the same argument
as above ${\frak B}$ is simple since we have

\begin{lemma}
For all finite closed boxes $V\subset\Z^2$ the algebras ${\frak B}(V)$ are
isomorphic to full matrix algebras.
\end{lemma}

\bsn
{\bf Proof:}\
We transform to ``unitary gauge'' by defining a unitary
$U:\,\calH_V\to\calH_{V_0}\ox\calH_{V_1}$ according to
$$
U\,|\vfi, \al\ket := |\vfi\ket\ox|\al-d\vfi\ket\quad.
$$
This gives
$$
U G(\chi) U^{-1}\,(|\vfi\ket\ox|\al\ket) = |\vfi+\chi\ket\ox|\al\ket
$$
and therefore $U{\frak G}(V)U^* = {\frak M}(V)\ox\UM$,
where ${\frak M}(V)$ consists of all shift operators $|\vfi\ket\mapsto
|\vfi+\chi\ket $ on
$\calH_{V_0}=\ell^2(\cc^0(V))\equiv\ell^2(\Z_N^{|V_0|})$.
Now  ${\frak M}(V)$ being maximal Abelian in $\End\calH_{V_0}$ and
${\frak A}(V)$ being the commutant of ${\frak G}(V)$ in ${\frak F}(V)
\equiv\End\calH_V$ we conclude
$$
\ba{rcl}
U{\frak A}(V)U^* &=& ({\frak M}(V)'\cap\End\calH_{V_0})\ox\End\calH_{V_1}\\
&=&{\frak M}(V)\ox\End\calH_{V_1}
\ea
$$
from which ${\frak B}(V)\cong\End\calH_{V_1}$ follows.
\Fullbox

\bsn
Note that the above transformation also gives
$$
\ba{rcl}
UP_G(\beta)U^*\,|\vfi\ket\ox |\al\ket &=& |\vfi\ket\ox
|\al+\beta\ket \\
UQ_{GH}(\beta)U^*\,|\vfi\ket\ox |\al\ket &=&
|\vfi\ket\ox e^{i\la\al,\,\beta\ra}|\al\ket
\ea
$$
showing that $\BB$ is generated by the field operators $P_G + \JJ$ and
$Q_{GH} +\JJ$, the later being defined in (\ref{Q_GH}).

We conclude this subsection with introducing the concept of charge
conjugation.
On ${\frak F}_{loc}$ we define the charge conjugation $ i_C$ to be the
involutive $*$-automorphism given by
\bear
i_C(Q_{H/G}(\chi)) & = & Q_{H/G}(-\chi)
\\
i_C(P_{H/G}(\alpha)) & = & P_{H/G}(-\alpha)
\eear
Then $i_C$ extends to $\FF$ by continuity.
The generators of gauge transformations satisfy
$i_C (G (x))= G(x)^*$ and therefore the observable
algebra is invariant under $ i_C$:
\bma
	     i_C({\frak A}) = {\frak A} .
\ema
Also note that the restriction of $i_C$ to $\FF(V)$ is unitarily
implemented on $\calH_V$ by
$|\vfi,\,\al\ket\to|-\vfi,\,-\al\ket$.

\subsection{Local Transfer Matrices}

Let us now introduce the dynamics
by defining suitable finite volume transfer matrices. Infinite volume
transfer matrices will be defined later.
The form of the transfer matrix is justified by its finite volume
ground state giving rise
to the classical expectation associated with the euclidean
$\Z_N$-Higgs model (\ref{EuCliDFoRm}).

We will consider local transfer matrices
$T_{V}\in{\frak A}_{loc}$, for finite ${V}\subset \Z^2$,
defined by:
\be
T_{V}=e^{A_{V}/2}e^{B_{V}}e^{A_{V}/2}       ,
\label{KPXKO}
\ee
with
\be
 A_{V}:=
\frac{1}{\sqrt{N}}
\sum_{p\in V_2}
\sum_{n=0}^{N-1}\bege (n) \, \left[\delta Q_G\right](p)^{n} +
\frac{1}{\sqrt{N}}
\sum_{b\in V_1}
\sum_{n=0}^{N-1}\beha(n) \, Q_{GH}(b)^{n} ,
\label{defAV}
\ee
and
\be
 B_{V}:=
\frac{1}{\sqrt{N}}
\sum_{b \in V_1}
\sum_{n=0}^{N-1}\gage(n) \,  P_G(b )^{n} +
\frac{1}{\sqrt{N}}
\sum_{x \in V_0}
\sum_{n=0}^{N-1}\gaha(n) \,  P_H(x )^{n}\ .
\label{defBV}
\ee

Here $ \beta_{g/h}$ and $ \gamma_{g/h}$ are even and real
valued functions on
$ \Z_N$, such that $ T_{V} $ is in fact positive,
has positive matrix elements and is invertible.
Moreover, this also implies that $T_V$ is charge conjugation
invariant, $i_C(T_V)=T_V$.

In the basis $|\vfi,\al\ket\in\calH_V$ these operators have matrix elements
given by
\be
e^{A_V}|\vfi,\al\ket
= \exp -\left[\sum_{p\in V_2}S_g(d\al(p))
+\sum_{b\in V_1}S_h(d\vfi(b)-\al(b))\right]
|\vfi,\al\ket
\ee
and
\be
\bra\vfi',\al'|\,e^{B_V}\,|\vfi,\al\ket =
\prod_{b\in V_1} \r_g(\al(b)-\al'(b)) \prod_{x\in V_0} \r_h(\vfi(x)-\vfi'(x))
\ee
where $S_g$ and $S_h$ are the euclidean actions (\ref{Essesebetas}) 
and where $\r_g$ and
$\r_h$ are positive functions on $\Z_N$ determined by
$$
\ba{rcl}
\sum_n \r_g(n)P_G(b)^n &=& \exp\left[\frac{1}{\sqrt{N}}\sum_n
\ga_g(n)P_G(b)^n\right] \\
\sum_n \r_h(n)P_H(x)^n &=& \exp\left[\frac{1}{\sqrt{N}}\sum_n
\ga_h(n)P_H(x)^n\right] \ .
\ea
$$
With a suitable choice of the choice
of the $\ga_{g,\,h}$'s as functions of the
$\beta_{g,\,h}$'s one can arrange (see \cite{FlorianBarataI})
$$
\r_{g,h}= e^{-S_{g,h}}
$$
and hence  $\mbox{Tr }T_{V}^n$ equals the partition function
appearing in the denominator of (\ref{EuCliDFoRm}) for a volume
$\Lambda\equiv V_n:=V \times \{0, \ldots , \, n-1\} \subset \Z^3$
with periodic boundary conditions
in ``time'' direction (free boundary conditions are also possible,
see \cite{FredMarcu}, \cite{FlorianBarataI} or below).

The euclidean dynamics is given by the strong limit of the non-$ *$
automorphisms of $ {\frak F}$ generated by the local transfer
matrices:
\be
     \alpha (A ) := \lim_{V\uparrow \Z^2} \alpha (A)_{V},
\qquad A \in {\frak F} ,
\label{dgjjfhjffhfhhhhlkq}
\ee
where $ \alpha (\cdot )_{V}$ is the automorphism of $ {\frak F}$
defined through
\be
     \alpha (A)_{V} := T_{V} A T_{V}^{-1} ,
\qquad A \in {\frak F}.
\ee
For $ A\in {\frak F}_{loc}$
the limit in (\ref{dgjjfhjffhfhhhhlkq}) is already reached at finite
$ V$.
In \cite{FlorianBarataI} and \cite{FredMarcu} the notation
$ \alpha_i$ in place of $\al$ was used,
due to the interpretation of $ \alpha_i\equiv\al$ as the
generator of a translation by one unit in {\em imaginary} (euclidean) time
direction.
Clearly, $\al$ commutes with the action of lattice
translations and rotations and with the charge conjugation
$i_C$.

In this work we will consider the classical expectations
(\ref{EuCliDFoRm}) in the so called ``free charge phase''
of the $\Z_N$-gauge Higgs model.
Let us introduce the functions $g$, $h: \; \Z_N \mapsto \R_+$ defined by
(see \cite{FlorianBarataI})
\bear
g(n) & := & \exp (-S_g (n)) , \\
h(n) & := & \calF\left[\exp (-S_h) \right] (n) ,
\eear
for $n\in \Z_N$,
where $\calF$ is the Fourier transform of functions on $ \Z_N$.
>From now on we will fix the additive constants $\beta_g (0)$ and
$ \beta_h (0)$ through the conditions
$ g(0)= h(0)=1$.
In \cite{FlorianBarataI} the infinite volume limits of
the classical expectations
of local observables (\ref{EuCliDFoRm}) have been shown to be analytic
functions of the couplings
$g(1), \ldots , g(N-1)$, $ h(1), \ldots , h(N-1)$
whenever
\bear
g_c & := & \mbox{max } \{ |g(1)|, \ldots , |g(N-1)| \} \leq e^{-k_g}
\; \mbox{ and}
\label{RI}
\\
h_c & := & \mbox{max } \{ |h(1)|, \ldots , |h(N-1)| \} \leq e^{-k_h} ,
\label{RII}
\eear
with $k_g$, $k_h > 0$, large enough. For this a
polymer and cluster expansion has been used.
The above region of analyticity (for real couplings)
is contained in the ``free charge region''
of the phase diagram. This phase is characterized by the absence of screening
and confinement.
All results of our present work, specifically those
concerned with the existence and the
properties of the charged states are valid for $ g_c$ and
$ h_c$ sufficiently small.

\subsection{Ground States}
\setcounter{theorem}{0}

In this subsection we recall
the important definition of a ground
state and discuss some of its basic features. We start with
introducing two useful concepts.

First, the adjoint $\gamma^*$ of an automorphism $\gamma$
of a unital $*$-algebra $\frak C$ is defined through
$\gamma^* (A):=(\gamma (A^*))^*$,
$A\in {\frak C}$. We have $ \gamma ^{* *}=\gamma$
and $\gamma$ is a $*$-automorphism
iff $\gamma = \gamma^*$.
For the composition of  automorphisms one has
$(\alpha \circ \beta)^*=\alpha^* \circ \beta^*$
and consequently $\alpha^{* \, -1}=\alpha^{-1 \, *}$. For an invertible
element $A\in {\frak C}$ one also has
$(\mbox{Ad }A )^* = \mbox{Ad } {A^{* \, -1 }} $.
For $\gamma=\alpha$ this in particular implies
$\alpha^*=\alpha^{-1}$.
Finally, if $\omega $ is a $\gamma$-invariant state on $\frak C$
then trivially it is also $\gamma^*$-invariant.

Second, we say that a state $\omega $ on a $*$-algebra $\frak C$
has the cluster property for the automorphism $\gamma$ if, for all
$A$, $B\in {\frak C}$, one has
\be
\lim_{n\to\infty}\omega (A\gamma^n (B))=\omega (A)\omega (B) .
\label{clusprop}
\ee

We now come to a central definition which has first been
introduced in \cite{FredMarcu}.

\begin{definition}
Let $\gamma$ be a (not necessarily $*$-preserving)
automorphism of a unital $*$-algebra $\frak C$.
A state $\omega$
on $\frak C$ is called a ``ground state'' with respect to $\gamma $
and $\frak  C $ if it is $\gamma$-invariant and if
\be
0 \leq  \omega (A^* \gamma (A)) \leq \omega (A^* A), \qquad
\forall A \in {\frak C}  .
\label{JKJKL}
\ee
Actually $ \gamma$-invariance follows from (\ref{JKJKL})
(see, e.g. \cite{FlorianBarataI}). $ \EndofStatement$
\label{DeFgrondSTATE}
\end{definition}

We will motivate this abstract definition below when we
discuss the ground state of the finite volume transfer matrix.

The following Lemma will be very useful for proving that
certain states are
ground states with respect to our euclidean dynamics $\alpha$
(or suitable modifications of $\alpha$ to be introduced in Section 3.1).
This Lemma was already implicitly used in \cite{FredMarcu}.

\begin{lemma}
\label{ClustereGroundstate}
Let $\gamma$ be an automorphism on a $*$-algebra $\frak C$
satisfying $\gamma^* = \gamma^{-1}$ and let $\omega $ be a $\gamma$-invariant
state on $\frak C$ which has the cluster property for $\gamma$.
(Actually one just needs that, for each $A\in {\frak C}$, the sequence
$\omega(A^* \gamma^a (A))$, $a\in \N$, is bounded).
Then, for all $A\in {\frak C}$,
\be
\left|
\omega (A^* \gamma (A))
\right|
\leq \omega (A^* A). 
\ee
\end{lemma}
$\EndofStatement$

The proof is easy.  See \cite{FlorianBarataI}.

We will now exhibit translation invariant ground states of the automorphism
$\alpha$. First, let us explain in this context the heuristic
motivation of Definition \ref{DeFgrondSTATE}.
If $\Omega_{V}\in \calH_V$ is the Frobenius
eigenvector of the finite volume
transfer matrix $T_{V}$ with eigenvalue
 $\|T_{V}\|_{\calH_{V}}$ then, in face of the positivity
of $T_{V}$, the inequalities
\be
0 \leq
\left(
\Omega_{V}, \;
A^* T_{V} A T_{V}^{-1}
\Omega_{V}
\right)
\leq
\left(
\Omega_{V}, \;
A^*  A
\Omega_{V}
\right)
\ee
obviously hold for any $A \in {\frak F} (V)$.

This motivates to consider the relation (\ref{JKJKL}) with
$\gamma = \alpha$ as a characterization of a state
replacing the vector states $\Omega_{V}$ for infinite
volumes. One should notice here that the usual characterization of a
ground state of a quantum spin system as a state $\omega$ for which
$\lim_{V \uparrow \Z^2} \omega(A^* [H_{V}, \, A]) \geq 0$ for all
local $A$ is inadequate for transfer matrix systems, due to the highly
non-local character of the
Hamilton operator $H_{V} \equiv - \ln T_{V}$
(at least in more than two space-time dimensions).

Following \cite{FredMarcu}, a translation invariant ground state for
$\alpha$ with respect to the algebra ${\frak F}$ can be obtained by
\be
\omega_0 (B) =
\lim_{V \uparrow \Z^2}
\lim_{n\to\infty}
\frac{
\mbox{Tr}_{\calH_{V}}
\left(
T_{V}^n
B
T_{V}^n E^0_{V}
\right)
}{
\mbox{Tr}_{\calH_{V}}
\left(
T_{V}^{2n} E^0_{V}
\right)
}
,
\label{rrerqewtqwoieuujjhh}
\ee
for $B \in {\frak F}_{loc}$,
where $E^0_{V}$ is an in principle arbitrary operator with positive
matrix elements.
Before the limits $V \uparrow \Z^2$ and $n\to\infty$
are taken the expression in the right-hand side of (\ref{rrerqewtqwoieuujjhh})
is identical to the classical expectation (\ref{EuCliDFoRm}) in a volume
$V \times \{ -n , \ldots , n\} \subset \Z^3$ of a suitable
classical observable $B_{cl}$ associated with $B$.
A convenient choice for $E^0_{V}$ giving free boundary
conditions in time-direction is (see \cite{FredMarcu})

\be
E_V^0 := e^{A_V/2}\, F_V^0\, e^{A_V/2}
\label{EV0}
\ee
where
\be
F_V^0 := \sum_{(\varphi, \, \al),\, (\varphi ' , \, \al' )}
|\vfi,\al\ket\bra\vfi',\al'|
\label{Xtres}
\ee

Here the sum in (\ref{Xtres}) goes over all classical time-zero
configurations with support in $ V$,
i.e. 0-cochains $ \varphi$, $ \varphi ' \in\cc^0(V)$
and 1-cochains $ \al$, $ \al' \in\cc^1(V)$ .

The existence of the thermodynamic limit in
 (\ref{rrerqewtqwoieuujjhh})
can be established using Griffiths' inequalities or the cluster expansions.
The cluster expansions also provide a
way to prove the translation and rotation invariance of the
limit state $\omega_0$. Another important fact derived from the cluster
expansion is that the restriction of
$\omega_0$ to $\AA$ has the cluster property for
 the automorphism $\alpha$.

One of the most useful aspects of the Definition \ref{DeFgrondSTATE} of
a ground state is the possibility to define infinite volume transfer
matrices. Indeed, if $\omega_0$ is a ground state of $\alpha$ with
respect to ${\frak F}$ and
$
(
\pi_{0}({\frak F}), \;
\Omega_{0} , \;
\calH_{0}
)
$ is the GNS-triple associated with $\om_0$, we define,
following \cite{FredMarcu}, the infinite volume
transfer matrix $T_{0}$
as the element of $\calB (\calH_{0})$ given
on the dense set $ \pi_{0} ({\frak F})\Omega_{0}$ by
\be
	T_{0} \pi_{0} (A)\Omega_{0} :=
\pi_{0} (\alpha (A)) \Omega_{0} ,
\label{T_0}
\ee
$A \in {\frak F}$.
One checks that this is a well-defined positive operator with
$
0 \leq
T_{0}
\leq 1
$.
If moreover $\om_0$ satisfies the cluster property (\ref{clusprop}),
then $\Om_0\in\calH_0$ is the unique (up to a phase)
eigenvector of $T_0$ with eigenvalue 1.
Hence we call $\pi_0$ the {\em vacuum representation} of $(\FF,\,\al)$.

Since $\omega_0 $ is translation invariant we can also define a
unitary representation of the translation group in the vacuum sector through
\be
	U_{0}(x ) \pi_{0} (A)\Omega_{0} :=
\pi_{0} (\tau_{x} (A)) \Omega_{0} ,
\ee
$A \in {\frak F}$, $x \in \Z^2$.
The momenta in the vacuum sector are therefore defined by
$U_{0}(x ) = \exp (i{\Bbb P} \cdot x)$, with
$\mbox{sp } {\Bbb P} \in [-\pi, \, \pi]^2$
and the vacuum $\Om_0$ is also the unique (up to a phase)
translation invariant vector in
$\calH_0$.

Next we remark that $\om_0$ is also charge conjugation
invariant. This can be seen from (\ref{rrerqewtqwoieuujjhh}) by
using $i_C(T_V)=T_V$, $i_C(E_V^0)=E_V^0$ and the fact that
$i_C\restriction\,\FF(V)$ is unitarily implemented on $\calH_V$.
Hence, charge conjugation is implemented as a symmetry of
the vacuum sector by the unitary operator $C_0\in\calB(\calH_0)$ given
on $\pi_0(\FF)\Om_0$ by
\be
C_0\,\pi_0(A)\Om_0 := \pi_0(i_C(A))\Om_0
\label{C_0}
\ee
When constructing charged states $\om_\r$ in Section 3
this will no longer hold, i.e. there we will have $\om_\r\circ i_C =
\om_{-\r}$.

\subsection{External Charges \label{2.4}}

Let $\pi$ be a representation of the field algebra $\FF$ on
some separable Hilbert space $\calH_\pi$ and for
$q\in\cc^0$ let $\calH_\pi^q\subset\calH_\pi$ be
the subspace of vectors $\psi\in\calH_\pi$ satisfying
\be
\pi(G(x))\psi = e^{\frac{2\pi i}{N}q(x)}\psi
\ee
for all $x\in\Z^2$. According to common terminology we call
$\calH_\pi^q$ the subspace with {\em external electric charge}
$q$, since the operators $\pi(G(x))$ implement the gauge
transformations on $\calH_\pi$.
Clearly, since $\AA$ commutes with $\GG$ we have
\be
\pi(\AA)\,\calH_\pi^q\subset \calH_\pi^q\ .
\ee
Moreover, using $Q_H(x)G(x) = e^{2\pi i/N}G(x)Q_H(x)$ for all
$x\in\Z^2$ we have
\be
\pi(Q_H(q'))\calH_\pi^q = \calH_\pi^{q-q'}
\label{extch}
\ee
for all $q'\in\cc_{loc}^0$. In particular, since $Q_H(q)\AA Q_H(q)^* =
\AA$, the representations of $\AA$ on $\calH_\pi^q$ and on
$\calH_\pi^{q-q'}$ are unitarily equivalent for all $q'$ with
finite support.
\footnote{Note, however, that this equivalence does not respect
the dynamics, since $\Ad Q_H(q)$ does not commute with
$\al$.}

Now $\calH_\pi^q$ might be zero for general representations
$\pi$ and general external charge distributions $q$ (e.g. for
the vacuum representation $\pi_0$ and $q$ with infinite
support).
In this work we are  only interested in representations $\pi$
containing a non-trivial subspace $\calH_\pi^0\neq 0$ of {\em zero} external
charge (and therefore, by (\ref{extch}), also  $\calH_\pi^q\neq
0$ for all external
charges with finite support). Moreover,
$\calH_\pi^0$ will always be
cyclic under the action of $\pi(\FF_{loc})$ and therefore we will always have
\footnote{Use that $\FF_{loc} =\oplus_q \AA_{loc} Q_H(q)$
is a grading labeled by
$q\in\cc_{loc}^0$, i.e. the irreducible representations of $\GG$ in
$\FF_{loc}$.}
\be
\calH_\pi = \bigoplus_{q\in\cc_{loc}^0} \calH_\pi^q
\ee
In fact, such a decomposition is always obtained for
GNS-representations $(\pi_\om,\,\Om_\om,\,\calH_\om)$ associated
with states $\om$ on $\FF$, provided $\Om_\om\in\calH_\om^q$
for some $q\in\cc_{loc}^0$ or, equivalently,
\be
\om(F\,G(x)) = e^{2\pi i\,q(x)/N} \,\om(F)
\ee
for all $F\in\FF$ and all $x\in(\Z^2)_0$. In this case we call
$\om$ an {\em eigenstate} (with external charge $q\in\cc_{loc}^0$)
of the gauge group algebra $\GG$.
Note that $\om$ is an eigenstate of $\GG$ with external charge
$q$ if and only if $\om\circ\Ad Q_H(q')$ is an eigenstate with
external charge $q+q'$. Correspondingly,
$\pi_\om(Q_H(q')^*)\,\Om_\om\in\calH_\om^{q+q'}$ will also be a
cyclic vector for $\pi_\om(\FF)$.
Hence, without loss, we may restrict ourselves to eigenstates $\om$
of $\GG$ with zero external charge.
In fact, we will only be studying the
restrictions of
representations $\pi(\AA)$ to $\calH_\pi^0$ as the
``physical'' subspace of $\calH_\pi$, i.e. the subspace on which
``Gauss' Law'', $\pi(G(x))=\UM$, holds.

We emphasize that this notion of external (or ``background'')
charge is not to be confused with the concept of {\em
(dynamically) charged states} and the associated {\em charged
representations} of $\AA$. By this we mean representations of
$\AA$ with zero external charge,
which are inequivalent to the vacuum representation (at least
when extended to a suitable ``dynamic closure''
$\hat\AA\supset\AA$), e.g. by the appearance of different mass
spectra or the absence of a time and space translation invariant
``vacuum'' vector.

By analogy with the terminology of quantum field theory we call an
equivalence class of such charged representations a
{\em superselection sector}.
They are the main interest of our work.
Hence, from now on by charged states we will always mean
dynamically charged states in this later sense.

\section{The Construction of Dyonic Sectors}
\label{TheConstructionofDyonicSectors}
\zerarcounters

In \cite{FlorianBarataI}
with the help of the cluster expansions
(see Appendix \ref{SegundoApendicehlkjhglkdgdg})
we were able to show
the existence of electrically and of magnetically charged
sectors in the ``free charges'' phase of the $\Z_N$-Higgs model.
We also proved the existence of massive charged 1-particle states in these
sectors.

This section is devoted to the construction of states and the
associated sectors which are at the
same time electrically and magnetically charged. Following the common
use we call
them {\em dyonic states}. Hereby we generalize and improve ideas from
\cite{Gaebler},
where such states have been first constructed for the $\Z_2$-Higgs
model.
In \cite{FBIII}
we will continue this analysis by constructing scattering states in
the dyonic sectors and identifying these to constitute an
``anyonic Fock space''
over the above mentioned 1-particle states.

We start in Section 3.1 with reformulating the Fredenhagen-Marcu
construction for charged states $\om_\r$ by defining them as
the thermodynamic limit of ground states of
{\em modified local transfer matrices} $T_V(\r)$. For
$\r=(\varepsilon,\,\mu)$ these modified transfer
matrices correspond to
modified Hamiltonians, where the kinetic term in the Higgs fields
is replaced by
$$
\frac{1}{2} (\pi_H , \, \pi_H)
\rightarrow
 \frac{1}{2} (\pi_H + \varepsilon  , \, \pi_H + \varepsilon)
$$
and the magnetic self energy is replaced by
$$
\frac{1}{2} (d\al , \, d\al)
\rightarrow
\frac{1}{2} (d\al+\mu , \, d\al+\mu)  .
$$
In the functional integral these states are represented by
euclidean expectations in the background of infinitely long
vertical Wilson and vortex lines sitting above sites
$x\in(\Z^2)_0$ and plaquettes $p\in(\Z^2)_2$, respectively, of
the time-zero plane $\Z^2$. Their $\Z_N$-values are given
by the the values of the electric charges $\varepsilon(x)$ and
the magnetic charges $\mu(p)$, respectively.

In Section 3.2 we construct the charged representations
$\pi_\r$ of $\AA$ associated with the states $\om_\r$ and the
global transfer matrices implementing the euclidean dynamics in
$\pi_\r$.

In Section 3.3 we prove that $\pi_\r$ and $\pi_{\r'}$ are
dynamically equivalent whenever their total charges coincide,
$q_\r = q_{\r'}$. We also conjecture that otherwise they are
dynamically inequivalent and give some criteria for a proof.

In Section 3.4 we show that the infimum of the energy spectrum
in the dyonic sectors may be uniquely normalized by imposing
charge conjugation symmetry and the requirement of decaying
interaction energies between two charge distributions
in the limit of infinite spatial separation.

We recall once more that without mentioning explicitly all
results of this section
are valid in the free charge phase, i.e. for $g_c$ and $h_c$ (defined
in (\ref{RI}, \ref{RII})) sufficiently small.

\subsection{Dyonic States}
\setcounter{theorem}{0}

We first recall
the idea behind the Fredenhagen-Marcu (FM) string
operator and its use in the construction of electrically charged
states \cite{FredMarcu}. Starting from the usual method to create localized
electric dipole states by applying a Mandelstam string operator to the
vacuum
\be
    \Omega_{x, \, y} := \phi(x)\phi(y)^* e^{i\int_x^y A_i dz^i} \Om ,
\label{Mandelstam}
\ee
Fredenhagen and Marcu \cite{FredMarcu} proposed a
modification so as to keep the energy
of the dipole state $\Om_{x, \, y}$ finite as $y\to \infty$.
Using our lattice notation the FM-proposal reads
\footnote{By a convenient abuse of notation we often drop the symbol $\pi_0$
when referring to the vacuum representation.}
\be
\Omega_{x,y}^{FM}:= \lim_{n\to\infty}
c_n Q_H(x) Q_H(y)^* T_0^n Q_G(s_{xy}) \Om   .
\label{hpdfzpbsdb}
\ee

Here $s_{x,y}$ is an arbitrary path  connecting
$x$ to $y$
in our spatial lattice, $T_0=e^{-H}$
is the global transfer matrix in the vacuum
sector and $c_n>0$ is a normalization constant to get
$\|\Om_{x,y}^{FM}\|=1$.
Note that $\Om_{x,y}^{FM}$ still lies in the vacuum sector and has zero external
charge.
In a second step one may then send one of the charges to infinity
\footnote{Actually, in \cite{FredMarcu} the two limits,
(\ref{hpdfzpbsdb}) and (\ref{idufzbpsiudb}), were performed simultaneously.}
to obtain dynamically charged states as expectation values on ${\frak A}_{loc}$
\be
\om_x (A) := \lim_{y \to \infty}
\left(
\Omega_{x,y}^{FM}, \, \, A \Omega_{x,y}^{FM}
\right),
\qquad A \in {\frak A}_{loc} .
\label{idufzbpsiudb}
\ee

Note that as a limit of eigenstates of $\GG$ with zero external charge,
$\om_x$ is also an eigenstate of $\GG$ with zero external charge.
Using duality transformations, an analogous procedure for the
construction of magnetically charged states has been given in
\cite{FlorianBarataI}.

In order to be able to discuss dyonic states within a common
formalism, we now pick up an observation of \cite{FlorianBarataI}
to reformulate the
above construction as follows.
First we recall from \cite{FredMarcu} that in our range of couplings we have
\be
	\Omega_{x,y}^{FM} = Q_H(x) Q_H (y)^* \psi_{x, \, y} ,
\label{ddsdoosugjjjjf}
\ee
where $\psi_{x,y}\in\calH_0^{\del_x-\del_y}$
is the unique (up to a phase) ground state vector of the
{\em restricted} global transfer matrix
$T_0\restriction \, \calH_{0}^{\delta_x-\delta_y}$, where
$\calH_{0}^{\delta_x-\delta_y}\sub \calH_{0}$ is the subspace
of an \underline{external} electric charge-anticharge pair sitting at $x$ and
$y$, respectively. In fact, by looking at (\ref{hpdfzpbsdb}) we have
\be
	\psi_{x, \, y}  = s-\lim_{n\to\infty} c_n T_0^n Q_G (s_{x, \, y})
	\Om
\label{IIOiuoiugdfgdfdfdfvdiun}
\ee
and using our cluster expansion it is easy to check that the limit
exists independently of the chosen string $s_{x,y}$ connecting $x$ to $y$.

To avoid the use of external charges (for which we do not have a
 magnetic analogue) we now equivalently reformulate this by saying
that $\Omega_{x,y}^{FM}$ is the unique (up to a phase) ground state
vector of the {\em modified transfer matrix} $T_0(\delta_x-\delta_y)$
defined by
\be
T_0(\delta_x-\delta_y):=
Q_H(x)Q_H(y)^* T_{0} Q_H(y)Q_H(x)^* \, \res \; \calH_{0}^0
\label{ttgfdghhjsjkdkmnmnm}
\ee
where $\calH_{0}^0\sub\calH_{0}$
is the subspace without external charges.
Since this modified transfer matrix generates a modified dynamics given by
$$
\al'= \Ad (Q(x)Q(y)^*)\circ\al\circ\Ad (Q(y)Q(x)^*)
$$
we conclude that the state $A\mapsto(\Omega_{x,y}^{FM},\,A\,\Omega_{x,y}^{FM})$
is a ground state of $\al'$  when restricted to the observable
algebra, $A\in\AA$. A similar
statement holds for magnetic dipole states and magnetically modified
transfer matrices \cite{FlorianBarataI}, see also below.

To treat electric and magnetic charges simultaneously we now
generalize this construction as follows.
Let $\calD^E\equiv\cc_{loc}^0$ and $\calD^M\equiv\cc_{loc}^2$ denote the set of
$\Bbb Z_N$-valued 0-cochains ($\equiv$ electric charge distributions)
and 2-cochains
($\equiv$ magnetic charge distributions), respectively, with finite
support in
our spatial lattice $\Z^2$, and let $\calD=\calD^E\x\calD^M $.
For $\rho=(\e,\mu)\in\calD $,
$\mbox{supp }\rho \subset V_0\x V_2$,
we generalize (\ref{ttgfdghhjsjkdkmnmnm}) and
define the modified local transfer matrices in a finite volume
$V \subset \Z^2$ as the element of ${\frak A} (V)$ given by
\be
T_{V} (\rho ) :=
Z (\mu)^{1/2} Q_H (\varepsilon ) T_{V}
Q_H(\varepsilon )^* Z(\mu )^{1/2}
,
\label{yyuyiuiusydybcdj}
\ee
where we have used the notations
\be
Q_H (\varepsilon ) := \prod_{x \in (\Z^2)_0} Q_H (x)^{\varepsilon (x)}
\ee
and
\be
Z(\mu) := \prod_{p\in (\Z^2)_2} Z^{(\mu(p))}(p)
.
\ee
Here
the gauge invariant operator
$Z^{(n)}(p)$, $n\in \{ 0, \ldots , N-1 \}$,
is defined  by (see also \cite{FlorianBarataI})
\be
Z^{(n)} (p) = \exp
\left\{
\frac{1}{\sqrt{N}}
\sum_{j=0}^{N-1} \bege (j)
\left(
\exp
\left(
     {\frac{2\pi i }{N}  j n }
\right) -1
\right)
\left(
\delta Q_G (p)
\right)^j
\right\}
\ .
\label{defdeZ}
\ee
On our local Hilbert spaces $\calH_V$ it acts by
\be
Z^{(n)}(p)\,|\vfi,\,\al\ket =
\frac{e^{-S_g(d\al(p)+n)}}{e^{-S_g(d\al(p))}}\,|\vfi,\,\al\ket\ .
\ee
Hence, this operator can be interpreted as the operator creating a
vortex with magnetic charge $n$ at the plaquette $p$.
The definition (\ref{yyuyiuiusydybcdj})
is also motivated by the fact that under duality
transformations we roughly have $T_V(\e,0)\leftrightarrow
T_{V^*}(0,\e^*)$, see \cite{FlorianBarataI} for the precise statement.
Also note that $i_C(T_V(\r)) = T_V(-\r)$ and that
$T_V (\rho)$ is still gauge invariant, i.e. $T_V (\rho) \in
{\frak A}_{loc}$.

In a formal continuum notation, these modified transfer
matrices correspond to
modified Hamiltonians, where the kinetic term in the Higgs fields
is replaced by
\be
\frac{1}{2} (\pi_H , \, \pi_H)
\rightarrow
 \frac{1}{2} (\pi_H + \varepsilon  , \, \pi_H + \varepsilon)
\ee
and the magnetic self energy is replaced by
\be
\frac{1}{2} (d\al , \, d\al)
\rightarrow
\frac{1}{2} (d\al+\mu , \, d\al+\mu)  .
\ee
Together with these modified transfer matrices we also have a modified
euclidean dynamics $\al_\rho$
given by the automorphism of ${\frak F} $
\be
\al_{\rho} (A) := \lim_{V \uparrow \Z^2}
T_{V}(\rho)AT_{V}(\rho)^{-1} ,
\qquad A \in {\frak F} ,
\label{jfhdsidnoieunvw}
\ee
such that $ \al_0 \equiv \al$.
As in the case of $ \al_0$, for each $ A\in {\frak F}_{loc}$
the limit above is already reached at
finite $ V$. Moreover $\al_\rho ({\frak A}) = {\frak
A}$ and
\be
i_C\circ\al_\r = \al_{-\r}\circ i_C\ .
\label{cdelro}
\ee

We emphasize that we introduce these modifications not as a substitute
of our original ``true'' dynamics, but for technical
reasons only.
>From (\ref{yyuyiuiusydybcdj}) we conclude
\be
\al_{\rho} = \mbox{Ad }K_\rho \circ \al_0 \circ \mbox{Ad }K_\rho^*
= \mbox{Ad } L_\rho \circ \al_0 ,
\label{uuuxxxuuuiuziubuvbs}
\ee
where
\be
	K_\rho := Z(\mu )^{1/2}Q_H (\epsilon )
\qquad \in {\frak F}_{loc}
\label{iiiiuhiudhfsdgs}
\ee
 and
\be
	L_\rho := K_\rho \al_0 (K_\rho^*) =
	T_V (\rho) T_V (0)^{-1}
\qquad \in {\frak A}_{loc} .
\label{iiiiuhiudhfsdgs1}
\ee
the last equality holding for $V$ large enough. Note also that
for $\rho = (\mu , \, \epsilon)$ and
$\rho ' = (\mu ' , \, \epsilon ')$ such that
$\mbox{supp }  \mu  \cap \mbox{supp } \mu ' = \emptyset$
one has
\be
K_{\rho + \rho '}  =  K_\rho K_{\rho '}.
\ee
If the distance between $\mbox{supp }\rho$ and $\mbox{supp }\rho '$
is large enough one also has
\be
L_{\rho + \rho '}  =  L_\rho L_{\rho '} .
\ee
Next, we define the {\em total charge}, $q_{\rho} \in \Z_N\x \Z_N$ ,
of a distribution $\rho=(\epsilon, \, \mu )\in\calD $  as
\be
	q_\rho=
\left( e_\epsilon , \; m_\mu \right)  :=
\left( \sum_x \e (x), \; \sum_p\mu(p) \right)
\label{difgoeubjygbduyw}
\ee
and put
\be
     \calD_q:=\{\rho\in\calD\: | \quad q_{\rho}=q\} .
\ee
Then, for all $\rho=(\e, \, \mu)\in\calD_0$, (i.e. with vanishing total
charge), there exist 1-cochains, $s_{\e}$ and $s_{\mu}$, with finite
support, such that
\be
 d^*s_{\e}=-\e \qquad \mbox{and} \qquad ds_{\mu}=-\mu
\label{hnepubwpeug}
\ee
This allows to generalize the FM-construction and define, for
$ \rho=(\e, \, \mu)\in\calD_0$, the family of
states $\omega_{\rho}$ on ${\frak A}$ as ground states of the
modified dynamics $\al_\r$ by the following method.

As already observed in \cite{FredMarcu} and \cite{FlorianBarataI}, ground
states for the modified transfer matrices can be obtained by taking the
thermodynamic limit of the state defined by the formula
\be
\omega_{V, \, \rho } ( A ) = \lim_{n\rightarrow \infty}
\omega_{V, \, \rho }^n ( A ) ,
\label{Ryfgsjdfghlsdkjfhgls}
\ee
where
\be
   \omega_{V, \, \rho }^n ( A ) =
\frac{Tr_{{\cal H}_{V}}
\left(
T_{V}(\rho )^n
A
T_{V}(\rho )^n E_{V}^\rho
\right)}{Tr_{{\cal H}_{V}} \left( T_{V}(\rho )^{2n}
E_{V}^\rho  \right)}
, \qquad A \in {\frak F}(V ) ,
\label{Xum}
\ee
and where $E_{V}^\rho$ is some suitably chosen matrix to adjust
the boundary conditions.
There are two possibilities we will discuss. The first one is
$E_{V}^\rho= \UM $, thus getting periodic boundary conditions in euclidean time
direction for the classical expectations associated with
$   \omega_{V, \, \rho }^n ( A ) $.
In order to understand what happens algebraically  one
checks that, for $\mu=0$, the desired ground state
 simply becomes $\omega_0 \circ \mbox{Ad }{Q_H (\epsilon )}$,
which is a state with
{\em external} electric charge.
Hence, the resulting representation of $\AA$ would be
equivalent to the vacuum representation.
A similar statement holds for $\mu\neq 0$.
Since this is not the kind of state we are interested in let us look at
the second case.

There we choose $E_{V }^\rho$ in such a way that we get free boundary
conditions in the euclidean time direction
for the classical expectations associated with
$   \omega_{V,\, \rho }^n ( A ) $,
together with  horizontal
electric and magnetic strings ``conducting'' the charges among the points of the
the support of $\epsilon $ and $\mu$, respectively,
and located in the highest and in the lowest
time-slices of $V_n := V \times \{-n , \ldots , n \}$.
Hence we choose  1-cochains $ s_\epsilon$ and $s_\mu$
obeying (\ref{hnepubwpeug}) and put
\be
E_{V}^\rho :=
\left[
Z(\mu)^{-1/2}M(s_\mu)Q_{GH}(s_\varepsilon )
\right]
E^0_{V}
\left[
Z(\mu)^{-1/2}M(s_\mu)Q_{GH}(s_\varepsilon )
\right]^*\ .
\label{mmmmmsdiogfudzgfsf}
\ee
Here $E_{V}^0$ has been defined in (\ref{EV0}), $Q_{GH}(b)$
are the gauge invariant link operators (\ref{Q_GH}) and
\be
M(s_\mu) :=e^{-A_V / 2} e^{-B_V} P_G(s_\mu)
e^{B_V} e^{A_V / 2}\equiv
e^{-A_V / 2}  P_G(s_\mu) e^{A_V / 2}
\label{M}
\ee
Note that $M(s_\mu)\in\AA(V)$ for all $V$ containing
$\supp\mu\cup\supp s_\mu$ and that $M(s_\mu)$ is actually
independent of $V$ provided the distance between $\supp s_\mu$
and the boundary $\partial V$ is $\ge 2$.
The motivation for this definition comes from its effect in
the euclidean functional integral, where $Q_{GH}(s_\varepsilon)$
produces a superposition of electric Mandelstam strings
connecting the charges described by $\ve$ along the
support of $s_\ve$.
Similarly, the operator $Z(\mu)^{-1/2}M (s_\mu )$
creates magnetic Mandelstam strings
joining the plaquettes of $\mbox{supp }\mu$ via the
support of $s_\mu$.
In the functional integral this magnetic string will
appear in the form of shifted (or frustrated) vertical plaquettes
placed between the first two and the last two time-slices
of the space-time volume $V_n$.

This can be seen by looking at the product
$T_V(\r)Z(\mu)^{-1/2}M (s_\mu )$ appearing in (\ref{Xum}) due to the definition
(\ref{mmmmmsdiogfudzgfsf}).
There, the factor
$Z (\mu )^{-1/2}e^{-A_V / 2} e^{-B_V}$ gets
canceled by a corresponding factor in $T_V(\r)$ and the
matrix elements of $ P_G(s_\mu)e^{B_V}$ (coming next according to
(\ref{M})) are given by
\be
\bra\vfi',\,\al'|\, P_G(s_\mu)e^{B_V}\,|\vfi,\,\al\ket =
\exp\left[-\sum_{b\in V_1} S_g(\al(b)-\al'(b)+s_\mu(b)) -
\sum_{x\in V_0} S_h(\vfi(x)-\vfi'(x))\right]
\label{magstring}
\ee
After transforming the functional integral expression for
(\ref{Xum}) to unitary gauge the shift $s_\mu(b)$ in
(\ref{magstring}) appears on the vertical plaquette spanned by
the horizontal bond $b$ and the time-like bond $\bra
t_{n-1},\,t_n\ket$ (and similarly, but with opposite sign at
$\bra t_{-n},\,t_{-n+1}\ket$).

The above construction of charged states can be
understood as analogous to the construction of Fredenhagen and Marcu,
except that, at finite volume, the horizontal Mandelstam strings
are already located at the highest (respectively lowest) time-slices.

Given the definition (\ref{mmmmmsdiogfudzgfsf}),
the limit (\ref{Ryfgsjdfghlsdkjfhgls}) is actually independent of
the choice of the horizontal strings $s_\varepsilon$ and $s_\mu$ satisfying
(\ref{hnepubwpeug}) and we can take the thermodynamic limit to obtain
\be
	\omega_\rho (A) := \lim_{V \uparrow \Z^2}
			   \omega_{V, \, \rho}(A) .
\label{ppopopppidiusugaasa}
\ee
Note that $G(x)E_V^0=E_V^0$ implies $G(x)E_V^\r=E_V^\r$ for all
$x\in V$ and therefore
$\om_\r$ provides an eigenstate of $\GG$ with zero external
charge. Also note that the boundary conditions
(\ref{mmmmmsdiogfudzgfsf}\,/\,\ref{M}) now imply
\be
\om\r\circ i_C = \om_{-\r}
\label{omch}\ .
\ee

Rewriting (\ref{ppopopppidiusugaasa})
in terms of euclidean expectation values in the unitary
gauge we get an expectation of a classical function $A_{cl, \, \rho}$
associated with $A$ in the presence of both, ``electric'' Wilson loops
$\calL_E(s_{\e},n)$ living on bonds and ``magnetic'' vortex
loops $\calL_M(s_{\mu},n)$ living on plaquettes, i.e.
\be
\om_\rho (A)=
\la
A_{cl, \, \rho}
\ra_{\rho }
:=
\lim_{V} \lim_{n} Z^{-1}_{\calL_E , \, \calL_M}
\int da \, A_{cl, \, \rho} (a)
e^{- S_G (da + \calL_M) -S_H (a)}
 e^{\frac{2\pi i}{N} (a, \, \calL_E)}
\label{wwsrdhfdcvsjydobcfsled}
\ee
where $a$ denotes the euclidean lattice $\Bbb Z_N$-gauge field
$a := d\varphi - \al$ (unitary gauge)
and $Z_{\calL_E , \, \calL_M}$ is the normalization
such that $\omega_\rho (\UM )=1$.
Here the vertical parts of the loops $\calL_E$ and $\calL_M$
run from
euclidean time $t=-n$ to $t=+n$ and are spatially located at the
supports of $\e$ and $\mu$, respectively.
The horizontal parts of $\calL_E$ and $\calL_M$ are
given by the 1-cochains $\pm s_{\e/\mu}$
(\ref{hnepubwpeug}), shifted to the euclidean
time slice $\pm n$, respectively.
To be more precise, $\calL_E(s_\ve,\,n)\in\cc^1(V_n)$ is the
unique 1-cochain on the euclidean space-time lattice $\Z^3$
satisfying $d^*\calL_E(s_\ve,\,n)=0$ together with
the condition that its horizontal
part is nonzero only on the time slices
$\pm n$, where it coincides with $\pm s_\ve$.
Similarly,  $\calL_M(s_{\mu},n)\in\cc^2(V_n)$ is the unique
2-cochain on $\Z^3$ satisfying  $d\calL_M(s_{\mu},n)=0$ plus
the condition that the horizontal part of its dual 1-cochain is non-vanishing
only on the time slices
$\pm (n-1/2)$, where it  coincides with the dual of $\pm s_\mu$.

We will refer to such expectations
by the following symbolic picture
\be
\om_\rho (A)= \lim_{n, \, V}
\frac{
\mbox{
\setlength{\unitlength}{0.0025in}
\begingroup\makeatletter\ifx\SetFigFont\undefined
\def\x#1#2#3#4#5#6#7\relax{\def\x{#1#2#3#4#5#6}}
\expandafter\x\fmtname xxxxxx\relax \def\y{splain}
\ifx\x\y
\gdef\SetFigFont#1#2#3{%
  \ifnum #1<17\tiny\else \ifnum #1<20\small\else
  \ifnum #1<24\normalsize\else \ifnum #1<29\large\else
  \ifnum #1<34\Large\else \ifnum #1<41\LARGE\else
     \huge\fi\fi\fi\fi\fi\fi
  \csname #3\endcsname}%
\else
\gdef\SetFigFont#1#2#3{\begingroup
  \count@#1\relax \ifnum 25<\count@\count@25\fi
  \def\x{\endgroup\@setsize\SetFigFont{#2pt}}%
  \expandafter\x
    \csname \romannumeral\the\count@ pt\expandafter\endcsname
    \csname @\romannumeral\the\count@ pt\endcsname
  \csname #3\endcsname}%
\fi
\fi\endgroup
\begin{picture}(920,670)(25,185)
\thinlines
\multiput(150,225)(8.98649,0.00000){75}{\makebox(0.1111,0.7778){\SetFigFont{5}{6}{rm}.}}
\multiput(150,800)(8.98649,0.00000){75}{\makebox(0.1111,0.7778){\SetFigFont{5}{6}{rm}.}}
\multiput(150,225)(0.00000,8.98438){65}{\makebox(0.1111,0.7778){\SetFigFont{5}{6}{rm}.}}
\multiput(815,225)(0.00000,8.98438){65}{\makebox(0.1111,0.7778){\SetFigFont{5}{6}{rm}.}}
\put(115,185){\line(-1, 4){ 84.118}}
\put(115,855){\line(-1,-4){ 84.118}}
\put(855,185){\line( 1, 4){ 84.118}}
\put(855,855){\line( 1,-4){ 84.118}}
\thicklines
\put(455,800){\line( 0,-1){575}}
\put(455,225){\line( 0, 1){  5}}
\put(690,800){\line( 0,-1){575}}
\thinlines
\multiput(340,800)(0.00000,-8.04196){72}{\line( 0,-1){  4.021}}
\put(515,800){\makebox(0.1111,0.7778){\SetFigFont{5}{6}{rm}.}}
\multiput(510,800)(0.00000,-8.04196){72}{\line( 0,-1){  4.021}}
\multiput(740,800)(0.00000,-8.04196){72}{\line( 0,-1){  4.021}}
\thicklines
\put(315,550){\line( 0,-1){190}}
\put(315,360){\line( 1, 0){230}}
\put(545,360){\line( 0, 1){ 90}}
\put(545,450){\line(-1, 0){160}}
\put(385,450){\line( 0, 1){190}}
\put(385,640){\line(-1, 0){ 70}}
\put(315,640){\line( 0,-1){ 90}}
\thinlines
\put(385,635){\line(-1,-1){ 70}}
\put(390,575){\line(-1,-1){ 70}}
\put(390,510){\line(-1,-1){ 70}}
\put(390,445){\line(-1,-1){ 70}}
\put(440,445){\line(-1,-1){ 85}}
\put(500,445){\line(-1,-1){ 85}}
\put(545,415){\line(-1,-1){ 60}}
\put(840,515){\makebox(0,0)[lb]{\smash{\SetFigFont{12}{14.4}{rm}$V_n$}}}
\put(905,205){\makebox(0,0)[lb]{\smash{\SetFigFont{12}{14.4}{rm}euclid.}}}
\put(240,395){\makebox(0,0)[lb]{\smash{\SetFigFont{12}{14.4}{rm}$A_{cl}$}}}
\end{picture}
}
}{
\mbox{
\setlength{\unitlength}{0.0025in}%
\begingroup\makeatletter\ifx\SetFigFont\undefined
\def\x#1#2#3#4#5#6#7\relax{\def\x{#1#2#3#4#5#6}}%
\expandafter\x\fmtname xxxxxx\relax \def\y{splain}%
\ifx\x\y
\gdef\SetFigFont#1#2#3{%
  \ifnum #1<17\tiny\else \ifnum #1<20\small\else
  \ifnum #1<24\normalsize\else \ifnum #1<29\large\else
  \ifnum #1<34\Large\else \ifnum #1<41\LARGE\else
     \huge\fi\fi\fi\fi\fi\fi
  \csname #3\endcsname}%
\else
\gdef\SetFigFont#1#2#3{\begingroup
  \count@#1\relax \ifnum 25<\count@\count@25\fi
  \def\x{\endgroup\@setsize\SetFigFont{#2pt}}%
  \expandafter\x
    \csname \romannumeral\the\count@ pt\expandafter\endcsname
    \csname @\romannumeral\the\count@ pt\endcsname
  \csname #3\endcsname}%
\fi
\fi\endgroup
\begin{picture}(920,670)(25,185)
\thinlines
\multiput(150,225)(8.98649,0.00000){75}{\makebox(0.1111,0.7778){\SetFigFont{5}{6}{rm}.}}
\multiput(150,800)(8.98649,0.00000){75}{\makebox(0.1111,0.7778){\SetFigFont{5}{6}{rm}.}}
\multiput(150,225)(0.00000,8.98438){65}{\makebox(0.1111,0.7778){\SetFigFont{5}{6}{rm}.}}
\multiput(815,225)(0.00000,8.98438){65}{\makebox(0.1111,0.7778){\SetFigFont{5}{6}{rm}.}}
\put(115,185){\line(-1, 4){ 84.118}}
\put(115,855){\line(-1,-4){ 84.118}}
\put(855,185){\line( 1, 4){ 84.118}}
\put(855,855){\line( 1,-4){ 84.118}}
\thicklines
\put(455,800){\line( 0,-1){575}}
\put(455,225){\line( 0, 1){  5}}
\put(690,800){\line( 0,-1){575}}
\thinlines
\multiput(340,800)(0.00000,-8.04196){72}{\line( 0,-1){  4.021}}
\put(515,800){\makebox(0.1111,0.7778){\SetFigFont{5}{6}{rm}.}}
\multiput(510,800)(0.00000,-8.04196){72}{\line( 0,-1){  4.021}}
\multiput(740,800)(0.00000,-8.04196){72}{\line( 0,-1){  4.021}}
\put(840,515){\makebox(0,0)[lb]{\smash{\SetFigFont{12}{14.4}{rm}$V_n$}}}
\put(905,205){\makebox(0,0)[lb]{\smash{\SetFigFont{12}{14.4}{rm}euclid.}}}
\end{picture}
}
} \qquad .
\ee
The dotted box indicates the space-time volume $V_n$. The
shaded region in the numerator indicates the support of the classical
function $A_{cl}$ associated with an observable $A$. The
bold face vertical lines indicate the vertical part of the loops
$\calL_M$, i.e.
the stacks of horizontal
plaquettes whose projection onto the time-zero plane is given by $\mu$.
They are the euclidean realization of magnetic vortices located at
$\mu$. The dashed vertical lines indicate the vertical part of the
loops $\calL_E$. Their projection onto the time-zero plane is given
by $\varepsilon$.
They are the euclidean realization of electric charges located at
$\varepsilon$. For finite $n$ all vertical lines actually close
to loops via the strings $\pm s_{\ve/\mu}$ running in (or just
inside of) the horizontal boundary of $V_n$. In the limit
$n\to\infty$ the choice of these horizontal parts becomes
irrelevant. In fact,
applying our cluster expansion techniques \cite{FlorianBarataI}
we have

\begin{proposition}
\label{estadosKK1}
For all $\rho\in\calD_0$ the limit
(\ref{ppopopppidiusugaasa}) provides a well defined state
$\omega_{\rho}$ on
${\frak F}$ which is independent of the choice of 1-chains $(s_{\e},s_{\mu})$
satisfying (\ref{hnepubwpeug}).
Moreover, $\omega_{\rho}\res {\frak A}$ provides a ground state
with respect to the modified dynamics $\al_{\rho}$
fulfilling the cluster property. $\EndofStatement$
\end{proposition}
Generalizing equation (\ref{hpdfzpbsdb}) we will
show in Theorem \ref{Phiroro'}
below, that for all
$\rho\in\calD_0$, i.e. with vanishing total charge,
the states  $\omega_{\rho}$ are actually induced by
vectors in the vacuum sector $\calH_0^0$.

To get charged states $\om_\rho$ for $\rho\in\calD_q, q\neq 0$, we now
have to move a counter-charge to infinity. To this end let
$\rho\cdot a$
denote the translate of $\rho$ by $a\in \Z^2$.
Then $\rho-\rho'\cdot a \in\calD_0$ for all $\rho,\rho'\in\calD_q$.

\begin{proposition}
\label{estadosKK2}
For $A\in\FF_{loc},\
0\neq q\in \Z_N \times \Z_N$ arbitrary and $\rho,\rho'\in\calD_q$ let
\be
\omega_{\rho}(A):= \lim_{a\to\infty} \om_{\rho - \rho'\cdot a}(A) .
\label{siuebiuybwpeuyb}
\ee
Then the limit exists independently of the chosen sequence
$a\to\infty$
and it is independent of $\rho'$. Moreover $\omega_{\rho}\res {\frak A}$
provides a ground state
with respect to the modified dynamics $\al_{\rho}$
fulfilling the cluster property.
$\EndofStatement$
\label{lidufhgpsiudfygps}
\end{proposition}
Clearly, (\ref{siuebiuybwpeuyb}) implies that now (\ref{omch})
also holds for $q_\r\neq 0$.

Next, we note that the symmetry group $\calS$ of our
spatial square lattice (i.e. consisting of translations by
$a\in \Z^2$ and rotations by $\frac{1}{2} n \pi$) acts naturally
from the right on
$\calD_q$ by $(\rho\cdot g)(x ):=\rho(g\cdot x)$ and we
obviously have
$\omega_{V , \, \rho} \circ \tau_g =
\omega_{g^{-1} V , \, \rho\cdot g}$ for all $g \in \calS$. Hence we get the

\begin{corollary}
For all $g\in \calS$ and all $\rho \in \calD$ we have
  $\om_{\rho\cdot g} = \om_{\rho} \circ \tau_g$,
where $\tau_g$ denotes the natural action of $\calS$ on ${\frak F}$.
$\EndofStatement$
\end{corollary}

We now give an interpretation of these
states as charged states in the following sense.
For $ V \subset \Z^2$, finite,
define the charge measuring operators
\be
\Phi^E (V):= \prod_{x \in V}\delta^* P_G (x )
\qquad
\mbox{ and }
\qquad
\Phi^M (V_2):= \prod_{p\in V_2}\delta Q_G (p ),
\ee
These operators are lattice analogues of the continuum expressions
$\exp -i\int\nabla E$ and $\exp i\int dA$, respectively, i.e. they
measure the total electric charge inside $V$ and the total magnetic flux
through $V$, respectively. As in Theorem 6.2 of \cite{FlorianBarataI}
we have

\begin{proposition}
\label{mageele}
If $V\subset\Z^2$ is e.g. a square
centered at the origin, one has, under the conditions of Proposition
\ref{lidufhgpsiudfygps}:
\bear
\lim_{V \uparrow\Z^2}
\frac{\om_\rho (\Phi^E (V))}{\omega_0 (\Phi^E (V))}
=
\exp \left(  \frac{2\pi i e_\epsilon }{N} \right)  ,
\label{zeroetres}
\\
\lim_{V \uparrow\Z^2}
\frac{\om_\rho (\Phi^M (V_2 ))}{\omega_0 (\Phi^M (V_2 ))}
=
\exp \left(  \frac{2\pi i m_\mu}{N} \right) .
\label{doiseum}
\eear
where $e_\epsilon$ and $m_\mu$ were defined in
(\ref{difgoeubjygbduyw}). $\EndofStatement$
\end{proposition}

We omit the proof here, since it is analogous to the equivalent one
found in \cite{FlorianBarataI}.

Finally, we check that our charges have Abelian composition
rules. Since, as opposed to the DHR-theory of superselection sectors
\cite{DHR4}, the states $\om_\rho$ are not given in terms of localized
endomorphisms, we take the following statement as a substitute for
this terminology.

\begin{proposition}
Let $\rho,\rho'\in\calD$ and for each $a\in \Z^2$ let $\om_{\rho}$
and $\om_{\rho+\rho'\cdot a}$ be given by Propositions \ref{estadosKK1} or
\ref{estadosKK2}, respectively.
Then
\bma
\lim_{a\to\infty}\om_{\rho+\rho'\cdot a}=\om_{\rho} . \EndofStatement
\ema
\label{ieurygpbsiudybvpdiubv}
\end{proposition}

Proposition \ref{ieurygpbsiudybvpdiubv} is actually a special case of
the following more general factorization property, which will be
important in the construction of scattering states in
\cite{FBIII}.

\begin{theorem}
Let $\rho_1$, $\rho_2 \in \calD$ and for $a\in \Z^2$ put
$\rho (a) := \rho_1 + \rho_2\cdot a$. Then, for all $A$,  $A'$, $B$ and
$B' \in {\frak A}_{loc}$ and all $n\in \N_0$
\bear
  \displaystyle \lim_{|a|\to\infty}
\omega_{\rho(a)}
\left(
  \tau_a^{-1}(B)A \al_{\rho (a)}^{n}(A' \tau_a^{-1}(B'))
\right) & = &
\nonumber \\
\omega_{\rho_1}
\left(
  A \al_{\rho_1 }^{n}(A' )
\right)
\omega_{\rho_2}
\left(
  B \al_{\rho_2 }^{n}(B' )
\right).
 & & 
\label{factorize}
\eear
\label{uuiupdsypspYUIYipkjhlkfhg}
$\EndofStatement$
\end{theorem}

A sketch of the
proofs of Propositions \ref{estadosKK1}, \ref{estadosKK2} and
Theorem \ref{uuiupdsypspYUIYipkjhlkfhg}
is given in the Appendices
\ref{ProofofPropositionprotectestadosKK1}-
\ref{ProofofTheoremuuiupdsypspYUIYipkjhlkfhg}.

\subsection{Global Transfer Matrices}
\setcounter{theorem}{0}

Given the family of
dyonic states $\om_{\rho}$ on $\frak A$ we denote by
$(\pi_{\rho},\calH_{\rho},\Om_{\rho})$ the associated GNS representation
$\pi_{\rho}$ of $\frak A$
on the Hilbert space $\calH_{\rho}$ with cyclic
 vector $\Om_{\rho}\in\calH_{\rho}$. (From now on we will no longer
  consider external charges, and therefore we simplify the notation by
  putting $\calH_{\rho}\equiv\calH_{\rho}^0$). On
  $\calH_{\rho}$ we define a {\em modified} global transfer matrix $T_\rho$
by putting for $A\in {\frak A}$
\be
T_\rho\pi_\rho(A)\Om_\rho=
e^{-E_\rho}\pi_\rho(\al_\rho(A))\Om_\rho ,
\label{udybuvdywp}
\ee
where $E_\rho\in \R$ is an additive normalization
constant, which will be determined later to
appropriately adjust the self energies of the charge distributions
$\rho$ relative to each other.

\begin{theorem}
For all $\r\in\calD$
equation (\ref{udybuvdywp}) uniquely defines a positive
bounded operator on $\calH_\rho$,
\be
0\le T_\rho\le e^{-E_\rho}
\ee
 which is invertible on
$\pi_\rho({\frak A})\Om_\rho$ and implements the modified dynamics,
i.e. $Ad \, T_\rho\circ\pi_\rho = \pi_\rho\circ \al_\rho$. Moreover,
${\Bbb C}\,\Om_\rho$ is the unique eigenspace of $T_\rho$ with maximal
eigenvalue $e^{-E_\rho } \equiv \|T_\rho \| $. $\EndofStatement$
\label{occooiajjduyyy}
\end{theorem}
As a warning we recall that the charged states
$\om_\r$ are ground states only with
respect to the {\em modified euclidean dynamics} $\al_\r$ and
only when restricted to the observable algebra
$\AA\subset\FF$.
Hence, although $\pi_\r$ and $T_\r$ could easily be extended to
$\FF$ and $\pi_\r(\FF)\Om_\r$, respectively,
$\Om_\r$ would not necessarily be a ground state of $T_\r$ in
this enlarged Hilbert space containing external charges.
However, since we will never look at external charges we will
not be bothered by such a possibility.

{\bf Proof of Theorem \ref{occooiajjduyyy}.  }
First $T_\rho$ is well defined on $\pi_\rho ({\frak A}) \Om_\rho$,
since $\pi_\rho (A)\Om_\rho =0$ implies $\omega_\rho (BA)=0$ for all
$B\in{\frak A}$ and therefore
$
e^{2E_\rho} \| T_\rho \pi_\rho (A) \|^2
=
\omega_\rho \left( \al_\rho (A)^*  \al_\rho (A)\right)
=
\omega_\rho \left( \al_\rho^{-1}(\al_\rho (A)^*) A \right)
=
0
$
by $\al_\rho$-invariance of $\omega_\rho$. Since $\al_{\rho}$
is invertible, $T_\rho$ is invertible on $\pi_\rho ({\frak A}) \Om_\rho$.
Moreover, by the ground state property we have
\be
\left(
\pi_\rho (A) \Om_\rho , \; T_\rho \pi_\rho (A) \Om_\rho
\right)
=
e^{-E_\rho}
\omega_\rho (A^* \al_\rho (A))
\leq
e^{-E_\rho}
\omega_{\rho} (A^*A)  ,
\ee
which implies $0 \leq T_\rho \leq e^{-E_\rho}$.
The identity
$
\mbox{Ad }T_\rho \circ \pi_\rho = \pi_\rho \circ
\al_\rho
$
is obvious. Finally, let $P_\rho \in \calB (\calH_\rho)$ be the
projection onto the eigenspace of $T_\rho$ with eigenvalue $e^{-E_\rho}$
and
$
\stackrel{\circ}{P}_\rho = P_\rho -
\left| \Om_\rho \rangle \langle\Om_\rho\right|
$.
Then, for all
$ A$, $B\in {\frak A}$,
\bear
\displaystyle \lim_n \omega_\rho (A^* \al_\rho^n (B)) & = &
\displaystyle \lim_n e^{nE_\rho}
(
\pi_\rho (A) \Om_\rho , \, T_\rho^n \pi_\rho (B) \Om_\rho
)
\nonumber \\
 & = &
(
\pi_\rho (A) \Om_\rho , \, P_\rho \pi_\rho (B) \Om_\rho
)
\nonumber \\
 & = &
\omega_\rho (A^*)\omega_\rho (B) +
(
\pi_\rho (A) \Om_\rho , \, \stackrel{\circ}{P}_\rho \pi_\rho (B) \Om_\rho
) .
\label{2.43}
\eear
Hence the cluster property implies $\stackrel{\circ}{P}_\rho =0$. $\Fullbox$

Given the modified global transfer matrices $T_\rho$ we now use
relation (\ref{uuuxxxuuuiuziubuvbs}) to define on $\calH_\rho$
the family of transfer matrices
\be
 T_\rho(\rho'):=\pi_\rho(L_{\rho'}L_\rho^{-1})T_\rho   ,
\label{ooqqqppapjdbfsiuh}
\ee
where $L_\rho$ has been given in (\ref{iiiiuhiudhfsdgs1}).
Hence, $T_\rho\equiv T_\rho(\rho)$ and we have
\begin{corollary}
For all $\rho, \, \rho'\in\calD$ the operators $T_\rho (\rho')$ are
positive and bounded on $\calH_\rho$ and
\be
Ad\,T_\rho(\rho')\circ\pi_\rho=\pi_\rho\circ\al_{\rho'} . 
\label{zzguzgujhzgla}
\ee
\label{cccopodfjjnsbhhuha}
$\EndofStatement$
\end{corollary}

{\bf Proof of Corollary \ref{cccopodfjjnsbhhuha}.}
For the purpose of this proof we consider temporarily
the enlarged GNS-triple
$(\pi_\rho ({\frak F}), \, \Omega_\rho , \, \calH_\rho)$,
where $\calH_\rho$ now is the closure of
$\pi_\rho ({\frak F})\Omega_\rho$
and contains vector states with external electric charges.
Equation (\ref{zzguzgujhzgla}) follows immediately from
(\ref{uuuxxxuuuiuziubuvbs}). To see that $T_\rho (\rho ')$ is positive
we use
$L_\rho = K_\rho \al_0 (K_\rho^*)$, for $K_\rho$ defined in
(\ref{iiiiuhiudhfsdgs}), to conclude
\bear
T_\rho (\rho') & = & \pi_\rho (K_{\rho'})T_{\rho}(0)\pi_{\rho}(K_{\rho'}^*)
\nonumber \\
 & = &
\pi_\rho (K_{\rho '}K_{\rho}^{-1}) T_{\rho}(\rho)
\pi_\rho(K_{\rho }^{* \, -1}K_{\rho '}^{*}) ,
\eear
which is positive since $T_\rho (\rho) \equiv T_\rho$ is positive. $\Fullbox$

In view of this result we call $T_\rho(0)$ the {\em unmodified} (``true'')
global transfer matrix, since it implements the original dynamics
$\al\equiv\al_0$. We also have the following
important

\begin{corollary}
Let ${\calA}_\rho\subset\calB(\calH_\rho)$ be the $*$-algebra generated by
$\pi_\rho({\frak A})$ and $T_\rho(0)$. Then the commutant of
${\calA}_\rho$ is trivial: ${\calA}_\rho' = \C\, \UM$. $\EndofStatement$
\label{tttrttrrusfkjhdfgus}
\end{corollary}

{\bf Proof of Corollary \ref{tttrttrrusfkjhdfgus}.}
Since $T_\rho (\rho)\in\calA_\r$ we have
$[B, \, T_\rho (\rho)]=0$ for all $B \in {\calA}_\rho '$ which implies
$B\Om_\rho = \lambda \Om_\rho$ for some
$\lambda \in \C$ by the uniqueness of the ground state vector $\Om_\r$
of $T_\r(\r)$. Hence
$
B\pi_\rho (A)\Om_\rho = \pi_\rho (A) B \Om_\rho
=
\lambda \pi_\rho (A) \Om_\rho
$
for all $A \in {\frak A}$ and therefore $B = \lambda \UM$.
$\Fullbox$

In view of Corollary \ref{tttrttrrusfkjhdfgus} we may consider
$\pi_\r$ as an irreducible representation of an extension
$\hat\AA\supset\AA$ such that $\calA_\r =
\pi_\r(\hat\AA)$.
Algebraically $\hat\AA$ is defined to be the crossed
product of $\AA$ with the semi-group ${\Bbb N_0}$ acting by
$n\mapsto\al^n\in\Aut\AA$. Described in terms of
generators and relations $\hat\AA$ is the $*$-algebra
generated by $\AA$ and a new selfadjoint element ${\bf t}\equiv
{\bf t}(0)$ with commutation relation $\ttt A =\al(A)\ttt$
for all $A\in\AA$.
Note that the construction (\ref{T_0}) shows that any
GNS-representation $\pi_\om$ of $\AA$ from an $\al$-invariant state
$\om$ naturally extends to a representation of $\AAA$.
Moreover, $T_\om\:=\pi_\om(\ttt)$ is positive provided the first
inequality in (\ref{JKJKL}) holds.
Using (\ref{iiiiuhiudhfsdgs1}) we may also define
\be
\ttt(\r) := L_\r\,\ttt\ \in\AAA .
\ee
Then $\ttt(\r)=\ttt(\r)^*$ (since $\al(L_\r^*)=L_\r)$) and
$\ttt(\r)\,A = \al_\r(A)\,\ttt(\r)$ for all $A\in\AA$.
By (\ref{ooqqqppapjdbfsiuh}) this implies that also $\pi_\r$ extends
to a representation of $\hat\AA$ by defining for fixed $\r$ and all $\r'$
\be
\pi_\r(\ttt(\r')):=c_\r\,T_\r(\r')\
\label{tt}
\ee
where $c_\r>0$ may be chosen arbitrarily.

\subsection{Global Charges}
\setcounter{theorem}{0}
 In this section we show that two charged representations of $\AA$,
$\pi_\r$ and $\pi_\r'$, are {\em dynamically equivalent} if their total charges
coincide, $q_{\r_1}=q_{\r_2}$. Here we take as an appropriate notion of
equivalence the following

\begin{definition}\label{3.3.1}
Two representations, $\pi$ and $\pi'$, of $\AA$ are called dynamically
equivalent, if
they both extend to representations of $\AAA$ and if there exists a
unitary intertwiner
$U:\,\calH_\pi\to\calH_{\pi'}$ such that
$U\,\pi(A) = \pi'(A)\,U$
for all $A\in\AA$ and $U\,\pi(\ttt) = c\,\pi'(\ttt)\,U$
for some constant $c>0$. $ \EndofStatement$
\end{definition}

The use of the constant $c$ is to allow for the
possibility of different zero-point normalizations of the
energy. Clearly, by rescaling the global transfer matrices one
can always achieve $c=1$. To prove that for $q_\rho=q_\r'$ the
representations $\pi_\r$ and $\pi_\r'$ are dynamically
equivalent in this sense we first show

\begin{proposition}\label{Phiroro'}
Let $q\in\Z_N\x\Z_N$ be fixed and let $\r,\r'\in\calD_q$. Then
there exists a unique (up to a phase) unit vector
$\Phi_{\r,\r'}\in\calH_\r$ such that
$ T_\r(\r')\,\Phi_{\r,\r'} = \|T_\r(\r')\|\,\Phi_{\r,\r'}$.
Moreover, $\Phi_{\r,\r'}\in\calH_\r$ is cyclic for $\pi_\r(\AA)$
and it induces the state $\om_{\r'}$, i.e.
\beq
(\Phi_{\r,\r'},\,\pi_\r(A)\Phi_{\r,\r'}) =
  \om_{\r'}(A),\quad\forall A\in\AA .
\eeq
$\EndofStatement$
\end{proposition}

\bsn{\bf Proof:}
We adapt the proof of Theorem 6.4 of \cite{FredMarcu} to our setting. Hence
we pick 1-cochains $\ell_\ve,\,\ell_m\in\cc_{loc}^1$ such
that $(d^*\ell_e,\,d\ell_m) = \r-\r'\in\calD_0$ and define
\be
\Phi_{\r,\r'}^n := \frac{T_\r(\r')^n \Om_{\r,\r'}}
{\|T_\r(\r')^n \Om_{\r,\r'}\|}
\label{3.47}
\ee
where
\bear
\Om_{\r,\r'} &:=& \pi_\r(A_{\r,\r'}(\ell_e,\ell_m))\,\Om_\r
\label{Omroro'}
\\
A_{\r,\r'}(\ell_e,\ell_m) &:=&
Z(\mu')^{-1/2}M(\ell_m)Q_{GH}(\ell_e)Z(\mu)^{1/2}
\label{Aroro'}
\eear
and where $\mu,\,\mu'$ denote the magnetic components in
$\r,\,\r'$.
Using the definitions
(\ref{Ryfgsjdfghlsdkjfhgls}-\ref{ppopopppidiusugaasa}) we check in
Appendix \ref{CompletingtheProofofPropositionPhiroro}
that the sequence of states
$(\Phi^n_{\r,\r'}\,,\,\pi_\r(A)\Phi^n_{\r,\r'})$ converges to
$\om_{\r'}(A)$ for all $A\in\AA_{loc}$.
We also show, that $\Phi^n_{\r,\r'}$ is
actually a Cauchy sequence in $\calH_\r$ and therefore
\be
\Phi_{\r,\r'} :=\lim_n\Phi^n_{\r,\r'}\ \in\calH_\r
\label{3.50}
\ee
exists. By construction it is an eigenvector of $T_\r(\r')$ with
eigenvalue
\be
\lambda_{\r'}  = \lim_n\;  \frac{\|T_\r(\r')^{n+1}
\Om_{\r,\r'}\|}{\|T_\r(\r')^n \Om_{\r,\r'}\|}
 =  \lim_n
(\Phi^n_{\r,\r'}\, , \,T_\r(\r')^2\,\Phi^n_{\r,\r'})^{1/2}
\label{eigvalue}
\ee
Let now $\calH_{\r,\r'} :=
\overline{\pi_\r(\AA)\Phi_{\r,\r'}}\subset\calH_\r$.
Then $T_\r(\r')\calH_{\r,\r'}\subset\calH_{\r,\r'}$ and
therefore $\pi_\r(\AAA)\calH_{\r,\r'}\subset\calH_{\r,\r'}$ by
(\ref{tt}). Hence $\calH_{\r,\r'}=\calH_\r$ by Corollary
\ref{tttrttrrusfkjhdfgus}.
By the cluster property of $\om_{\r'}$ with respect to
$\al_{\r'}$ we conclude $\lambda_{\r'}= \|T_\r(\r')\|$ and
similarly as in (\ref{2.43}) the associated eigenspace must
be 1-dimensional. 

$\Fullbox$

For later purposes we emphasize that Proposition \ref{Phiroro'}
in particular implies that the choice of $(\ell_e,\,\ell_m)$ in
(\ref{Omroro'}) only influences the phase of the limit vector
$\Phi_{\r,\r'}$ in (\ref{3.50}).

Using Proposition \ref{Phiroro'} we now get the
equivalence of charged representations whenever their total charge coincides.

\begin{theorem}
\label{sectors}
Let $\r,\,\r'\in\calD$ and $q_{\r} = q_{\r'}$. Then
$\pi_\r$ and $\pi_{\r'}$ are dynamically equivalent and
\be
\frac{\|T_{\r}(0)\|}{\|T_{\r'}(0)\|} =
\frac{\|T_{\r}(\r'')\|}{\|T_{\r'}(\r'')\|}\quad\forall\r''\in\calD
\label{normratio}
\ee
\end{theorem}

\bsn{\bf Proof:}
Let $U:\,\calH_{\r'}\to\calH_\r$ be given on
$\pi_{\r'}(\AA)\Om_{\r'}$ by
\be
U\pi_{\r'}(A)\Om_{\r'} := \pi_\r(A)\Phi_{\r,\r'}\ .
\label{gyobyuierteower}
\ee
Then $U$ extends to a unitary intertwining $\pi_{\r'}$ and $\pi_\r$.
Moreover, since $T_{\r'}(\r') $ and $T_{\r}(\r')$ implement
$\al_{\r'}$ on $\calH_{\r'}$ and $\calH_\r$, respectively,
one immediately concludes
\be
UT_{\r'}(\r')\|T_{\r'}(\r')\|^{-1} = T_\r(\r')\|T_\r(\r')\|^{-1}U
\ee
Using (\ref{ooqqqppapjdbfsiuh}) this implies
\be
UT_{\r'}(\r'')\|T_{\r'}(\r')\|^{-1} = T_\r(\r'')\|T_\r(\r')\|^{-1}U
\quad\forall\r''\in\calD
\label{equtr}
\ee
and therefore $\pi_\r$ and $\pi_{\r'}$
are dynamically equivalent in the sense of Definition \ref{3.3.1}.
Finally, (\ref{equtr}) immediately implies (\ref{normratio}).
\Fullbox

We now recall that fixing $\|T_\r(\r)\|\equiv e^{-E_\r}$ for a
given $\r\in\calD$ amounts to fixing $\|T_\r(\r'')\|$ for any
other $\r''\in\calD$. Hence,
Theorem \ref{sectors} shows that the energy normalization
$E_\r,\, \r\in\calD_q,$ in (\ref{udybuvdywp}) may be fixed up to a constant
depending only on the total charge $q$
by requiring ${\|T_{\r}(\r'')\|} = {\|T_{\r'}(\r'')\|}$ for all
$\r,\r'\in\calD_q$ and some (and hence all) $\r''\in\calD$.
In particular we now take $\r''= 0$ and require
\be
\|T_\r(0)\| = e^{-\epsilon(q)},\qquad \quad\forall\r\in\calD_q
\label{IonEn}
\ee
for some - as yet undetermined - function $\epsilon(q)\in{\Bbb R}$
describing the ``minimal self-energy'' in the sector
with total charge $q\in\Z_N\x\Z_N$.
In the next subsection we will appropriately fix $\epsilon(q)$
such that
$2\epsilon(q)\equiv\epsilon(q)+\epsilon(-q)$ gives the minimal
energy needed to create a pair of
charge-anticharge configurations of total charge $\pm q$ from
the vacuum and separate them apart to infinite distance.

We close this subsection by remarking that we expect conversely
(at least for a generic choice of couplings in the free charge
phase) that the representations $\pi_\r$ and $\pi_{\r'}$ are
dynamically inequivalent if $q_\r\neq q_{\r'}$. Inequivalence
to the vacuum sector for $q_\r\neq 0$ can presumably be shown by
analogue methods as in \cite{FredMarcu}, i.e. by proving the
absence of a translation invariant vector in $\calH_\r$ (see
Section 4.3 for the implementations of translations in the
charged sectors).
More generally, one could try to prove
that $\|T_\r(\r')\|$ is not an eigenvalue of $T_\r(\r')$ if
$q_\r\neq q_{\r'}$. To this end one would have to show that
$q_\r\neq q_{\r'}$ implies
\beq
\mbox{s}-\lim_n\frac{T_\r(\r')^n}{\|T_\r(\r')\|^n}\,\pi_\r(A)\Om_\r = 0
\eeq
for all $A\in\AA_{loc}$.
As in Proposition 5.5 of \cite{FredMarcu} a sufficient
condition for this would be
\beq
\lim_{n\to\infty}\frac{\pi_\r(A)\Om_\r,\,T_\r(\r')^n \pi_\r(A)\Om_\r)^2}
{\pi_\r(A)\Om_\r,\,T_\r(\r')^{2n} \pi_\r(A)\Om_\r)} = 0
\label{stringratio}
\eeq
Using (\ref{ooqqqppapjdbfsiuh}) this would amount to finding
the asymptotics of the vertical string expectation value
\beq
\om_\r\left(A^*\prod_{k=0}^{n-1}\al_\r^k(L_{\r'}L_\r^{-1})\al_\r^n(A)\right).
\label{BriFr}
\eeq
Similarly as in the Bricmont-Fr\"ohlich criterion for the
existence of free charges \cite{BriFr} equation (\ref{stringratio}) would follow
provided (\ref{BriFr}) would decay like
$n^{-\al}e^{-n\cdot{const}}$ for some $\al>0$ (in
\cite{BriFr} $\al = d/2$). Since we have not tried to prove
this we only formulate the

\begin{conjecture}\label{conj1}
If $q_\r\neq q_{\r'}$ then $\|T_\r(\r')\|$ is not an eigenvalue
of $T_\r(\r')$.
\end{conjecture}

Together with Proposition \ref{Phiroro'} and Theorem
\ref{sectors} the Conjecture \ref{conj1} would imply that
$\pi_\r$ and $\pi_{\r'}$ are dynamically equivalent if {\em and only
if} their total charges coincide.

\subsection{The Dyonic Self Energies}
\setcounter{theorem}{0}

In this subsection we determine the ``dyonic selfenergies"
$\epsilon(q)$ introduced in (\ref{IonEn}).
In order to get charge conjugation invariant self energies it will be useful
to start with
implementing the charge conjugation as an intertwiner
$C_\rho$: $ \calH_\rho \to \calH_{-\rho}$ by putting for
$A\in {\frak A}$ and $ \rho \in \calD$
\be
     C_\rho \pi_{\rho} (A) \Om_\rho := \pi_{-\rho}(i_C(A))\Om_{-\rho} .
\ee
\begin{lemma}\label{OOPOIoiuiugflsfdgsfdgsdfsffsdfsdf}
 $C_\rho$ extends to a well defined unitary
$\calH_{\rho} \to \calH_{-\rho}$ such that
\begin{enumerate}
\item
  $\mbox{Ad }C_\rho \circ \pi_\rho = \pi_{-\rho} \circ i_C$.
\item
  $C_\rho^* = C_{-\rho}$.
\item
  $C_\rho T_\rho (\rho ' ) e^{E_\rho}=
   T_{-\rho}(-\rho ')e^{E_{-\rho}}C_{\rho}$. $\EndofStatement$
\end{enumerate}
\end{lemma}

\bsn{\bf Proof:}
By equation (\ref{omch}) $C_\rho$ is well defined, unitary
and obeys i). Part ii) follows from $i_C^2 = \mbox{id}$.
Finally, in the case $ \rho'=\rho$ part iii) follows from equation
(\ref{cdelro}).
By (\ref{ooqqqppapjdbfsiuh}) the case
$\rho' \neq \rho$ follows from
$i_C(L_\rho)=L_{-\rho}$ $. \Fullbox$

By charge conjugation symmetry we would of course like to have
$E_\rho = E_{-\rho}$. We now fix the self-energies $\epsilon(q)$
in (\ref{IonEn}) of the sectors $q\in\Z_N\x\Z_N$ by first putting the vacuum
energy to zero, $\epsilon(0)=0$, i.e.
\be
   \|T_\rho (0)\| = 1 , \quad \forall\rho \in \calD_0\ .
\label{XXfgdfjytfkjgkdgsld}
\ee
This fixes
$
e^{-E_\rho}\equiv \|T_\rho (\rho)\|=
\|\pi_\rho(L_\rho)T_\rho (0)\|
$
for all $\rho \in \calD_0$ and in particular implies
$E_\r=E_{-\r},\ \forall\r\in\calD_0$. In fact, we have

\begin{lemma}
For fixed $q\in\Z_N\x\Z_N$ we have
$\epsilon(q)=\epsilon(-q)$ if and only if $E_\r=E_{-\r},\
\forall\r\in\calD_q$. $\EndofStatement$
\end{lemma}

\bsn{\bf Proof:}
Putting $\r'=0$ in
Lemma \ref{OOPOIoiuiugflsfdgsfdgsdfsffsdfsdf} iii) we see that
$\|T_\rho (0)\|=\|T_{-\rho} (0)\|,\ \forall\r\in\calD_q$, is
equivalent to $E_\r=E_{-\r},\
\forall\r\in\calD_q$, since $C_\r$ is unitary. \Fullbox

We now show that under the conditions $\ep(0)=0$ and
$\ep(q)=\ep(-q)$ the self-energies of the charged sectors are
completely determined
if we adjust them such that
$2\ep(q)\equiv\ep(q)+\ep(-q)$ becomes equal to the
minimal energy needed to create a pair of
charge-anticharge configurations of total charge $\pm q$ from
the vacuum and separate them apart to infinity.
Physically this means that for infinite separation we can
consistently normalize the interaction energy between two charges to zero.
Mathematically this is expressed by a factorization
formula similar to Theorem \ref{uuiupdsypspYUIYipkjhlkfhg}
for matrix elements of the global transfer matrix . As it will turn out,
this is further equivalent to an additivity property of the modified
ground state energies, i.e. $E_{\r+\r'\cdot a}\to E_\r+E_{\r'}$ as
$|a|\to\infty$.

\begin{theorem}\label{selfenergy}
There exists precisely one assignment of self-energies
$\Z_N\x\Z_N\ni q\to\ep(q)\in\R$ such that
$\|T_\r(0)\| = e^{-\ep(q)},\quad\forall\r\in\calD_q$
and such that the following
conditions hold
\bear
	&
	\ba{rrcl}
		\hspace{-1.5cm}{\sf i)} &\qquad\qquad \ep(0) & = & 0\\
		{\sf ii)} &\qquad\qquad  \ep(q) & = & \ep(-q)\\
		{\sf iii)}&
		\multicolumn{3}{l}{\mbox{
		The factorization formula (\ref{factorize})
		extends to transfer matrices, i.e.}}
	\ea
	&
	\nonumber\\
	  &
	\displaystyle \lim_{|a|\to\infty}\;
	\left(
	\pi_{\rho (a)}(A \tau_{-a} (B )) \Om_{\rho (a)}
	\,, \;
	T_{\rho (a)}(0)^n
	\pi_{\rho (a)}(A' \tau_{-a} (B' )) \Om_{\rho (a)}
	\right)
	&
	\nonumber\\
	&
	 =  \left(
	\pi_{\rho_1}(A) \Om_{\rho_1}
	\,, \;
	T_{\rho_1}(0)^n
	\pi_{\rho_1}(A' ) \Om_{\rho_1}
	\right)\;
	\left(
	\pi_{\rho_2}(B ) \Om_{\rho_2}
	\,,\;
	T_{\rho_2}(0)^n
	\pi_{\rho_2}( B') \Om_{\rho_2}
	\right)
	&
	\label{Tfactorize}
\eear
where we used the notation of Theorem
\ref{uuiupdsypspYUIYipkjhlkfhg}.
$\EndofStatement$
\end{theorem}

To prove Theorem \ref{selfenergy} and in particular the
cluster property \ref{Tfactorize} we have to control norm ratios
of (modified) transfer matrices, for which we have to invoke our polymer
expansions. First we have

\begin{lemma}
\label{Xroro'}
Let $\r,\,\r'\in\calD_q$ for some $q\in\Z_N\x\Z_N$ and choose 1-cochains
$\ell_e,\,\ell_m\in\cc_{loc}^1$ such that
$(d^*\ell_e,\,d\ell_m) = \r-\r'$. Let
$A_{\r,\r'}\equiv A_{\r,\r'}(\ell_e,\,\ell_m)\in\AA_{loc}$ be defined
as in
(\ref{Aroro'})  and put
\be
X_{\r,\r'}^{(\nu)}:=A_{\r,\r'}^*
\left[\prod_{k=0}^{\nu-1}\al_\r^k(L_{\r'}L_\r^{-1})\right]
\al_\r^\nu(A_{\r,\r'})\ .
\ee
Then we have
\be
\frac{\|T_\r(\r')\|}{\|T_\r(\r)\|} =
\lim_{\nu\to\infty}
\frac{\om_\r(X_{\r,\r'}^{(2\nu+1)})}{\om_\r(X_{\r,\r'}^{(2\nu)})}.
\label{3.61}
\ee
$\EndofStatement$
\end{lemma}

\bsn{\bf Proof:}
By Proposition \ref{Phiroro'} and equations (\ref{3.47}) - (\ref{3.50}) we have for
all $q\in\Z_N\x\Z_N$ and all $\r,\,\r'\in\calD_q$
\be
\|T_\r(\r')\| = (\Phi_{\r,\r'}\,,\,T_\r(\r')\Phi_{\r,\r'})
= \lim_{\nu\to\infty}\frac{( \Om_{\r,\r'},\;T_\r(\r')^{2\nu+1}
\Om_{\r,\r'})}{ (\Om_{\r,\r'},\;T_\r(\r')^{2\nu} \Om_{\r,\r'})}
\label{3.62}
\ee
where
$$
\Om_{\r,\r'} := \pi_\r(A_{\r,\r'}(\ell_e,\ell_m))\,\Om_\r
$$
has been given in (\ref{Omroro'}). The Lemma follows by
using (\ref{ooqqqppapjdbfsiuh}) to rewrite
$$
T_\r(\r')^\nu =\left[\prod_{k=0}^{\nu-1}\al_\r^k(L_{\r'}L_\r^{-1})\right]
T_\r(\r)^\nu
$$
and recalling $T_\r(\r)\Om_\r=\|T_\r(\r)\|\Om_\r$ by Theorem
\ref{occooiajjduyyy}. \Fullbox

Note that the norm ratio (\ref{3.61}) is obviously independent of the choice
 of the scale $e^{E_\r}\equiv\|T_\r(\r)\|$.
 We also recall from Proposition \ref{Phiroro'} that the choice of $(\ell_e,\,
 \ell_m)$ only influences the phase of $\Phi_{\r,\r'}$ and hence the limit
 $\nu\to\infty$ in (\ref{3.61}) is in fact independent of this choice.
When expressing the r.h.s. of (\ref{3.61}) in terms of euclidean path integrals
 we get     a ratio of expectations of the form (\ref{wwsrdhfdcvsjydobcfsled}),
 where the classical function corresponding to $X_{\r,\r'}^{(\nu)}$ is itself
 a superposition of Wilson loops and vortex loops.
 These are built by the same recipes as before, i.e. with horizontal
 parts given by $\pm (\ell_e,\,\ell_m)$ in the time slices $t=\nu$ and $t=0$,
 respectively, and with vertical parts joining the endpoints of these strings.
 Hence we get the ratio of euclidean path integral expectations of two such
 loop configurations with time-like extension $\nu+1$ and $\nu$,
 respectively, in the limit $\nu\to\infty$.
Here these expectations have to be taken in the background determined
by $\r$, i.e. in the presence of the vertical parts of $\calL_E$ and
$\calL_M$ described in (\ref{wwsrdhfdcvsjydobcfsled}), where there the
limit $n\to \infty$ and $V\to\infty$ has to be taken first.
\footnote{This is why in our polymer expansion only the vertical parts
 of $\calL_E$ and $\calL_M$ matter, since the horizontal parts run in
 the time-like boundary of $V_n$ and since the contributions from
 polymers reaching from the support of a classical function $A_{cl}$
 to this boundary decay to zero for $n\to \infty$.}

Next we have to control certain factorization properties of the above
norm ratios in the limit of the charge distributions being separated
to infinity.

\begin{proposition}\label{Tratio}
Let $\r_1,\ \r_2\in\calD$
and for $a\in\Z^2$ put
$\r_1(a)=\r_1-\r_1\cdot a,\ \r_2(a)=\r_2-\r_2\cdot a$. Then

i) The limit
\be
c_{\r_1,\,\r_2}:=\lim_{|a|\to\infty}
\frac{\|T_{\r_1(a)}(\r_2(a))\|}{\|T_{\r_1(a)}(\r_1(a))\|}>0
\label{3.63}
\ee
exists independently of the sequence $a\to\infty$ and satisfies
$c_{\r_1\cdot g,\,\r_2\cdot g} = c_{\r_1,\,\r_2},\ \forall g\in \calS$.
Moreover, for $b\in\Z^2$ we have the factorization property
\be
\lim_{|b|\to\infty} c_{\r_1+\r'_1\cdot b,\,\r_2 +\r'_2\cdot b}=
c_{\r_1,\,\r_2}c_{\r'_1,\,\r'_2}
\label{3.64}
\ee

ii) If $q_{\r_1}=q_{\r_2}$ then
\be
c_{\r_1,\,\r_2}
=\frac{\|T_{\r_1}(\r_2)\|^2}{\|T_{\r_1}(\r_1)\|^2}. 
\label{3.65}
\ee
\end{proposition}
$\EndofStatement $

\bsn
{\bf Proof:}
Part i) is proven in detail in Appendix \ref{ProofofPropositionTratio}.
Here we just remark that it can roughly be understood from the
perimeter law of Wilson and vortex loop expectations,
which guarantees that ratios of loop expectations converge for
infinitely large loops
if the difference of their perimeters stays finite
(in the above case this difference is given by two lattice units, due
to the difference by one unit in time extension, see
(\ref{3.61})).

To prove ii) we pick in (\ref{3.62}) the choice
\be
\Om_{\r_1(a),\,\r_2(a)} =
\pi_{\r_1(a)}\left(A_{\r_1(a),\,\r_2(a)}(\ell_e(a),\,\ell_m(a))\right)
\Om_{\r_1(a)}
\ee
where we take
$$
(\ell_e(a),\,\ell_m(a))=(\ell_e-\ell_e\cdot a,\,\ell_m-\ell_m\cdot a)
$$
for some fixed configuration $(\ell_e,\,\ell_m)$ satisfying
$(d^*\ell_e,\,d\ell_m)=\r_1-\r_2$.
With this choice it is not difficult to check that for $|a|$ large
enough
\be
A_{\r_1(a),\,\r_2(a)}(\ell_e(a),\,\ell_m(a))=
A_{\r_1,\,\r_2}(\ell_e,\,\ell_m)
\tau_{-a}(A_{\r_1,\,\r_2}(\ell_e,\,\ell_m))
\ee
Similarly, for fixed $\nu$ and $|a|$ large enough
\be
X^{(\nu)}_{\r_1(a),\,\r_2(a)}=
X^{(\nu)}_{\r_1,\,\r_2}\,\tau_{-a}(X^{(\nu)}_{\r_1,\,\r_2})
\label{3.68}
\ee
When proving part i) in Appendix \ref{ProofofPropositionTratio}
we show that plugging
(\ref{3.68}) into (\ref{3.61}) the limit $\nu\to
\infty$ is uniform in $a$ and hence we may take the limit
$|a|\to\infty$ first. However, this takes us into the setting of the
factorization formula (\ref{factorize}) implying
\be
\lim_{|a|\to\infty}\om_{\r(a)}(X^{(\nu)}_{\r(a),\,\r'(a)})=
\om_{\r_1}(X^{(\nu)}_{\r_1,\,\r_2})^2\ ,
\ee
which proves part ii).
\Fullbox

\bsn{\bf Proof of Theorem \ref{selfenergy}:}
The idea of the proof consists of translating the physically
motivated conditions {\sf i)-iii)} into conditions on the family
of constants $E_\r,\ \r\in\calD$.
We have already remarked in (\ref{XXfgdfjytfkjgkdgsld}) that
{\sf i)} may be obtained by suitably fixing $E_\r$ for all
$\r\in\calD_0$.
In Lemma 3.4.2 we have noticed that {\sf ii)} is equivalent
to $E_\r=E_{-\r}$ for all $\r$.
We now show that {\sf iii)} holds if and only if for all
$\r_1,\r_2\in\calD$
\be
 \displaystyle \lim_{|a|\to\infty}  \;
E_{\r_1+\r_2\cdot a} = E_{\r_1} + E_{\r_2}
\label{Efactorize}
\ee
To this end we use
\be
T_\r(\r)^n = \left[ \pi_\r(L_\r)T_\r(0)\right]^n
=\pi_\r(Y_\r^{(n)})T_\r(0)^n
\ee
where
\be
Y_\r^{(n)}:= \prod_{k=0}^{n-1}\al_0^k(L_\r)
\ee
Furthermore, for $|a|$ large enough and $\r(a)=\r_1+\r_2\cdot a$ we
have
\be
L_{\r(a)}=L_{\r_1}\tau_{-a}(L_{\r_2})=\tau_{-a}(L_{\r_2})L_{\r_1}\ .
\ee
Using that $\al_0$ commutes with $\tau_{-a}$ we get for
large enough $|a|$ and $A,A',B,B'\in\AA_{loc}$
\beanon
	&\left(
	\pi_{\rho (a)}(A \tau_{-a} (B )) \Om_{\rho (a)}
	\,, \;
	T_{\rho (a)}(0)^n
	\pi_{\rho (a)}(A' \tau_{-a} (B' )) \Om_{\rho (a)}
	\right)&
\\
	& = \omega_{\rho(a)}
	\left(
	\tau_{-a}(B^*[Y^{(n)}_{\r_2}]^{-1})
	A^*[Y^{(n)}_{\r_1}]^{-1}
	\al_{\rho (a)}^{n}(A' \tau_{-a}(B'))
	\right) e^{-nE_\r(a)}\ .
\eeanon
Here we have chosen $|a|$ large enough (depending on $n$)
such that $A^*$ commutes with
$
\tau_{-a}\left(Y_{\r_2}^{(n)}\right)^{-1}
$.
On the other hand, we get for $i=1,\,2$ and all
$A,A'\in\AA_{loc}$
\be
	\left(
	\pi_{\rho_i}(A) \Om_{\rho_i}
	\,, \;
	T_{\rho_i}(0)^n
	\pi_{\rho_i}(A' ) \Om_{\rho_i}
	\right) =
	\omega_{\rho_i}
	\left( A^*
	[Y_{\r_i}^{(n)}]^{-1}
	\al_{\rho_i }^{n}(A' )
	\right)
	 e^{-nE_{\r_i}}\ .
\ee
Thus, by Theorem \ref{uuiupdsypspYUIYipkjhlkfhg} the
factorization formula (\ref{Tfactorize}) is equivalent to
(\ref{Efactorize}).

Hence, there can be at most one solution $\ep$ of the
conditions {\sf i)-iii)} above. Indeed, equations (\ref{Efactorize})
together with $E_\r=E_{-\r}$ imply
\be
E_\r = \frac{1}{2} \lim_{|a|\to\infty} E_{\r-\r a}
\label{Edef}
\ee
for all $\r\in\calD$.  Since $\r-\r\cdot a\in\calD_0$ and since
condition {\sf i)}
completely fixes $E_\r=E_{-\r}$ for all $\r\in\calD_0$,
equation (\ref{Edef}) fixes $E_\r=E_{-\r}$ uniquely for all
$\r\in\calD$.

We are left to prove that given  $E_\r=E_{-\r}$ for all
$\r\in\calD_0$ via condition {\sf i)} the limit  (\ref{Edef}) indeed
exists for all $\r\in\calD$ and satisfies (\ref{Efactorize}).
However this is an immediate consequence of Proposition
\ref{Tratio}. Indeed, putting $\r_2=0$ in (\ref{3.63}) and
using $\|T_{\r_1(a)}(0)\|=1$ by condition {\sf i)} we get for
all $\r\in\calD$
\be
\lim_{|a|\to\infty}e^{E_{\r-\r\cdot a}} = c_{\r,\,0}
\ee
and hence (\ref{Edef}) implies
\be
E_\r = \frac{1}{2}\ln c_{\r,\,0}
\label{Erho}
\ee
Note that for $q_\r=0$ this is consistent with (\ref{3.65}).
The cluster property (\ref{Efactorize}) now is an immediate
consequence of (\ref{3.64}). This concludes the proof of
Theorem \ref{selfenergy}.
\Fullbox

We remark that we have not worked out an analytic expression
for the ratio $\frac{\|T_{\r}(0)\|}{\|T_{\r}(\r)\|}$ in the
case $q_\r\neq 0$, which is why we do not have further analytic
knowledge of the dyonic self-energies $\epsilon(q)$. In
particular we have not tried to confirm the natural ``stability
conjecture'' $\epsilon(q)>0$ for all $q\neq 0$. In the purely
electric (or purely magnetic) sectors the existence of 1-particle
states \cite{FlorianBarataI}
implies that $e^{-\epsilon (q_\r)} \equiv \|T_\r(0)\|$
lies in the spectrum of $T_\r(0)$, i.e. $\epsilon(q_\r)$ is
precisely the 1-particle self-energy at zero momentum.

\section{State Bundles and Intertwiner Connections}
\label{StateBundlesandIntertwinerConnections}
\zerarcounters
\setcounter{theorem}{0}

In this section we consider an appropriate analogue of what in the
DHR-theory of super selection sectors would be called the state
bundle. In our setting, by this we mean the collection of all GNS
triples $({\cal H}_\rho , \, \pi_\rho , \, \Om_\rho )$
of cyclic representation of $\hat{\frak A}$
obtained from the family of states $\omega_\rho$, $\rho \in {\cal D}$.
Modulo Conjecture \ref{conj1}, these representations fall into
equivalence classes labelled by their total charges $q_\rho$.
Hence we obtain, for each value of the global charge
$q \in \Z_N \times \Z_N$, a Hilbert-bundle $\calB_q$
over the discrete base space $\calD_q$, which is simply given as the
disjoint union
\be
		\calB_q := \dot{\bigcup_{\rho \in \calD_q}} \calH_\rho .
\ee
The fibers $ \calH_\rho $ of this bundle are all naturally isomorphic
as $ \hat{\frak A}$-modules by Theorem \ref{sectors}. Since
$
\pi_\rho (\hat{\frak A})
$
acts irreducibly on $ \calH_\rho$ by Corollary
\ref{tttrttrrusfkjhdfgus}, these isomorphisms are all uniquely
determined up to a phase.

When trying to fix this phase ambiguity one is naturally led
to the problem of
constructing an {\em intertwiner connection} on $ \calB_q$, i.e. a family of
intertwiners $ U(\Gamma ): \; \calH_\rho \to \calH_{\rho'} $ satisfying
$
\pi_{\rho '} = \mbox{Ad } U(\Gamma ) \circ \pi_\rho
$
and depending on paths $ \Gamma $ in $ \calD_q $ from $ \rho$ to
$\rho '$.
Here, by a path in $ \calD_q$ we mean a finite sequence
$(\rho_0 , \; \rho_1 , \; \ldots , \rho_n )$
of charge distributions $ \rho_i \in \calD_q$, such that
$ \rho_i $ and $ \rho_{i+1} $ are ``nearest neighbours'' in $ \calD_q$
in a suitable sense to be explained below.

In order to be able to formulate a
concept of locality for these intertwiners
we will also consider the Hilbert direct sum
\be
   {\Bbb H}_q \; := \; \bigoplus_{\rho \in \calD_q} \; \calH_\rho
\ee
on which we let $A\in\hat{\frak A}$ be represented by
\be
   \Pi_q(A) \; := \bigoplus_{\rho \in \calD_q} \; \pi_\rho(A)\ .
\ee
The above mentioned connection will then be given in terms of an
intertwiner $ \Z_N$-Weyl algebra $ {\frak W}_q$ acting on
${\Bbb H}_q$ and commuting with
$ \Pi_q (\hat{\frak A})$.
Locality in this framework is formulated by the statement that
${\frak W}_q$ is generated by elementary ``electric string operators''
$ \calE_q (b)$ and elementary ``magnetic string operators''
$ \calM_q (b)$ satisfying for all oriented bonds $ b,b'\in (\Z^2 )_1$
\be
     \calE_q (b)^N = \calM_q (b)^N = \UM ,
\label{OISHBNDIOEURBGPVIOW}
\ee
and the local $ \Z_N$-Weyl commutation relations
\beq
\begin{array}{lcl}
  \calE_q (b) \calE_q (b' ) & = & \calE_q (b') \calE_q (b ) \\[.2cm]
  \calM_q (b) \calM_q (b' ) & = & \calM_q (b') \calM_q (b ) \\[.2cm]
  \calE_q (b) \calM_q (b' ) & = & e^{\frac{2 \pi i}{N}\delta_{b, \; b'}}
  \calM_q (b') \calE_q (b ) .
\end{array}
\label{liseuvhbeuifbpsuie}
\eeq
These operators will map each fiber $ \calH_\rho $, $ \rho \in \calD_q$,
isomorphically onto a ``neighbouring one'', i.e.
\be
\calE_q (b) \calH_\rho \; = \; \calH_{\rho + (d^* \delta_b , \; 0)},
\label{ueyvcbowiwe1}
\ee
\be
\calM_q (b) \calH_\rho \; = \; \calH_{\rho + (0 , \; d \delta_b)},
\label{ueyvcbowiwe2}
\ee
where $ \delta_b \in \calC^1_{loc}$ denotes the 1-cochain taking value
one on $ b$ and zero else. Hence, one might think of
$ \calE_q (b)$ ($ \calM_q (b)$) as creating an electric (magnetic)
charge-anticharge pair sitting in the  boundary (coboundary) of the
bond $ b$.  Defining charge distributions in $ \calD_q$ to be
{\em nearest neighbours}
if they differ by either such an elementary electric or magnetic
dipole, the connection in $ \calB_q$ along a path
$(\rho_0 , \; \rho_1 , \; \ldots , \rho_n )$ in $ \calD_q$ is
now given in the obvious way as an associated
product of $ \calE_q (b)$'s and
$ M_q (b')$'s, mapping each fiber $ \calH_{\rho_i}$
isomorphically onto its successor $ \calH_{\rho_{i+1}}$
according to (\ref{ueyvcbowiwe1})-(\ref{ueyvcbowiwe2}).

We remark that in a DHR-framework one would expect these intertwiners
to be given on the dense subspace
$ \pi_\r ({\frak A }) \Omega_\rho \subset \calH_\r$ in terms of
unitary localized charge transporters $ S_{el} (b)$ and
$ S_{mag} (b) \in {\frak A}_{loc}$ by
\bea
\calE_q (b) \pi_\rho (A) \Omega_\rho &=&
\pi_{\rho + (d^* \delta_b , \; 0)}
(A\,S_{el} (b)) \Omega_{\rho + (d^* \delta_b , \; 0 )} \\
\calM_q (b) \pi_\rho (A) \Omega_\rho &=&
\pi_{\rho + (0, \; d\delta_b )}
(A\,S_{mag} (b)) \Omega_{\rho + ( 0, \; d\delta_b )}\ ,
\eea
which would be consistent and well defined if
\bea
\omega_{\rho + (d^* \delta_b , \; 0 ) } & = &
\omega_\rho \circ \mbox{Ad }S_{el} (b)
\label{ueyrgbtoweuirybvoeuiybw1}
\\
\omega_{\rho + ( 0, \; d\delta_b ) } & =&
\omega_\rho \circ \mbox{Ad }S_{mag} (b) .
\label{ueyrgbtoweuirybvoeuiybw2}
\eea
Intuitively one could think of $ S_{el}$ as a kind of ``electric''
Mandelstam string operator as in (\ref{Mandelstam}) and of
$ S_{mag}(b)$ as its dual ``magnetic'' analogue. Unfortunately, due to
the non-local energy regularization in (\ref{hpdfzpbsdb})
(and more generally, in (\ref{Xum})), the existence of such localized
charge transporters $ S_{el}$ and $ S_{mag}$ in $ {\frak A}_{loc}$
satisfying (\ref{ueyrgbtoweuirybvoeuiybw1})-(\ref{ueyrgbtoweuirybvoeuiybw2})
is very questionable.

We also remark that, as opposed to the DHR-framework, in our
lattice model the states $\om_\r$ are not given in the form
$\om_\rho = \omega_0 \circ \gamma_\rho$ for some localized
automorphism $\gamma_\rho$ on ${\frak A}$. This is why we do not have a
field algebra extension of $ \frak A$ carrying a global
$\Z^N \times \Z^N  $ symmetry, i.e. we are not able to define charged
fields intertwining
$ \pi_\rho \circ \gamma_{\rho '}$ and $ \pi_{\rho + \rho '}$.

It is therefore astonishing and, in our opinion, asks for further
conceptual explanation that in this model one is nevertheless able to
construct an intertwiner algebra with local commutation relations as
given in (\ref{liseuvhbeuifbpsuie}). We will come back to this
question in \cite{FBIII}, where the existence of the local intertwiner
algebra (\ref{OISHBNDIOEURBGPVIOW})-(\ref{liseuvhbeuifbpsuie}) will be
the basic input for the construction of Haag-Ruelle scattering states
in the charged sectors.

\subsection{The Local Intertwiner Algebra}

We now proceed to the construction of the intertwiner algebra
$ {\frak W}_q $ obeying
(\ref{OISHBNDIOEURBGPVIOW})-(\ref{liseuvhbeuifbpsuie}).

For $ \rho$, $ \rho ' \in \calD_q$, let
$
P_\rho(\rho') : \;\calH_\rho \to \calH_\rho
$
be the one-dimensional projection onto the ground state of
$ T_{\rho} (\rho ')$. Using the notation of
(\ref{Omroro'})-(\ref{Aroro'}) we then define for arbitrary 1-cochains
$ l \in \calC^1_{loc}$
\bea
\phi^{el}_{\rho} (l) & = &
\frac{
P_\rho(\rho - (d^* l , \; 0))
\pi_{\rho} (A_{\rho , \; \rho - (d^* l , \; 0)} (l, \; 0)) \Omega_\rho
}{
\left\|
P_\rho( \rho - (d^* l , \; 0))
\pi_{\rho} (A_{\rho , \; \rho - (d^* l , \; 0)} (l, \; 0)) \Omega_\rho
\right\|
}
\\
\phi^{mag}_{\rho} (l) & = &
\frac{
P_\rho(\rho - (0, \; dl))
\pi_{\rho} (A_{\rho , \; \rho - (0, \; dl)} (0, \; l)) \Omega_\rho
}{
\left\|
P_\rho(\rho - (0, \; dl))
\pi_{\rho} (A_{\rho , \; \rho - (0, \; dl)} (0, \; l)) \Omega_\rho
\right\|
}\ ,
\eea
which are well defined unit vectors in $ \calH_\rho$, by
Proposition \ref{Phiroro'}. We may then define unitary operators
$ \hat{\calE}_q (l)$, $ \hat{\calM}_q (l)$ on $ {\Bbb H}_q$ commuting
with $ \Pi_q (\hat{\frak A})$ by putting on the dense set
$ \pi_\rho (A)\Omega_\rho$, $ A\in {\frak A}$, $ \rho \in \calD_q$,
\bea
\hat{\calE}_q (l) \pi_\rho (A) \Omega_\rho &:= &
\pi_{\rho + (d^* l , \; 0)} (A)
\phi_{\rho + (d^* l , \; 0) }^{el} (l )
\\
\hat{\calM}_q (l) \pi_\rho (A) \Omega_\rho &:= &
\pi_{\rho + ( 0, \; dl)} (A)
\phi_{\rho + ( 0, \; dl ) }^{mag} (l ),
\eea
which are well defined as in (\ref{gyobyuierteower}).

It turns out that these intertwiners do not yet obey local
commutation relations, however the violation of locality can
be described by a kind of ``coboundary'' equation.

\begin{theorem}
There exists an assignment of phases
$ z^{el}_\rho (l)$, $ z^{mag}_\rho (l) \in U(1)$ such that for  all
$ l_1 $ and $ l_2 \in \calC^1_{loc}$ and all $ \rho \in \calD_q$
\bear
i) & &
\left(
\hat{\calE}_q (l_1 + l_2) \Omega_\rho , \; \;
\hat{\calE}_q (l_1 )\hat{\calE}_q ( l_2) \Omega_\rho
\right)
=
\frac{
 z^{el}_{\rho + (d^* l_2 , \; 0)} (l_1 )\,z^{el}_\rho (l_2 )
}{
z^{el}_{\rho } (l_1 + l_2 ) .
}
\label{BTOVCWEIUsjs1}
\\
ii) & &
\left(
\hat{\calM}_q (l_1 + l_2) \Omega_\rho , \; \;
\hat{\calM}_q (l_1 )\hat{\calM}_q ( l_2) \Omega_\rho
\right)
=
\frac{
 z^{mag}_{\rho + ( 0 , \; dl_2)} (l_1 )\,z^{mag}_\rho (l_2 )
}{
z^{mag}_{\rho } (l_1 + l_2 )
} .
\label{BTOVCWEIUsjs2}
\\
iii) & &
\left(
\hat{\calM}_q (l_1 ) \hat{\calE}_q (l_2) \Omega_\rho , \; \;
\hat{\calE}_q (l_2 )\hat{\calM}_q ( l_1) \Omega_\rho
\right)
=
e^{ i \bra l_1 , \; l_2\ket} \;
\frac{
 z^{el}_{\rho + ( 0 , \; dl_1)} (l_1 )\,z^{mag}_\rho (l_1 )
}{
z^{mag}_{\rho + (d^* l_2 , \; 0)} ( l_1 )\,z^{el}_{\rho } ( l_2 )
}.
\label{BTOVCWEIUsjs3}
\\
iv) & &
\mbox{If } d^* l = 0 \mbox{ then } z_\rho^{el} (l) = 1 .
\label{BTOVCWEIUsjs4a}
\\
 & &
\mbox{If } d l = 0 \mbox{ then } z_\rho^{mag} (l) = 1 .
\label{BTOVCWEIUsjs4b}
\\
v) & &
z_\rho^{el/mag} (l) = z_{\rho \cdot g}^{el/mag} (l\cdot g),
 \; \; \forall g \in \calS .
\label{BTOVCWEIUsjs5}
\eear
$\EndofStatement$
\label{puiDyrbpowuie}
\end{theorem}

The proof of this theorem and
the precise definition of the phases $ z_\rho^{el} (l)$ and
$ z_\rho^{mag} (l)$ will be given in Appendix \ref{uyhvbngsSuiry}.

We now use these phases to define on $ {\Bbb H}_q$ the unitaries
$ Z_q^{el} (l) $ and $ Z_q^{mag}(l)$, $ l\in \calC^1_{loc}$, by
\bear
 Z_q^{el} (l) \restriction \calH_\rho & := & z_\rho^{el} (l) \UM_{\calH_{\rho}},
\label{peuircbyvwe1}
\\
 Z_q^{mag} (l) \restriction \calH_\rho & := & z_\rho^{mag} (l) \UM_{\calH_{\rho}},
\label{peuircbyvwe2}
\eear
Then $Z_q^{el} (l) $ and $ Z_q^{mag} (l)$ clearly also commute with
$ \Pi_q (\hat{\frak A})$ and we may define new intertwiners
\bear
 \calE_q (l) & := & \hat{\calE}_q (l) Z_q^{el} (l)^{-1},
\label{siuyfvbwo1}
\\
 \calM_q (l) & := & \hat{\calM}_q (l) Z_q^{mag} (l)^{-1},
\label{siuyfvbwo2}
\eear
which now obey local commutation relations.

\begin{theorem}
For all $ l\in \calC^1_{loc} $ the operators
$ \calE_q (l) $ and
$ \calM_q (l) $ are unitaries in $\calB({\Bbb H_q})$ commuting with
$\Pi_q (\hat{\frak A}) $. They satisfy the Weyl algebra relations
\bear
\calE_q (l_1 ) \calE_q (l_2 ) & = &
\calE_q (l_1 + l_2) ,
\label{wetyrvw4ryvw1}
\\
\calM_q (l_1 ) \calM_q (l_2 ) & = &
\calM_q (l_1 + l_2) ,
\label{wetyrvw4ryvw2}
\\
\calE_q (l_1 ) \calM_q (l_2 ) & = &
e^{i\bra l_1 , \; l_2\ket}
\calM_q (l_2 ) \calE_q (l_1 )
\label{wetyrvw4ryvw3}
\eear
and restricted to each subspace $ \calH_\rho \subset {\Bbb H}_q$,
$ \rho \in \calD_q$, we have
\bear
     \calE_q (l) \calH_\rho & = & \calH_{\rho + ( d^* l , \; 0  )} ,
\label{MNuhcoweSkjsh01}
\\
     \calM_q (l) \calH_\rho & = & \calH_{\rho + ( 0     , \; dl )} .
\label{MNuhcoweSkjsh02}
\eear
$ \EndofStatement $
\label{SSDywevcbrgowuyet}
\end{theorem}

{\bf Proof.} It remains to prove the identities
(\ref{wetyrvw4ryvw1})-(\ref{wetyrvw4ryvw3}), for which it is enough to
check them when applied to
$ \Omega_\rho \in \calH_\rho$, $ \forall \rho \in \calD_q$. First
we get
\be
   \calE_q (l_1 + l_2 ) \Omega_\rho \; = \;
z_\rho^{el} (l_1 + l_2 )^{-1} \;
\phi^{el}_{\rho + (d^* l_1 + d^* l_2  , \; 0)}
\label{YTfriuywbocqwuitw}
\ee
which is a unit vector in the image of the one dimensional projection
$ P_{\rho + (d^* l_1 + d^* l_2 , \;0)} (\rho )$. On the other hand we
have
\be
\calE_q (l_1 ) \calE_q (l_2 ) \Omega_\rho =
z^{el}_\rho (l_2 )^{-1} \calE_q (l_1 )
\phi^{el}_{\rho + ( d^* l_2 , \; 0 ) } (l_2 ) .
\label{Ypiuybfpcwsgtw}
\ee
Since
$ \calE_q (l_1 ) \restriction \calH_{\rho'}$ intertwines
$ \pi_{\rho '} (\hat{\frak A})$ and
$ \pi_{\rho ' + (d^* l_1 , \; 0)} (\hat{\frak A})$
we have
\be
\calE_q (l_1) T_{\rho '} (\rho )\restriction \calH_{\rho '} =
T_{\rho ' + (d^* l_1 , \; 0)}(\rho )
\calE_q (l_1 ) \restriction \calH_{\rho '},
\ee
implying
\be
\calE_q (l_1) P_{\rho '} (\rho ) \restriction \calH_{\rho '} =
P_{\rho ' + (d^* l_1 , \; 0)}(\rho )
\calE_q (l_1 ) \restriction \calH_{\rho '} .
\ee
Hence (\ref{Ypiuybfpcwsgtw}) implies
\be
\calE_q (l_1 ) \calE_q (l_2 ) \Omega_\rho
\in P_{\rho ''} (\rho ) \calH_\rho ,
\label{Reyhwbnpv}
\ee
where $ \rho '' = \rho + (d^*l_1 + d^*l_2, \; 0)$. Comparing
(\ref{YTfriuywbocqwuitw}) and (\ref{Reyhwbnpv}) we conclude that
$ \calE_q (l_1 ) \calE_q (l_2 ) \Omega_\rho$ and
$ \calE_q (l_1 + l_2)\Omega_\rho$ only differ by a phase. This phase
must be one since, by (\ref{BTOVCWEIUsjs1}) and
(\ref{siuyfvbwo1})-(\ref{siuyfvbwo2})
\be
\left(
\calE_q (l_1 + l_2 ) \Omega_\rho , \; \;
\calE_q (l_1 ) \calE_q (l_2 ) \Omega_\rho
\right) \; = \; 1 .
\ee
Equations (\ref{wetyrvw4ryvw2})-(\ref{wetyrvw4ryvw3}) are proven similarly.
$ \Fullbox $

Putting $ \calE_q (b) \equiv\calE_q (\delta_b )$
and
$ \calM_q (b)\equiv \calM_q (\delta_b )$,
Theorem \ref{SSDywevcbrgowuyet} provides the local intertwiner algebra
$ {\frak W}_q $ announced in
(\ref{OISHBNDIOEURBGPVIOW})-(\ref{liseuvhbeuifbpsuie}).

\subsection{The Intertwiner Connection}

We now turn to the bundle theoretic point of view, where we consider
the above intertwiners as a connection on $ \calB_q $. As already
explained, we call a pair $ \rho$, $ \rho' \in \calD_q$ nearest
neighbours if they differ by an elementary electric or magnetic
dipole, i.e. if there exists a bond $ b\in(\Z^2)_1$ such that
$ \rho ' - \rho = (\pm d^*\delta_b , \; 0)$ or
$ \rho ' - \rho = (0, \; \pm d\,\delta_b)$.

In the first case we put
\be
     \calU_{\rho' , \, \rho} =
\calE_q (b)^{\pm 1}: \quad \calH_\rho \to \calH_{\rho '}
\ee
and in the second case we put
\be
     \calU_{\rho' , \, \rho} =
\calM_q (b)^{\pm 1}: \quad \calH_\rho \to \calH_{\rho '} .
\ee

If
$\Gamma = (\rho_0 , \; \ldots , \; \rho_n ) $ is a path in $ \calD_q
$, i.e. a finite sequence of nearest neighbour pairs, then we put
\be
\calU(\Gamma ) \; := \;
\calU_{\rho_n \, \rho_{n-1}}
\cdots
\calU_{\rho_1 \, \rho_{0}} : \quad \calH_{\rho_0} \to \calH_{\rho_n}
\ee
as the associated ``parallel transport''. If $ \Gamma $ is a closed
path, i.e.
$ \rho_n = \rho_0$, then $ \calU(\Gamma)$ must be a phase by the
ireducibility of $ \pi_{\rho_0} (\hat{\frak A})$.

In order to determine these ``holonomy phases'' one may use the
commutation relation (\ref{wetyrvw4ryvw3}) which allows to restrict
ourselves to purely electric or purely magnetic loops $ \Gamma $
(i.e. where $\calU (\Gamma )$ only consists of a product of
$ \calE_q (b)$'s or $ \calM_q (b)$'s, respectively). Using
(\ref{wetyrvw4ryvw1})-(\ref{wetyrvw4ryvw2}) such loop operators
$\calU (\Gamma ) $ are always of the form
$ \calU (\Gamma ) = \calE_q (l_e )\restriction \calH_\rho$
or
$ \calU (\Gamma ) = \calM_q (l_m )\restriction \calH_\rho$,
where the loop condition on $\Gamma$ implies $ d^* l_e = 0$ or
$ dl_m = 0$, respectively. Note that this is consistent with
(\ref{MNuhcoweSkjsh01})-(\ref{MNuhcoweSkjsh02}), i.e. these
loop operators must map $ \calH_\rho$ onto itself.

To compute the holonomy phases we now use the fact that
$ d^* l_e = 0  $ and $ dl_m =0$ implies that there exist uniquely
determined cochains $ s \in \calC^{2}_{loc}$ and
$ k \in \calC^{0}_{loc}$ such that
\be
    d^* s \; =\; l_e , \qquad dk \; = \; l_m .
\ee
We also recall that in these cases Theorem \ref{puiDyrbpowuie} {\em iv)}
implies
$\calE_q (l_e ) = \hat{\calE}_q (l_e )$
and
$\calM_q (l_m ) = \hat{\calM}_q (l_m )$.
Hence, being a phase when restricted to
$ \calH_\rho $, it is enough to apply these operators to $ \Omega_\rho$
where, by the definitions
(\ref{peuircbyvwe1})-(\ref{peuircbyvwe2}) and
(\ref{siuyfvbwo1})-(\ref{siuyfvbwo2}),
they yield
$ \phi_\rho^{el } (l_e ) $
and
$ \phi_\rho^{mag} (l_m ) $, respectively.

Thus, we get our holonomy phases from

\begin{proposition}

Let $ s \in \calC^{2}_{loc} $ and $ k \in \calC^{0}_{loc}$ such that
$ d^* s = l_e$ and $ dk = l_m$. Then, for
$ \rho = (\epsilon , \; \mu )$ we have
\bear
	\phi_\rho^{el} (l_e) & =& e^{i\bra\mu, \; s\ket}
\Omega_\rho
\label{holonomyphase1}
\\
	\phi_\rho^{mag} (l_m) & =& e^{-i\bra\epsilon, \; k\ket}
\Omega_\rho .
\label{holonomyphase2}
\eear
$ \EndofStatement $
\label{Tfifuygorfwiue}
\end{proposition}

By the above arguments Proposition \ref{Tfifuygorfwiue} implies
\beq
\begin{array}{rcl}
 \calE_q (l_e ) \restriction \calH_\rho &=& e^{i\bra\mu, \; s\ket}\\[.2cm]
 \calM_q (l_m ) \restriction \calH_\rho &=& e^{i\bra\epsilon, \; k\ket}\ .
\end{array}
\label{hol}
\eeq
If we think of $ s$ being  the characteristic function of a surface
encircled by an electric loop $ l_e $, then the holonomy phase
(\ref{holonomyphase1}) is just the magnetic flux through this
surface. Interchanging electric and magnetic (and passing to the dual
lattice), equation (\ref{holonomyphase2}) may be stated analogously.
This state of affairs will be the origin of anyon statistics of
scattering states in \cite{FBIII}. Proposition \ref{Tfifuygorfwiue}
will be proven in Appendix \ref{ProofofPropositionTfifuygorfwiue}.

\subsection{The Representation of Translations}
\label{TheRepresentationofTranslations}

Using our local intertwiner algebra $ {\frak W}_q$  we are now in the
position to define on each fiber $ \calH_\rho $ a unitary
representation $ D_\rho (a)$, $ a\in \Z^2$, of the group of the
lattice translation, such that
$ \mbox{Ad }D_\rho (a) \circ \pi_\rho = \pi_\rho \circ \tau_a$ and
such that $ D_\rho (a)$ commutes with
$ T_\rho (0)$, $ \forall a \in \Z^2$. Moreover, for fixed $ q$, these
representations are all equivalent, i.e.,
$ \calU(\Gamma )D_{\rho}(a)= D_{\rho '}(a) \calU (\Gamma )$ for all
paths $ \Gamma : \; \rho \to \rho '$.

First we need a lift of the natural action of translations on
$\calD_q$ to bundle automorphisms on $\calB_q$. For the sake of
generality let us formulate this by including also the lattice
rotations.

\begin{lemma} For
$ g\in \calS$ let $ V_\rho (g): \; \calH_\rho \to \calH_{\rho \cdot g}$
be given by
\be
    V_\rho (g) \pi_\rho (A)\Omega_\rho \; := \;
\pi_{\rho \cdot g} (\tau_{g}^{-1} (A)) \Omega_{\rho \cdot g}.
\label{YTytoiweuphwver}
\ee
Then $ V_\rho (g)$ is a well defined unitary intertwining
$ \pi_\rho \circ \tau_g $ with $ \pi_{\rho \cdot g}$
\footnote{
Note that the $ *$-automorphism $ \tau_g \in \mbox{Aut }{\frak A}$ may
be extended to a $*$-automorphism on $ \hat{\frak A}$ by putting
$ \tau_g ({\bf t}) = {\bf t}$.}
, i.e.
\bear
V_\rho (g)\pi_\rho (\tau_g (A)) &  = &
\pi_{\rho \cdot g} (A) V_\rho (g),
\label{ueybvrpwuieybrpSi1}
\\
V_\rho (g) T_\rho (\rho ') &  = &
T_{\rho \cdot g}(\rho ' \cdot g) V_\rho (g),
\label{ueybvrpwuieybrpSi2}
\eear
for all $ A\in {\frak A}$.
$ \EndofStatement$
\end{lemma}

{\bf Proof.} $ V_\rho (g)$ is well defined and unitary since
$ \omega_{\rho} \circ \tau_g = \omega_{\rho \cdot g}$.
Equations
(\ref{ueybvrpwuieybrpSi1})-(\ref{ueybvrpwuieybrpSi2}) follow from
$\al_\rho \circ \tau_{\rho \cdot g}= \al_{\rho \cdot g}$ and
$ E_{\rho \cdot g} = E_\rho$ (by (\ref{Erho}) and
$\calS$-invariance of $c_{\r,\,\r'}$).
$\Fullbox$

Note that the definition (\ref{YTytoiweuphwver}) implies the obvious identity
\be
 V_{\rho \cdot g}(h) V_\rho (g) \; = \;V_\rho (g h) .
\label{Vrhog}
\ee
for all $g,\,h\in\calS$, which means
that the family of fiber isomorphisms $ V_\rho (g)$,
$ \rho \in \calD_q$, may indeed be viewed as a lift of the natural right
action of $ \calS $ on $ \calD_q$ to an action by unitary bundle
automorphism on $ \calB_q$.
Equivalently, we now consider
\be
{\Bbb V}_q (g) \; := \;
\bigoplus_{\rho \in \calD_q} \; V_\rho (g)
\ee
as a unitary representation of
$ \calS$ on $ {\Bbb H}_q$, satisfying
\be
    \mbox{Ad }{\Bbb V}_q (g) \circ \Pi_q \; = \;
\Pi_q \circ \tau_{g^{-1}} .
\ee
Moreover, we have

\begin{lemma}
\bear
     {\Bbb V}_q (g)\calE_q (l) & = & \calE_q (l \cdot g) {\Bbb V}_q (g),
\\
     {\Bbb V}_q (g)\calM_q (l) & = & \calM_q (l \cdot g) {\Bbb V}_q (g).
\eear
$ \EndofStatement$
\label{wuvybwpervwt}
\end{lemma}

{\bf Proof.} By Theorem \ref{puiDyrbpowuie} {\em v)} we have
\be
     {\Bbb V}_q (g) Z_q^{el/mag} (l) \; = \;
 Z_q^{el/mag} (l\cdot g)  {\Bbb V}_q (g) .
\ee
Hence it is enough to prove the claim with $\calE_q$, $ \calM_q$
replaced by $ \hat{\calE}_q $, $ \hat{\calM}_q $. However this is a
straight forward consequence of the definitions and the fact that
by (\ref{ueybvrpwuieybrpSi2})
\be
V_\rho (g) P_\rho (\rho ') \; = \;
P_{\rho \cdot g}(\r'\cdot g)V_\rho (g)
\ee
and, therefore,
\be
    V_\rho (g) \phi_{\rho}^{el/mag}(l) \; = \;
\phi_{\rho\cdot g}^{el/mag}(l\cdot g)  .
\label{vnhperovbret}
\ee
$ \Fullbox$

We remark that Lemma \ref{wuvybwpervwt} implies that the connection
$ \calU (\Gamma )$ is $ \calS$-invariant, i.e.
\be
    V_{\rho'} (g)\calU (\Gamma ) =
\calU (\Gamma \cdot g) V_\rho (g)
\ee
for all paths $ \Gamma : \; \rho \to \rho '$ and all
$ g\in \calS$.

In order to arrive at an implementation of the translation group
$ \Z^2 \subset \calS $ mapping each fiber $\calH_\rho $ onto itself we
now have to compose $ V_\r (a)^*$, $ a\in \Z^2$, with a
parallel transporter $U_\r(a):\calH_\r\to\calH_{\r\cdot a}$
along a suitable path $\Gamma:\r\to\r\cdot a$.
To this end it is enough to consider the
cases $ a = e_i\,$, $ i=1, \; 2$, where $ e_1=(1,\; 0)$ and
$ e_2 = (0, \; 1)$ denote the generators of $ \Z^2$.

A minute's thought shows that for any electric charge distribution
$ \epsilon \in \calC^0_{loc}$ there exists a unique 1-cochain
$ \ell_e (\epsilon , \; i) \in \calC^1_{loc}$ with support only on bonds
in direction $i$, such that the translated image $\epsilon\cdot e_i$
of $ \epsilon$ by one 
unit in direction $ i$ satisfies
\be
    \epsilon \cdot e_i \; = \; \epsilon + d^* \ell_e (\epsilon , \; i).
\ee
Similarly, for a magnetic charge distribution $ \mu \in \calC^2_{loc}$
there exists a unique 1-cochain $ \ell_m (\mu , \; i)$ with support
only on the bonds perpendicular to the direction $ i$, such that
\be
    \mu \cdot e_i \; = \; \mu + d \ell_m (\mu , \; i).
\ee
For $ \rho = (\epsilon , \; \mu)\in \calD_q$ we now define
$
U_\rho (e_i ) : \; \calH_\rho \to \calH_{\rho \cdot e_i}
$
by putting
\bear
U_\rho (e_i ) & := & \calE_q (\ell_e (\epsilon , \; i))
			   \calM_q (\ell_m (\mu      , \; i))\restriction
			   \calH_\rho
\nonumber \\
 & = &    \calM_q (\ell_m (\mu , \; i))
			   \calE_q (\ell_e (\epsilon, \; i))\restriction
			   \calH_\rho,
\label{IpiuaybciuwybS}
\eear
where the second equality follows since, by construction,
$ \ell_m (\mu, \; i)$ and $ \ell_e (\epsilon , \; i)$ always have
disjoint support.

With this construction we now define the unitaries
$ D_\rho (e_i ): \; \calH_\rho \to \calH_\rho$, $ i=1, \; 2$, by
\be
    D_\rho (e_i ) \; := \;
V_\rho (e_i )^* \; U_\rho (e_i )  .
\label{YTuytowirtypwierW}
\ee

\begin{theorem}
The unitaries $ D_\rho (e_i)$, $ i=1, \; 2$, generate a representation
of $ \Z^2$ implementing the translation automorphism on
$ \hat{\frak A}$, i.e.,
\bear
	D_\rho (e_1 ) D_\rho (e_2 ) & = &    D_\rho (e_2 ) D_\rho (e_1 )
\label{yetrvcqyuwdtvqo1}
\\
  D_\rho (e_i) \pi_\rho (A) & = & \pi_\rho
(\tau_{e_i}(A))D_\rho (e_i),
\label{yetrvcqyuwdtvqo2}
\eear
for all $A\in \hat{\frak A}$.
\label{wiueytcLKGHvrbwoeyRGHEiurtvbowiebw}
$ \EndofStatement$
\end{theorem}

{\bf Proof.} Equation (\ref{yetrvcqyuwdtvqo2}) immediately follows from
(\ref{IpiuaybciuwybS}) and
(\ref{ueybvrpwuieybrpSi1})-(\ref{ueybvrpwuieybrpSi2}).
To prove (\ref{yetrvcqyuwdtvqo1}) we first use
Lemma \ref{wuvybwpervwt}  to conclude that
\bear
D_\rho (e_2)D_\rho (e_1) & = &
V_\rho (e_2 )^* V_{\rho\cdot e_2} (e_1 )^*
U_{\rho \cdot e_1} (e_2)
U_\rho (e_1) ,
\\
D_\rho (e_1)D_\rho (e_2) & = &
V_\rho (e_1 )^* V_{\rho\cdot e_1} (e_2 )^*
U_{\rho \cdot e_2} (e_1)
U_\rho (e_2) .
\eear
Since (\ref{Vrhog}) implies
\be
V_{\rho \cdot e_2} (e_1 )V_\rho (e_2 ) =
V_{\rho \cdot e_1} (e_2 )V_\rho (e_1 )=
V_\rho (e_1 + e_2 )    ,
\ee
we are left to check
\be
U_{\rho } (e_2)^{-1} U_{\rho \cdot e_2} (e_1)^{-1}
U_{\rho \cdot e_1} (e_2) U_{\rho } (e_1)=
\UM_{\calH_\rho}
.
\label{swkeyurtvcbowfey}
\ee
Using the definition (\ref{IpiuaybciuwybS}) and the Weyl algebra
relations
(\ref{wetyrvw4ryvw1})-(\ref{wetyrvw4ryvw3}), equation (\ref{swkeyurtvcbowfey})
is equivalent to
\be
\calM_q (L_m (\mu)) \, \calE_q (L_e (\epsilon ))
\, \restriction \, \calH_\rho \; = \;
(u_1 u_2 u_3)^{-1} ,
\label{ysdfgbvcowuyieee}
\ee
where
\bear
L_m (\mu) =
\ell_m (\mu , \; 1) + \ell_m (\mu \cdot e_1 , \; 2) -
\ell_m (\mu \cdot e_2, \; 1) - \ell_m (\mu  , \; 2) ,
\label{uieyvbwuege1}
\\
L_e (\epsilon) =
\ell_e (\epsilon , \; 1) + \ell_e (\epsilon \cdot e_1 , \; 2) -
\ell_e (\epsilon \cdot e_2, \; 1) - \ell_e (\epsilon  , \; 2) ,
\label{uieyvbwuege2}
\eear
and where $ u_i \in U(1)$ are the phases obtained by commuting
in (\ref{swkeyurtvcbowfey}) all factors of $\calE_q $'s to the
right of $ \calM_q$'s, i.e.,
\bear
u_1 {(\epsilon , \; \mu)} & = &
\exp
\left[
- i
\bra
\ell_e (\epsilon, \; 2), \; \;
\ell_m (\mu , \; 1) - \ell_m (\mu\cdot e_2, \; 1)
\ket
\right],
\label{ueyrbovctweiurbwo1}
\\
u_2 {(\epsilon , \; \mu)} & = &
\exp
\left[
- i
\bra
\ell_e (\epsilon \cdot e_2, \; 1), \; \;
\ell_m (\mu\cdot e_1, \; 2)
\ket
\right],
\label{ueyrbovctweiurbwo2}
\\
u_3 {(\epsilon , \; \mu)} & = &
\exp
\left[
- i
\bra
\ell_e (\epsilon \cdot e_1, \; 2), \; \;
\ell_m (\mu , \; 1)
\ket
\right] .
\label{ueyrbovctweiurbwo3}
\eear

To verify (\ref{ysdfgbvcowuyieee}) we decompose $ \epsilon$ and $ \mu$
into a sum over ``monopoles'', i.e., cochains supported on a single
site or a single plaquette, respectively. Using the obvious fact that
$ \ell_e$ and $ \ell_m$ are $ \Z_N$-module maps, i.e.,
\bear
\ell_e (n \epsilon_1 + \epsilon_2 , \; i) & = &
n\ell_e (\epsilon_1 , \; i) + \ell_e (\epsilon_2 , \; i) ,
\\
\ell_m (n \mu_1 + \mu_2 , \; i) & = &
n\ell_m (\mu_1 , \; i) + \ell_m (\mu_2 , \; i) ,
\eear
for all $ n\in \Z_N$, $ \epsilon_{1, \; 2} \in \calC^0_{loc}$ and
$ \mu_{1, \; 2} \in \calC^2_{loc}$,
we conclude that  (\ref{ysdfgbvcowuyieee}) holds if and only if it
holds for all pairs of monopole distributions
$
 \rho = (\epsilon , \; \mu) =
(\delta_x , \; \delta_p )
$,
$ x\in (\Z^2)_0$, $ p\in (\Z^2)_2$.
In fact, we have
$$
 u_i (\epsilon , \; \mu )
=
\prod_{x\in (\Z^2)_0} \prod_{ p\in (\Z^2)_2}
u_i (\delta_x , \; \delta_p)^{\epsilon (x) \mu (p)}
$$
and similarly
\be
\begin{array}{ll}
\displaystyle
\left(
\Omega_{(\epsilon , \; \mu)} , \; \;
\calM_q (\ell_m (\mu     ))
\calE_q (\ell_e (\epsilon))
\Omega_{(\epsilon , \; \mu)}
\right)
\; \; = &
\\
 &
\\
\displaystyle
\prod_{x\in (\Z^2)_0}\prod_{ p\in (\Z^2)_2}
\left(
\Omega_{(\delta_x , \; \delta_p)} , \; \;
\calM_q (\ell_m (\delta_p     ))
\calE_q (\ell_e (\delta_x ))
\Omega_{(\delta_x , \; \delta_p)}
\right)^{\epsilon (x) \mu(p)}
. &
\end{array}
\ee
Now, for $ \epsilon = \delta_x$ one easily verifies
\be
     \ell_e (\delta_x, \; j) \; = \;
- \delta_{\la x-e_j , \; x \ra },
\label{yqwuetvcwuyiercqvo}
\ee
implying
$
     \ell_e (\delta_x \cdot e_i , \; j) \; = \;
- \delta_{\la x-e_i-e_j , \; x -e_i\ra } .
$
For $ \mu=\delta_p$ and
$ p = \la y, \; y + e_1, \; y + e_1 + e_2 , \; y+e_2 \ra$
we get
\bear
     \ell_m (\delta_p , \; 1) \; = \;
 \delta_{\la y , \; y+e_2 \ra },
\label{TiuytwhjqfgeGOYow1}
\\
     \ell_m (\delta_p, \; 2) \; = \;
- \delta_{\la y , \; y+e_1 \ra } .
\label{TiuytwhjqfgeGOYow2}
\eear
Plugging this into
(\ref{ueyrbovctweiurbwo1})-(\ref{ueyrbovctweiurbwo3})
we get
\bear
u_1 & = &
\exp
\left[
\frac{2\pi i}{N}
\left(
	    \delta_{x, \; y+e_2} - \delta_{x, \; y}
\right)
\right] ,
\\
u_2 & = &
\exp
\left[
-\frac{2\pi i}{N}
	    \delta_{x, \; y+e_2}
\right] ,
\\
u_3 & = &
\exp
\left[
-\frac{2\pi i}{N}
	    \delta_{x, \; y+e_1+e_2}
\right] ,
\eear
implying
\be
(u_1 u_2 u_3)^{-1} \; = \;
\exp
\left[
\frac{2 \pi i}{N}
\left(
	  \delta_{x, \; y} + \delta_{x, \; y+e_1 + e_2}
\right)
\right] .
\label{TDFiuwerybvoeui}
\ee

Next we look at (\ref{uieyvbwuege1})-(\ref{uieyvbwuege2})
to compute
\bear
& L_m (\delta_p)
\; =\;
 \delta_{\la y, \; y+e_2 \ra }
-\delta_{\la y-e_1, \; y \ra }
-\delta_{\la y-e_2, \; y \ra }
+\delta_{\la y, \; y+e_1 \ra }
\; = \;
- d\delta_y &
\\
& L_e (\delta_x)
\; =\;
-\delta_{\la x-e_1     , \; x     \ra }
-\delta_{\la x-e_1-e_2 , \; x-e_1 \ra }
+\delta_{\la x-e_1-e_2 , \; x-e_2 \ra }
+\delta_{\la x-e_2     , \; x \ra }
\; = \;
- d^*\delta_q &\ ,
\eear
where $ q$ is the oriented plaquette
$
q \; = \;
\la
x-e_1-e_2 , \; x-e_2, \; x, \; x-e_1
\ra .
$
This implies
\be
\calM_q (L_m (\delta_p )) \calE_q (L_e (\delta_x )) \Omega_\rho
=
\exp
\left[
\frac{2 \pi i}{N}
\left(
\delta_{x-e_1 -e_2 , \; y} + \delta_{x, \; y}
\right)
\right]
\Omega_\rho .
\label{wejyrbtpweqwcfwevctrthrv}
\ee
Comparing (\ref{TDFiuwerybvoeui}) with
(\ref{wejyrbtpweqwcfwevctrthrv}) we have proven
(\ref{ysdfgbvcowuyieee}) and therefore Theorem
\ref{wiueytcLKGHvrbwoeyRGHEiurtvbowiebw}.
$ \Fullbox$

We remark that our definition (\ref{YTuytowirtypwierW}) is
consistent with the translation invariance of the vacuum $\Om_0$, since
$\r=0$ implies $U_0(e_i)=\UM$ and $V_0(e_i)\Om_0 = \Om_0$.

Next we show that our intertwiner connection also intertwines the
representations $ D_\rho$ of the translation group $\Z^2$ in
$ \calH_\rho $, $ \forall \rho \in \calD_q$.
This is formulated most economically by putting
\be
	  {\Bbb D}_q (e_i ) \;\;  := \;\;
\bigoplus_{\rho \in \calD_q} \; D_\rho (e_i) ,
\ee
implying
\be
{\Bbb D}_q (e_i ) \Pi_q (A) \; = \;
\Pi_q (\tau_{e_i} (A)) {\Bbb D}_q (e_i ) ,
\ee
for all $ A\in \hat{\frak A}$.

\begin{proposition}

For $ i=1, \; 2$ the intertwiner algebra $ {\frak W}_q$ commutes with
${ \Bbb D}_q (e_i )$.
$ \EndofStatement$
\end{proposition}

{\bf Proof.}
Putting
$ {\Bbb U}_q (e_i ):= {\Bbb V}_q (e_i ){\Bbb D}_q (e_i )$ and using
Lemma \ref{wuvybwpervwt} we have to show
\bear
{\Bbb U}_q (e_i )\calE_q (\ell ') & = &
\calE_q (\ell '\cdot e_i){\Bbb U}_q (e_i ) ,
\label{euyivtgoweb1}
\\
{\Bbb U}_q (e_i )\calM_q (\ell ') & = &
\calM_q (\ell '\cdot e_i){\Bbb U}_q (e_i ),
\label{euyivtgoweb2}
\eear
for all $ \ell ' \in \calC^1_{loc}$. Since
$ {\Bbb U}_q (e_i )$ provides an intertwiner mapping
$ \calH_\rho \to \calH_{\rho\cdot e_i}$ it is again enough to check
these identities on
$ \Omega_\rho$, $ \forall \rho \in \calD_q$.
Using (\ref{IpiuaybciuwybS}), (\ref{YTuytowirtypwierW}) and
commuting $ \calM_q (\ell_m (\mu, \; i))$ to the left, equation
(\ref{euyivtgoweb1}) is equivalent to
\be
\calE_q (\ell_e (\epsilon + d^* \ell ', \; i))
\calE_q (\ell ')
\Omega_{(\epsilon , \; \mu )} \; = \;
e^{i\bra\ell ' \cdot e_i , \; \ell_m (\mu , \; i)\ket}
\calE_q (\ell '\cdot e_i )
\calE_q (\ell_e (\epsilon , \; i ))
\Omega_{(\epsilon , \; \mu )} .
\label{TrREUjmoiwmnevtew}
\ee
To prove (\ref{TrREUjmoiwmnevtew}) we compute
\bear
L(\ell ' , \; i) & := & \ell_e (\epsilon + d^* \ell ', \; i) + \ell '
- \ell ' \cdot e_i - \ell_e (\epsilon , \; i)
\nonumber \\
 & = &
\ell_e (d^* \ell ', \; i) + \ell ' - \ell ' \cdot e_i ,
\eear
yielding $ d^* L(\ell ' , \; i)=0$.
Let $ S(\ell ' , \; i)\in \calC^1_{loc}$ be such that
$ d^* S(\ell ' , \; i)= L(\ell ' , \; i)$. Then, by (\ref{hol}),
equation (\ref{TrREUjmoiwmnevtew}) is equivalent to
\be
e^{i\bra\ell ' \cdot e_i , \; \ell_m (\mu , \; i)\ket}=
e^{i\bra S(\ell ' , \; i), \; \; \mu\ket}.
\label{wytuvctowvtyYYYYYYYYYYYO}
\ee
Similarly as in the proof of Theorem
\ref{wiueytcLKGHvrbwoeyRGHEiurtvbowiebw} it is enough to check
(\ref{wytuvctowvtyYYYYYYYYYYYO}) for all $ \ell ' = \delta_b$,
$ b\in (\Z^2)_1 $, and all $ \mu=\delta_p$, $p\in (\Z^2)_2 $.
Hence, let
$ b=\la x, \; x+e_j \ra$ and
$
p=
\la
y, \; y+e_1, \; y+e_1+e_2, \; y+e_2
\ra
$.
Then
$
d^* \delta_b = \delta_{x+e_j}-\delta_x
$
and (\ref{yqwuetvcwuyiercqvo}) gives
\bear
L(\delta_{\la x, \; x+e_j\ra} , \; i)
 & = &
\ell_e (\delta_{x+e_j}, \; i)-\ell_e(\delta_x , \; i)
+
\delta_{\la x, \; x+e_j \ra} -
\delta_{\la x-e_i, \; x+e_j-e_i \ra}
\nonumber \\
 & = &
 \delta_{\la x    , \; x+e_j \ra}
-\delta_{\la x+e_j-e_i, \; x+e_j \ra}
+\delta_{\la x-e_i, \; x \ra}
-\delta_{\la x-e_i, \; x+e_j-e_i \ra} .
\eear
Thus, if $ j=i$ then
$ L( \delta_{\la x , \; x+e_j \ra} , \; i)=0$
implying (\ref{wytuvctowvtyYYYYYYYYYYYO}), since
in this case $ \ell_m (\mu, \; i)$ is perpendicular to the direction $
j$. We are left to check the case $j\neq i$, which gives
$
    L(\delta_{\la x, \; x+e_j \ra}, \; i) =
(-1)^j d^* \delta_{q-e_i} ,
$
where
$
q=\la
x, \; x+e_1, \; x+e_1+e_2 , \; x+e_2
\ra
$.
Thus we get for $i\neq j$
\be
e^{i\bra S(\delta_{\la x, \; x+e_j\ra}), \; \; \delta_p\ket}
=
\exp (\frac{2\pi i}{N} (-1)^j \delta_{x-e_i , \; y})\ .
\label{iuvycbgpoewuip}
\ee
On the other hand,
equations (\ref{TiuytwhjqfgeGOYow1})-(\ref{TiuytwhjqfgeGOYow2})
give for $ i\neq j$
\be
     \ell_m (\delta_p , \; i) \; = \;
(-1)^j \delta_{\la y, \; y+e_j \ra}
\label{awoyuvtbowueirvby}
\ee
and hence (\ref{iuvycbgpoewuip})-(\ref{awoyuvtbowueirvby})
imply (\ref{wytuvctowvtyYYYYYYYYYYYO}) in the case $i\neq j$.
Thus we have proven (\ref{euyivtgoweb1}).
Equation (\ref{euyivtgoweb2}) is proven by similar methods.
$ \Fullbox $

\subsection{Conclusions}

In this work we have investigated the dyonic sector structure of
$2+1$-dimensional lattice $\Z_N$-Higgs models described by the
Euclidean action (\ref{ACAOum}) in a range of couplings
(\ref{RI})-(\ref{RII}) corresponding to the ``free charge phase'' of
the Euclidean statistical mechanics model (\ref{EuCliDFoRm}).

We have worked in the Hamiltonian picture by formulating the model in terms
of its observable algebra $\A$ generated by the time-zero fields.
A Euclidean (modified) dynamics $\alpha_\r = \lim_V\Ad T_V(\r)$ has
been defined on $\A$ in terms of local (modified) transfer matrices
$T_V(\r)$, where $\r=(\varepsilon,\mu)$ is a superposition of electric
and magnetic $\Z_N$-charge distributions with finite support on the
spatial lattice $\Z^2$.

Dyonic states $\omega_\r$ have been constructed as ground states of
the modified dynamics $\alpha_\r$ on $\A$.
The associated charged representations $(\pi_\r,\calH_\r,\Omega_\r)$ of
$\A$ extend to irreducible representations of the ``dynamic closure''
$\hat\A\supset\A$, where $\hat\A=\langle\A,{\bf t}\rangle$ is the abstract
$*$-algebra generated by $\A$ and a global positive transfer matrix
${\bf t}$ implementing the ``true'' (i.e. unmodified) dynamics
$\alpha_0$.
$\pi_\r$ and $\pi_{\r'}$ are equivalent as representations of
$\hat\A$, provided their total charges coincide, $q_\r=q_{\r'}$. 
We have conjectured that the total charges $q_\r\in\Z_N\times\Z_N$ indeed
label the sectors of the model, i.e. $\pi_\r\not\sim\pi_{\r'}$ if
$q_\r\neq q_{\r'}$.
The ground state energy $\epsilon(q)$ of each sector has been shown to
be uniquely fixed by the  conditions
$\epsilon(0)=0$ and $\epsilon(q)=\epsilon(-q)$ and the requirement of
decaying of interaction energies for infinite spatial separation.

In Section \ref{StateBundlesandIntertwinerConnections} we
have analyzed structural algebraic aspects of the state bundle
$\calB_q= \bigcup_{\r\in\calD_q} (\r,\calH_\r) $, $\calD_q:=\{\r\mid
q_\r=q\}$, by constructing on ${\Bbb H}_q=\oplus_{\r\in\calD_q}\calH_\r$
a local intertwiner algebra ${\frak W}_q$ commuting with 
$\oplus_{\r\in\calD_q}\pi_\r(\hat\A)$.
The generators of ${\frak W}_q$ are given by electric and magnetic
``charge transporters'', $\calE_q(b)$ and $\calM_q(b)$, localized on
bonds $b$ in $\Z^2$ and fulfilling local Weyl commutation relations
(Theorem \ref{SSDywevcbrgowuyet}).
In terms of these charge transporters we have obtained a unitary connection
$U(\Gamma):\calH_\r\to\calH_{\r'}$ intertwining $\pi_\r$ and
$\pi_{\r'}$ for any path $\Gamma:\r\to\r'$ in $\calD_q$.
The holonomy of this connection is given by
$\Z_N$-valued phases related to the electric and magnetic charges
enclosed by $ \Gamma$ (see (\ref{hol})).

Finally, the connection $\Gamma$ has been used to construct on each
$\calH_\r$ a unitary 
representation of the group of spatial lattice translations acting
covariantly on $\pi_\r(\hat\A)$ and being intertwined by $U(\Gamma)$.
We remark that the existence
of such representations in the charged sectors of our model is by no
means an obvious feature. 

In \cite{FBIII} the holonomy phases of our connection $\Gamma$ will be the main
ingredient for establishing the anyonic nature of multiparticle
scattering states of electrically and magnetically charged particles
whose existence has been shown in \cite{FlorianBarataI}.

\newpage

\begin{appendix}

	\begin{center}
		\begin{Large}
			{\bf Appendix}
		\end{Large}
	\end{center}

\section{A Brief Sketch of the Polymer and Cluster Expansions}
\zerarcounters
\label{SegundoApendicehlkjhglkdgdg}

\subsection{Expansions for the Vacuum Sector}
\setcounter{theorem}{0}

In this Appendix we
present the basics of the polymer and cluster expansion
developed in \cite{FlorianBarataI}. We intend to present here
only the most
relevant facts to make the main ideas in the proofs of this
Appendix understandable.
For more details see [1].

\begin{definition}
For a 1-cochains  $E$ with $d^*E = 0$ and a
2-cochain D with $dD =0$, both with finite support, we define the
``winding number of $E$ around $D$'' as

\be
[D: \, E] := \exp \left( \frac{2\pi i}{N}\la u^D , \, E  \ra \right) .
\ee
One easily sees that this definition does not depend of the choice of the
particular configuration $u^D$. $\EndofStatement$

\end{definition}

Let us now prepare the definition of our polymers and their activities.
Define the sets
\bma
{\cal P} =
\left\{
P \in (\Z^3)^+_2 : \;  P \; \mbox{is finite, co-connected and} \;
P = \supp D, \; \mbox{for some} \; D\in (\Z^3)^2 , \, dD=0 , \; D\neq 0
\right\}  ,
\ema
\bma
{\cal B} =
\left\{
M \in (\Z^3)^+_1 : \;   M \; \mbox{is finite, connected and} \;
M = \supp E, \;\mbox{for some} \;  E\in (\Z^3)^1 , \, d^* E=0 ,
\: E\neq 0
\right\} ,
\ema
where $(\Z^3)^+_1$ (respectively $(\Z^3)^+_2$) refers to the set of positively
oriented bonds (plaquettes) of $\Z^3$
and the sets
\bear
{\cal P}_{total} & = &
\left\{
P \in (\Z^3)^+_2 \; \mbox{finite, so that} \;
P = \supp D, \; \mbox{for some} \; D\in (\Z^3)^2 , \, dD=0
\right\} ,
\nonumber \\
{\cal B}_{total} & = &
\left\{
M \in (\Z^3)^+_1 \; \mbox{finite, so that}\;
M = \supp E, \;\mbox{for some} \;  E\in (\Z^3)^1 , \, d^* E=0
\right\}
.
\nonumber
\eear

Note that the sets ${\cal P}_{total}$ and ${\cal B}_{total}$
contain the empty set and that the non-empty elements of
${\cal P}_{total}$ and of ${\cal B}_{total}$
are build up by unions of co-disjoint
elements of ${\cal P}$, respectively, by unions of disjoint elements of
${\cal B}$.  One has naturally
${\cal P}\subset {\cal P}_{total}$
and
${\cal B}\subset {\cal B}_{total}$.

Each non-empty set
$P\in {\cal P}_{total} $ and  $M\in {\cal B}_{total}$
can uniquely be decomposed into  disjoint unions
$P=P_1 + \cdots + P_{A_P}$, $M= M_1 + \cdots + M_{B_M}$
(the symbol ``$+$'' indicates here disjoint union)
where $P_i\in {\cal P}$  and $M_j \in {\cal B}$.
Then, if $D\in (\Z^3)^2$ is such that
$\supp D = P$, there is a unique decomposition
$D=D_1 + \cdots + D_{A_P}$
with $D_i\in (\Z^3)^2$, $\supp D_i = P_i$. Moreover
if $E\in (\Z^3)^1$ is such that
$\supp E = M$ then there is a unique decomposition
$E=E_1 + \cdots + E_{B_M}$
with $E_i\in (\Z^3)^1$, $\supp E_i = M_i$.
One can also decompose $u=u^{D_1} + \cdots +u^{D_{A_P}}$ with
$u^{D_i} \in (\Z^3)^1$, $du^{D_i} = D_i$.

For $P\in {\cal P}_{total}$ and  $M\in {\cal B}_{total}$
we define the sets
\bear
{\cal D}(P) & := &
\{
D \in (\Z^3)^2 \mbox{ so that }   \supp D = P \mbox{ and } dD=0
\} ,
\nonumber
\\
{\cal E}(M) & := &
\{
E \in (\Z^3)^1 \mbox{ so that }   \supp E = M \mbox{ and } d^* E=0
\}  .
\nonumber
\eear

We consider now pairs $(P, \; D)$ with
$P\in {\cal P}_{total}$ and $D\in{\cal D}(P)$ and pairs
$(M, \; E)$ with
$M\in {\cal B}_{total}$ and $E\in{\cal D}(M)$ and define
$w((P, \; D) , \, (M, \; E))=
w((M, \; E), \, (P, \; D))\in \{0,\ldots ,N-1\}$ as
the ``$\Z_N$-winding number'' of $(M, \; E)$ around $(P, \; D)$:
\be
w((P, \; D) , \, (M, \; E))=
w((M, \; E), \, (P, \; D)):=
[D: \, E]
.
\ee
The pairs with
$P\in {\cal P}$ and
$M\in {\cal B}$
will be the building blocks of our polymers.

With the help of $w$ we can establish a connectivity relation
between pairs  $(P, \; D)$ with  $P\in {\cal P}$,
$D\in {\cal D}(P)$  and pairs
$(M, \; E)$ with $M\in {\cal B}$, $E\in {\cal E}(M)$:
we say that $(P, \; D)$ and $(M, \; E)$ are ``$w$-connected'' if
$w((P, \; D), \, (M, \; E))\neq 1$
and ``$w$-disconnected'' otherwise.
We arrive then at the following

\begin{definition}
A polymer $\gamma$ is formed by two pairs
\bma
\left\{
(P^{\gamma}, \; D^{\gamma} ), \; (M^{\gamma}, \; E^{\gamma } )
\right\}
,
\ema
with $P^{\gamma}\in {\cal P}_{total}(V)$,
$M^{\gamma}\in {\cal B}_{total}(V)$
and $D^{\gamma}\in {\cal D}(P^{\gamma })$,
$E^{\gamma}\in {\cal E}(M^{\gamma })$, so that the set
\be
\{
(P^{\gamma}_1, \, D^{\gamma}_1),
\ldots , (P^{\gamma}_{A_{\gamma}}, \, D^{\gamma}_{A_{\gamma}} ) ,
(M^{\gamma}_1, \, E^{\gamma}_1),
\ldots , (M^{\gamma}_{B_{\gamma}}, \, E^{\gamma}_{B_{\gamma}} )
\}
\ee
formed by the decompositions
$P^{\gamma}= P^{\gamma}_1 + \cdots + P^{\gamma}_{A_{\gamma}} $,
$M^{\gamma}= M^{\gamma}_1 + \cdots + M^{\gamma}_{B_{\gamma}} $ with
$P^{\gamma}_i\in {\cal P}(V)$,
$M^{\gamma}_j\in {\cal B}(V)$
and
$D^{\gamma}= D^{\gamma}_1 + \cdots + D^{\gamma}_{A_{\gamma}} $,
$E^{\gamma}= E^{\gamma}_1 + \cdots + E^{\gamma}_{B_{\gamma}} $
with
$D^{\gamma}_i\in {\cal D}(P^{\gamma}_i)$,
$E^{\gamma}_j\in {\cal E}(M^{\gamma}_j)$
is a $w$-connected set. $\EndofStatement$
\end{definition}

Below, when we write $(M, \, E)\in \gamma$
and $(P, \, D)\in \gamma$ we are intrinsically be assuming
that $M\in {\cal B}$ with $E \in {\cal E}(M)$ and that
$P\in {\cal P}$ with $D \in {\cal D}(P)$.

For a polymer
$
\gamma =
((P^{\gamma}, \; D^{\gamma} ), \; (M^{\gamma}, \; E^{\gamma } ))
$ we call the pair
$
\gamma_g :=
(P^{\gamma}, \; M^{\gamma})
$
the geometrical part of $\gamma$ and
the pair
$
\gamma_c :=
(D^{\gamma} , \; E^{\gamma} )
$ is the ``coloring'' of $\gamma $.
Of course the coloring determines uniquely the geometric part.
Each pair $(D, \; E)$,
$D\in{\cal D}(P)$, $E\in{\cal E}(M)$ with
$P\in{\cal P}$, $M\in{\cal B}$ is a color
for $(P, \; M)$.
Another important definition is the ``size'' of a polymer.
We define the size of $\gamma$ by
$|\gamma |=|\gamma_g|:=|P^{\gamma }|+|M^{\gamma}|$,
where $|P^{\gamma }|$ (respectively $|M^{\gamma}|$) is the number
of plaquettes (respectively bonds) making up
$P^{\gamma }$ (respectively $M^{\gamma}$).

The activity $\mu (\gamma)\in \C$ of a polymer $\gamma$ is defined to be
\be
\mu(\gamma ) :=
\left[ D^{\gamma} : \, E^{\gamma} \right]
\left\{
\prod_{i=1}^{A_{\gamma }}
\left[
\prod_{p\in P^{\gamma}_i} g (D^{\gamma}_i(p))
\right]
\right\}
\left\{
\prod_{j=1}^{B_{\gamma }}
\left[
\prod_{b\in M^{\gamma}_j} h (E^{\gamma}_j(b))
\right]
\right\}
,
\label{atividade}
\ee
with $\mu (\emptyset )=1$.

For a polymer model we need the notions of
``compatibility'' and ``incompatibility''  between pairs of polymers.
This is defined in the following way.
Two polymers $\gamma $ and $\gamma '$ are said to be incompatible,
$\gamma \not \sim \gamma '$,
if at least one of the following conditions hold:
\begin{enumerate}
\item
There exist $M^{\gamma}_a \in \gamma_g$ and
$M^{\gamma' }_b \in \gamma_g '$,
so that
$M^{\gamma}_a $ and $M^{\gamma' }_b $ are connected
(i.e., there exists at least one lattice point $x$ so that
$x\in \partial b$ and $x\in \partial b'$ for some bonds
$b \in M^{\gamma}_a$ and $b' \in M^{\gamma '}_b$);
\item
There exists $P^{\gamma}_a \in \gamma_g$ and
$P^{\gamma' }_b \in \gamma_g '$,
so that
$P^{\gamma}_a $ and $P^{\gamma' }_b $ are co-connected
(i.e., there exists at least one cube $c$ in the lattice  so that
$p\in \partial c$ and $p' \in \partial c$ for some plaquettes
$p \in P^{\gamma}_a$ and $p' \in P^{\gamma '}_b$);
\item
There exists $(M^{\gamma}_a , \, E^{\gamma}_a )\in \gamma$ and
$(P^{\gamma' }_b , \, D^{\gamma '}_b )\in \gamma '$,
so that
$(M^{\gamma}_a , \, E^{\gamma}_a )$ and
$(P^{\gamma' }_b , \, D^{\gamma '}_b )$ are $w$-connected.
Or the same with $\gamma $ and $\gamma '$ interchanged.
\end{enumerate}
They are said to be compatible, $\gamma \sim \gamma '$, otherwise.

We will denote by ${\cal G}(V)$  the set of all polymers in $V\subset \Z^3$ and
by ${\cal G}_{com} (V)$ the set of all finite sets of compatible polymers.

We want to express the vacuum expectation of classical observables
in terms of our polymer expansion.
We consider the following
\begin{definition}
\label{defdeBalbet}
Let $\alpha$  be a 1-cochain and $\beta$ 2-cochain, both with finite support. Define
the classical observable
\be
B (\alpha , \, \beta) :=
\exp
\left[
-\frac{2\pi i}{N} \la\alpha , \, u \ra
\right]
\prod_{p}\frac{g ((du+\beta)(p))}{g (du(p))}
\qquad .
\label{bvubgywuibvpiubvwyeubw}
\ee
Any classical observable can be written as a linear combination of such
functions. $\EndofStatement$
\end{definition}

For a finite volume $V\subset \Z^3$ (say, a cube)
 we have the following\footnote{For simplicity we will neglect some
boundary terms, which can be controlled with more work.}

\be
\la B (\alpha , \, \beta ) \ra_V =
\frac{1}{Z^{1}_{V}}
\sum_{D\in V^2  \atop d(D-\beta)=0 }
\sum_{E\in V^1  \atop d^*(E-\alpha)=0 }
\left[ D-\beta : \, E-\alpha\right]
\left[
\prod_{p\in \mbox{\footnotesize{supp}} D} g (D(p))
\right]
\left[
\prod_{b\in \mbox{\footnotesize{supp}} E} h (E (b))
\right]
.
\label{oeybweyinweiuneriueorne}
\ee

Here the normalization factor $Z^1_V$ is given by

\be
Z_V^1  =\sum_{\Gamma \in  {\cal G}_{com} (V)} \mu^{\Gamma }
,
\label{partpoli}
\ee
in multi-index notation, i.e.,
$\displaystyle \mu^{\Gamma}:= \prod_{\gamma \in \Gamma}\mu(\gamma)$.
We will often identify the elements of
${\cal G}_{com}$ with their characteristic functions.

The cochains $D$  appearing in the sums in (\ref{oeybweyinweiuneriueorne})
can uniquely be decomposed  in such a way that
$D=D_0 + D_1$ with $d(D_0 - \beta)=0$ and $dD_1=0$ and so that
$\mbox{supp}\, D_0 $ is co-connected and co-disconnected from
$\mbox{supp }D_1 $.
If $d\beta =0$ we choose $D_0 =0$.
Analogously,
the cochains $E$ appearing in the sums in (\ref{oeybweyinweiuneriueorne})
can be decomposed uniquely in such a way that
$E=E_0 + E_1$ with $d^*(E_0 - \alpha )=0$ and $d^* E_1=0$ and so that
$\mbox{supp}\, E_0 $ is connected and disconnected from $\mbox{supp }E_1 $.
If $d^*\alpha =0$ we choose $E_0 =0$.

We denote by ${\cal C}_1 (\alpha ) $ the set of the supports of all such
$E_0$'s, for a given $\alpha$
and by ${\cal C}_2 (\beta ) $ the set of the supports of all such
$D_0$'s, for a given $\beta$.
For $d^*\alpha =0$ we have ${\cal C}_1 (\alpha ) =\emptyset$
and for $d\beta =0$ we have ${\cal C}_2 (\beta ) =\emptyset$.
We  define the sets of pairs
\be
\mbox{Conn}_1 (\alpha, \, V ) :=
\left\{
(M, \; E), \mbox{ so that } M\in{\cal C}_1 (\alpha ) \mbox{ and }
E\in V^1 , \mbox{ with } \supp E =M \mbox{ and } d^*E= d^* \alpha
\right\}  ,
\ee
\be
\mbox{Conn}_2  (\beta , \, V)   :=
\left\{
(P, \; D), \mbox{ so that } P\in{\cal C}_2 (\beta ) \mbox{ and }
D\in V^2 , \mbox{ with } \supp D =P \mbox{ and } dD=d\beta
\right\} .
\ee

We then write
\bear
\la B(\alpha , \, \beta ) \ra_V & = &
\sum_{(M, \, E)\in\mbox{\footnotesize{Conn}}_1 (\alpha , \, V) \atop
(P, \, D)\in\mbox{\footnotesize{Conn}}_2 (\beta , \, V ) }
\left[ D-\beta : \, E-\alpha \right]
\nonumber \\
 & &
\times
\left[
\prod_{p\in P} g (D(p))
\right]
\left[
\prod_{b\in M}  h (E(b))
\right]
\frac
{
\displaystyle
\sum_{\Gamma \in {\cal G}_{com}}
a_{(M, \, E), \, \alpha }^{\Gamma}
b_{(P, \, D), \, \beta }^{\Gamma}
\:
\mu^{\Gamma}
}
{
\displaystyle
\sum_{\Gamma \in {\cal G}_{com}}
\mu^{\Gamma}
} ,
\label{jardimperi}
\eear
for
\be
a_{(M, \, E), \, \alpha }(\gamma ) :=
\left\{
\begin{array}{ll}
0, & \mbox{if } M^{\gamma } \mbox{ is connected with } M,  \\
\left[ D^{\gamma} : \, E-\alpha \right],
	 & \mbox{otherwise}
\end{array}
\right.         ,
\ee
and
\be
b_{(P, \, D), \, \beta  }(\gamma ) :=
\left\{
\begin{array}{ll}
0, & \mbox{if } P^{\gamma } \mbox{ is co-connected with } P  \\
\left[ D-\beta : \, E^{\gamma} \right],         & \mbox{otherwise}
\end{array}
\right.         .
\ee

It is for many purposes useful to write the normalization factor
$Z^1_V $ in the form
\be
Z^1_V =
\exp
\left\{
\sum_{\Gamma \in {\cal G}_{clus}(V) }c_{\Gamma }\mu^{\Gamma}
\right\}       .
\label{cappv}
\ee
Let us explain the symbols used above. Our notation is close to that of
\cite{FredMarcu}.
${\cal G}_{clus}(V)$ is the set of all finite
clusters of polymers in $V$, i.e.,
an element $\Gamma \in {\cal G}_{clus} $  is a finite set of (not necessarily
distinct) polymers building a connected ``incompatibility graph''. An
incompatibility graph is a graph which
has polymers as vertices and where two vertices are connected by
a line if the corresponding polymers are incompatible.
We will often identify an elements $\Gamma\in{\cal G}_{clus}$ with a function
$\Gamma$: ${\cal G} \rightarrow \N$, where $\Gamma (\gamma )$ is the
multiplicity of $\gamma $ in $\Gamma \in {\cal G}_{clus}$.
The coefficients $c_{\Gamma }$ are the ``Ursell functions'' and are of purely
combinatorial nature.
They are defined (see \cite{FredMarcu} and \cite{Seiler}) by
\be
   c_{\Gamma}:=\sum_{n=1}^{\infty}\frac{(-1)^{n+1}}{n} {\cal N}_n (\Gamma ) ,
\ee
where ${\cal N}_n (\Gamma ) $ is the number of ways of writing
$\Gamma $ in the form $\Gamma= \Gamma_1 + \cdots +\Gamma_n$ where
$0\neq \Gamma_i \in {\cal G}_{com}$, $i=1, \ldots , n$.

Relation (\ref{cappv}) makes sense provided the sum over clusters
is convergent.
As discussed in \cite{FredMarcu} and \cite{FlorianBarataI}
a sufficient condition for this is $\|\mu \|\leq\|\mu \|_c$,
where
$
\displaystyle
\|\mu\|:= \sup_{\gamma \in {\cal G}} |\mu (\gamma )|^{1/|\gamma |}
$,
and $\| \mu \|_c$ is a constant defined in \cite{FredMarcu}.
By (\ref{atividade}),
\be
|\mu (\gamma )|\leq
\left[
\max \{g(1), \ldots , g(N-1), h(1), \ldots, h(N-1)\}
\right]^{|\gamma |}
,
\ee
what justifies the conditions
(\ref{RI})-(\ref{RII}).

Calling
$
\mbox{Conn}_1 \, (\alpha ) := \mbox{Conn}_1 \, (\alpha , \,
\Z^3)
$,
$
\mbox{Conn}_2 \, (\beta ) := \mbox{Conn}_2 \, (\beta , \,
\Z^3)
$
and $\calG_{clus} := \calG_{clus} \, (\Z^3)$, we can also write
the thermodynamic limit of
$\la B (\alpha , \, \beta ) \ra_V $ as

\bear
\la B (\alpha , \, \beta ) \ra & = &
\sum_{(M, \, E)\in\mbox{\footnotesize{Conn}}_1 (\alpha ) \atop
(P, \, D)\in\mbox{\footnotesize{Conn}}_2 (\beta ) }
\left[ D-\beta : \, E-\alpha \right]
\left[
\prod_{p\in P} g (D(p))
\right]
\left[
\prod_{b\in M}  h (E(b))
\right]
\nonumber \\
 & &
\times
\exp
\left(
\sum_{\Gamma \in {\cal G}_{clus}}
c_{\Gamma }
\left(
a_{(M, \, E), \, \alpha }^{\Gamma} \,
b_{(P, \, D), \, \beta }^{\Gamma}
-1
\right)
\mu^{\Gamma}
\right) .
\label{formulamestra}
\eear

The presence of the phases
$\left[ D-\beta : \, E-\alpha \right]$ is an important feature
of this last expression and is responsible for the
emergence of the anyonic statistics. Note that for $\alpha $ and
$\beta $ such that $d\beta =0$ and $d^*\alpha =0$ the expectation
$\la B(\alpha , \, \beta)\ra$ is proportional to
$\left[ \beta : \, \alpha \right]$, i.e. to the winding number of
$\beta $ around $\alpha$.

Let $\Gamma$ be a cluster of polymers. We say that a polymer $\gamma$
is incompatible with $\Gamma$, i.e., $\gamma\not\sim\Gamma$, if there is
at least one $\gamma ' \in \Gamma$ with $\gamma\not\sim\gamma'$.
For two clusters $\Gamma$, $\Gamma ' $ we have $\Gamma \not \sim \Gamma '$
if there is at least one $\gamma\in\Gamma$ with $\gamma\not\sim\Gamma '$.

For the polymer system discussed in this work we have the following
result:

\begin{theorem}
\label{clusterTEORDA}
There is a convex, differentiable, monotonically decreasing function
$F_0$: $(a_0, \, \infty)\rightarrow \R_+$, for some $a_0\geq0$, with
$\lim_{a\rightarrow \infty}F_0(a)=0$ such that, for all sets of polymers
$\Gamma$, and for all $a>a_0$,
\be
   \sum_{\gamma\not\sim\Gamma}e^{-a|\gamma |}
\leq
   F_0(a) \, \|\Gamma \|,
\label{clusterUM}
\ee
where $\|\Gamma \|=\sum \Gamma (\gamma ') |\gamma '|$,
$\Gamma (\gamma ') $ being the multiplicity of $\gamma '$ in $\Gamma$.

Once inequality (\ref{clusterUM}) has been established,
it has been proven in \cite{FredMarcu},
Appendix A.1, that the two following results hold:
\be
   \sum_{ \Gamma \in {\cal G}_{clus} \atop \Gamma \not\sim\Gamma_0 }
|c_{\Gamma }| \, |\mu^{\Gamma }|
\leq
F_1 (-\ln \|\mu \|) \, \|\Gamma_0 \| ,
\label{clusterDOIS}
\ee
\be
	\sum_{
\Gamma \in {\cal G}_{clus} \atop
{\Gamma \not\sim\Gamma_0 \atop \|\Gamma \|\geq n }}
|c_{\Gamma }| \, |\mu^{\Gamma }|
\leq
\left( \frac{\|\mu \|}{\|\mu_c \|} \right)^n \, \|\Gamma_0 \| F_0 (a_c )
,
\label{clusterTRES}
\ee
where $a_c$ and $\|\mu_c\| >0$ are constants defined in \cite{FredMarcu},
$F_1$: $(a_c+F_0 (a_c), \, \infty)\rightarrow \R_+$ is the solution of
$F_1 (a+F_0 (a))=F_0 (a)$ and
$\|\mu\|:=\sup_{\gamma}|\mu(\gamma)|^{1/|\gamma|} $.
 $\EndofStatement $
\end{theorem}

For a proof we refer the reader to \cite{FlorianBarataI} and
\cite{FredMarcu}.
The inequalities (\ref{clusterDOIS}) and (\ref{clusterTRES})
are of central importance in the theory of cluster expansions
and are often used for proving theorems. For instance,
(\ref{clusterTRES}) tells us that the sums like
$\sum_{\Gamma} |c_\Gamma| |\mu^\Gamma |$ involving only
clusters with size larger than a certain $n$
(and which are incompatible with some $\Gamma_0 $ fixed)
decay exponentially with $n$.

\subsection{Expansions for the Dyonic Sectors}
\setcounter{theorem}{0}
\label{ExpansionsfortheDyonicSectors}

Let us now present the corresponding expansions for the states
$\omega_\rho$ for $\rho \in \calD_0$.

To each $B\in {\frak F}_0$ we can associate a classical observable
$ B_{cl , \, \rho } = B_{cl , \, \rho } (d\varphi - A) $
(see (\ref{wwsrdhfdcvsjydobcfsled})). A possible but non-unique choice
is (see \cite{FlorianBarataI} and \cite{FredMarcu})

\be
B_{cl, \, \rho} =
\frac{
Tr_{{\cal H}_{\undV}}
\left(
F_{(\varphi (0),\, A(0)) , \, (\varphi (1) ,\, A (1))   }
B \, T_{\undV} (\rho)
\right)
}
{
Tr_{{\cal H}_{\undV}}
\left(
F_{(\varphi (0),\, A(0)) , \, (\varphi (1) ,\, A (1))   }
  \, T_{\undV} (\rho)
\right)
}
\ee
where
\be
F_{(\varphi (0),\, A(0)) , \, (\varphi (1) ,\, A (1))   }
=
\sum_{(\varphi (0),\, A(0)) , \, (\varphi (1) ,\, A (1))   }
|\varphi (0),\, A(0) \ra \la \varphi (1),\, A(1) |
\ee
and
where $\varphi (k) , \;  A (k) $  refers to the variables in the $k$-th
euclidean time plane.

Since any such classical observable can be written as a finite linear
combination of the functions $B (\alpha , \, \beta)$ previously
introduced (with coefficients eventually depending on $\rho$) we
concentrate on expectations of such functions.

Proceeding as in the previous sections we can express
$\la B (\alpha , \, \beta) \ra_{\rho}$ defined in
(\ref{wwsrdhfdcvsjydobcfsled})
as
\bear
\la B (\alpha , \, \beta ) \ra_\rho  & = &
\sum_{(M, \, E)\in\mbox{\footnotesize{Conn}}_1 (\alpha ) \atop
(P, \, D)\in\mbox{\footnotesize{Conn}}_2 (\beta ) }
\left[ D-\beta : \, E-\alpha \right]
\left[ D-\beta : \, - \tilde{ \epsilon } \right]
\left[ - \tilde{\mu } : \, E-\alpha \right]
\left[
\prod_{p\in P} g (D(p))
\right]
\nonumber \\
 & &
\times
\left[
\prod_{b\in M}  h (E(b))
\right]
\exp
\left(
\sum_{\Gamma \in {\cal G}_{clus}}
c_{\Gamma }
\left(
a_{(M, \, E), \, \alpha }^{\Gamma} \,
b_{(P, \, D), \, \beta }^{\Gamma}
-1
\right)
a_\epsilon^\Gamma
b_\mu^\Gamma
\mu^{\Gamma}
\right) ,
\label{formulamestracomrho}
\eear
where $\tilde{\mu}$ is the 2-cochain on $\Z^3$ defined by
$\tilde{\mu}((x, \, n)) := \mu (x)$ for all $x \in (\Z^2)_2$ and $n\in \Z$
and $\tilde{\epsilon}$ in the 1-cochain on $\Z^3$ defined by
$\tilde{\epsilon}((y  , \, n+ 1/2)) := \epsilon (y)$,
for all $y \in \Z^2$ and $n\in \Z$. Here $(y , \, n + 1/2 )$
indicates the vertical bond in $(\Z^3)_1^+$ whose projection onto
$\Z^2$ is $y$ and is located between the euclidean time-planes $n$ and
$n+1$.
Beyond this we defined
\bear
a_{ \epsilon } (\gamma )  & := &
[D^\gamma : \, - \tilde{\epsilon}] ,
\\
b_{ \mu } (\gamma ) & := &
[-\tilde{\mu}: \, E^\gamma ] .
\eear
The cochains $\tilde{\mu}$ and $\tilde{\epsilon}$ do not have finite
support but, since the polymers are finite, the right hand side of the
last two expressions
can be defined using some limit procedure, for instance, by
closing $\tilde{\mu}$ and
$\tilde{\epsilon}$ at infinity by adding, before the thermodynamic
limit is taken, the
cochains $s_\mu$ and $ s_\epsilon$ to them. Since the polymers are
finite, the limit does not depend on the particular $ s_\mu$ and
$ s_\epsilon$ chosen..
The same can be said about the winding numbers
$\left[ D-\beta : \, - \tilde{ \epsilon } \right]
$ and
$
\left[ - \tilde{\mu } : \, E-\alpha \right]
$ in (\ref{formulamestracomrho}).

Concerning the sum over clusters in (\ref{formulamestracomrho}), the
following estimate can be established

\begin{proposition}
For
$(M, \, E)\in\mbox{Conn}_1 (\alpha ) $ and
$(P, \, D)\in\mbox{Conn}_2 (\beta ) $
one has
\be
\left|
\sum_{\Gamma \in {\cal G}_{clus}}
c_{\Gamma }
\left(
a_{(M, \, E), \, \alpha }^{\Gamma} \,
b_{(P, \, D), \, \beta }^{\Gamma}
-1
\right)
a_\epsilon^\Gamma
b_\mu^\Gamma
\mu^{\Gamma}
\right|
\leq
 c_0
\left(
(|M| + |\mbox{sup } \alpha |) + (|P| + |\mbox{sup } \beta |)
\right) ,
\label{udibvtygoeref}
\ee
where $ c_0$ is a positive constant.
 $\EndofStatement $
\label{peuvbhpretbvyer}
\end{proposition}

{\bf Proof.} One has
\bear
\left|
\sum_{\Gamma \in {\cal G}_{clus}}
c_{\Gamma }
\left(
a_{(M, \, E), \, \alpha }^{\Gamma} \,
b_{(P, \, D), \, \beta }^{\Gamma}
-1
\right)
a_\epsilon^\Gamma
b_\mu^\Gamma
\mu^{\Gamma}
\right|
 & \leq &
\sum_{\Gamma \in {\cal G}_{clus}}
| c_{\Gamma } |
\left|
a_{(M, \, E), \, \alpha }^{\Gamma} \,
b_{(P, \, D), \, \beta }^{\Gamma}
-1
\right| \;
|\mu^{\Gamma} |
\nonumber \\
 & \leq &
\sum_{{\Gamma \in {\cal G}_{clus}} \atop \Gamma \not\sim \gamma_1}
| c_{\Gamma } | \; |\mu^{\Gamma} |
+
\sum_{{\Gamma \in {\cal G}_{clus}} \atop \Gamma \not\sim \gamma_2}
| c_{\Gamma } | \; |\mu^{\Gamma} |
\nonumber \\
 & \leq &
F_1 (-\ln \|\mu \|) (\|\gamma_1 \|+\|\gamma_2 \|),
\eear
by (\ref{clusterDOIS}), where
$ \gamma_1 =  (\mbox{sup }( E + \alpha ), \; E + \alpha)$ and
$ \gamma_2 =  (\mbox{sup }( D + \beta), \; D + \beta)$.
Clearly,
$
\|\gamma_1\|
\leq
|M| + |\mbox{sup } \alpha |
$ and
$
\|\gamma_2\|
\leq
|P| + |\mbox{sup } \beta |
$.
$ \Fullbox$

Proposition \ref{peuvbhpretbvyer} has a simple corollary

\begin{proposition}
If $ g_c$ and $ h_c$ are small enough and if
$ \min \{ \alpha_l , \; \beta_l \} \geq n$, where
\bear
\alpha_l & := & \inf \{ |M|, \quad (M, \, E)\in\mbox{Conn}_1 (\alpha ) \} ,
\\
\beta_l & := &  \inf \{ |P|, \quad (P, \, D)\in\mbox{Conn}_2 (\beta ) \} ,
\eear
then
\be
| \la B(\alpha , \; \beta )\ra_\rho | \leq
c_a e^{- c_b \, n},
\ee
for positive constants $ c_a $ and $ c_b$.
$ \EndofStatement$
\label{propdecayexpn}
\end{proposition}

{\bf Proof.} Using the representation (\ref{formulamestracomrho}) of
$\la B(\alpha , \; \beta )\ra_\rho$ in terms of cluster expansions
and the estimate (\ref{udibvtygoeref}), one gets
\be
| \la B(\alpha , \; \beta )\ra_\rho | \leq
\sum_{(M, \, E)\in\mbox{\footnotesize{Conn}}_1 (\alpha ) \atop
(P, \, D)\in\mbox{\footnotesize{Conn}}_2 (\beta ) }
g_c^{|P|} \; h_c^{|M|} \;
\exp
\left\{
 c_0
\left(
(|M| + |\mbox{sup } \alpha |) + (|P| + |\mbox{sup } \beta |)
\right)
\right\}
,
\ee
By standard arguments one has
\be
\sum_{(M, \, E)\in\mbox{\footnotesize{Conn}}_1 (\alpha ) }
\; ( h_c \, e^{c_0} )^{|M|}
\leq const. \, e^{-c_a \, n /2}
,
\ee
for some positive $ c_a$,
provided $ h_c$ is small enough and, analogously,
\be
\sum_{(P, \, D)\in\mbox{\footnotesize{Conn}}_2 (\beta ) }
\; ( g_c \, e^{c_0} )^{|P|}
\leq const. \, e^{-c_a \, n/2}
,
\ee
provided $ g_c $ is small enough.
This proves the proposition. $ \Fullbox $

An important particular case of (\ref{formulamestracomrho}) occurs when
$ d\beta = d^* \alpha =0$. In this case we get simply
\be
\la B (\alpha , \, \beta ) \ra_\rho   =
\left[ \beta : \, \alpha \right]
\left[ \beta : \,  \tilde{ \epsilon } \right]
\left[  \tilde{\mu } : \, \alpha \right]
\exp
\left(
\sum_{\Gamma \in {\cal G}_{clus}}
c_{\Gamma }
\left(
a_{(\emptyset, \, 0), \, \alpha }^{\Gamma} \,
b_{(\emptyset, \, 0), \, \beta  }^{\Gamma}
-1
\right)
a_\epsilon^\Gamma
b_\mu^\Gamma
\mu^{\Gamma}
\right) .
\label{formulamestracomrho2}
\ee
Notice the presence of the $\Z_N $-factors
$
\left[ \beta : \, \alpha \right]
\left[ \beta : \,  \tilde{ \epsilon } \right]
\left[  \tilde{\mu } : \, \alpha \right]
$ related to winding numbers involving $ \alpha$, $ \beta $ and the
background charges $ \rho$.

\section{The Remaining Proofs}
\zerarcounters
\label{PrimeiroApendicehlkjhglkdgdg}

\subsection{Proof of Proposition \protect\ref{estadosKK1}}
\setcounter{theorem}{0}
\label{ProofofPropositionprotectestadosKK1}

Let $ E_G$ be the projection ${\frak F}_{loc} \to
{\frak  A}_{loc}$. Since $ \omega_{V, \; \rho}$ is gauge invariant, it
is enough to prove the existence of
$\displaystyle \lim_{V\uparrow \Z^2} \omega_{V, \; \rho} \restriction
{\frak  A}_{loc}$.
Expectations like $\omega_{V, \; \rho} (A)$ for
$ A \in {\frak  A}_{loc}$ can be written as finite linear combinations
of the previously introduced
classical expectations $ \la B(\alpha , \; \beta) \ra_{V, \; \rho}$,
whose thermodynamic limit was described in subsection
\ref{ExpansionsfortheDyonicSectors}.

To show that $\omega_\rho\restriction {\frak  A}_{loc}$ is a ground
state with respect to $ \alpha_\rho$ we first notice that, by the
representation of  $\omega_{V, \; \rho} (A)$, $ A\in {\frak  A}_{loc}$
in terms of cluster expansions we can write, in analogy to
(\ref{Xum}),
\be
   \omega_{V, \, \rho }( A ) \; = \;
\lim_{n\to\infty}
\frac{Tr_{{\cal H}_{V}}
\left(
T_{V}(\rho )^n
A
T_{V}(\rho )^{n-1} E_{V}^\rho
\right)}{Tr_{{\cal H}_{V}} \left( T_{V}(\rho )^{2n-1}
E_{V}^\rho  \right)}
, \qquad A \in {\frak F}(V ) ,
\label{uypgwegwrtewef}
\ee
and, hence, for $ V $ large enough, one has
\be
   \omega_{V, \, \rho }( A^* \alpha_\rho (A) ) \; = \;
\lim_{n\to\infty}
\frac{Tr_{{\cal H}_{V}}
\left(
T_{V}(\rho )^n
A^* T_{V}(\rho )
T_{V}(\rho )^{n} E_{V}^\rho
\right)}{Tr_{{\cal H}_{V}} \left( T_{V}(\rho )^{2n-1}
E_{V}^\rho  \right)}
, \qquad A \in {\frak F}(V ) .
\label{Sdhdvbgdfb}
\ee
Now, by (\ref{mmmmmsdiogfudzgfsf})
and (\ref{EV0})-(\ref{Xtres}), $ E_{V}^\rho $ is a positive operator
and so,
the numerator in (\ref{Sdhdvbgdfb}) is clearly positive.
This proves that $ \omega_\rho (A^* \alpha_\rho (A)) \geq 0$.

To show that $\omega_\rho (A^* \alpha_\rho (A)) \leq
\omega_\rho (A^*A) $ we can make use of Lemma \ref{ClustereGroundstate}
and show that $\omega_\rho$ fulfills the cluster property with respect
to $ \alpha_\rho$. We can represent
$\omega_\rho (A^* \alpha_\rho^n (A))$ in terms of classical
expectations of the classical functions associated to the operator
$ A^* \alpha_\rho^n (A)$. These classical expectations can be written
as a finite linear combination of expectations like
$ \la B(\alpha_n , \; \beta_n) \ra_\rho$, where
the local cochains $ \alpha_n$ and $ \beta_n$ can be written, for $ n$ large
enough,
as sums
$ \alpha_n = \alpha (0) + \alpha (n)$ and
$ \beta_n = \beta (0) + \beta (n)$,
where the local cochain $ \alpha (n)$ (respect. $\beta (n)$)
is the complex conjugate of the translate of $ \alpha (0)$
(respect., of $\beta (0)$) by $n$ units in euclidean time
direction.

Recalling now the representation (\ref{formulamestracomrho}) of
$\la B(\alpha , \; \beta)\ra_\rho$ in terms of cluster expansions we
notice that, by Proposition \ref{propdecayexpn},
the contributions of sets
$
(M, \, E)\in\mbox{\footnotesize{Conn}}_1 (\alpha )
$
connecting the support of $ \alpha (0)$ to the support of $ \alpha (n)$
decay exponentially with $ n$, the same happening with
the contribution of the sets
$
(P, \, D)\in\mbox{\footnotesize{Conn}}_2 (\beta )
$
connecting the support of $ \beta (0)$ to the support of $ \beta (n)$.
The only surviving terms, after taking the limit $ n\to\infty$
correspond to sets
$
(M, \, E)\in\mbox{\footnotesize{Conn}}_1 (\alpha )
$
and sets
$
(P, \, D)\in\mbox{\footnotesize{Conn}}_2 (\beta )
$
connecting
the supports of   $ \alpha (0)$,  $ \alpha (n)$,
$ \beta (0)$ and $ \beta (n)$ with themselves.
The contributions of these last terms converges to
the product
$
\la B(\alpha (0), \; \beta (0))\ra_\rho
\la B(\alpha (n), \; \beta (n))\ra_\rho
$.
This implies that $ \omega_\rho (A^* \alpha_\rho^n (A))\to
\omega_\rho (A^*) \omega_\rho (A)
$, $ n\to\infty$, thus proving the ground state property.
The general cluster property
$ \omega_\rho (A \alpha_\rho^n (B))\to
\omega_\rho (A) \omega_\rho (B)
$, $ A$, $ B\in {\frak A}_{loc}$,
follows from the same arguments.

$\Fullbox$

\subsection{Proof of Propositions \protect\ref{estadosKK2}
and \protect\ref{ieurygpbsiudybvpdiubv}
}
\label{ProofofPropositionsestadosKK2andieurygpbsiudybvpdiubv}
\setcounter{theorem}{0}

In order to prove Proposition \ref{estadosKK2} we have to study
$
\displaystyle \lim_{a\to\infty}
\la B(\alpha , \; \beta ) \ra_{\rho - \rho' a}
$.
We recall the representation (\ref{formulamestracomrho}) of
$\la B(\alpha , \; \beta ) \ra_{\rho - \rho' a}$
and notice that, since
\be
\left|
c_{\Gamma }
\left(
a_{(M, \, E), \, \alpha }^{\Gamma} \,
b_{(P, \, D), \, \beta }^{\Gamma}
-1
\right)
a_{\epsilon  -\epsilon ' a}^\Gamma
b_{\mu - \mu ' a}^\Gamma
\mu^{\Gamma}
\right|
\leq
\left|
c_{\Gamma }
\left(
a_{(M, \, E), \, \alpha }^{\Gamma} \,
b_{(P, \, D), \, \beta }^{\Gamma}
-1
\right)
\mu^{\Gamma}
\right|
\ee
which is summable, we can write
\bear
\lim_{a\to\infty}
\sum_{\Gamma \in {\cal G}_{clus}}
c_{\Gamma }
\left(
a_{(M, \, E), \, \alpha }^{\Gamma} \,
b_{(P, \, D), \, \beta }^{\Gamma}
-1
\right)
a_{\epsilon  -\epsilon ' a}^\Gamma
b_{\mu - \mu ' a}^\Gamma
\mu^{\Gamma}
  & = &
\nonumber \\
\sum_{\Gamma \in {\cal G}_{clus}}
c_{\Gamma }
\left(
a_{(M, \, E), \, \alpha }^{\Gamma} \,
b_{(P, \, D), \, \beta }^{\Gamma}
-1
\right)
(\lim_{a\to\infty}
a_{\epsilon  -\epsilon ' a}^\Gamma
b_{\mu - \mu ' a}^\Gamma
)
\mu^{\Gamma} .
 & &
\eear
But, clearly,
$
\displaystyle
\lim_{a\to\infty}
a_{\epsilon  -\epsilon ' a}^\Gamma
b_{\mu - \mu ' a}^\Gamma
=
a_{\epsilon}^\Gamma
b_{\mu }^\Gamma
$
for every cluster $ \Gamma$, since the polymers are finite.
The limit does not depend on the particular way as $ a\to\infty$.
This shows that the representation (\ref{formulamestracomrho}) holds
also for $\rho \in \calD_q$, $q\neq 0$, and can be used to describe
$\omega_\rho (A)$, $ A\in {\frak A}_{loc}$
with  $\rho \in \calD_q$, $q\neq 0$. The cluster property, and
consequently the ground state property for $ A\in {\frak A}_{loc}$,
can be proven in the same way
as in the previous case.
The proof of Proposition \ref{ieurygpbsiudybvpdiubv} is analogous to
the proof of Proposition \ref{estadosKK2} and does not need to be
repeated but in the next subsection we present the proof of the more
general Theorem \ref{uuiupdsypspYUIYipkjhlkfhg}.
$ \Fullbox$

\subsection{Proof of Theorem \protect\ref{uuiupdsypspYUIYipkjhlkfhg} }
\label{ProofofTheoremuuiupdsypspYUIYipkjhlkfhg}
\setcounter{theorem}{0}

Let us consider
$
\omega_{\rho(a)}
\left(
  \tau_a^{-1}(B)A \alpha_{\rho (a)}^{n}(A' \tau_a^{-1}(B'))
\right)
$
for $A$, $B$, $A'$ and  $B' \in {\frak A}_{loc}$.
According to (\ref{uuuxxxuuuiuziubuvbs}) one has,
for $|a|$ large enough,
$ \alpha_{\rho (a)}^n (A') = \alpha_{\rho_1}^n (A')$
and
$
\alpha_{\rho (a)}^{n}( \tau_a^{-1}(B')) =
\alpha_{\rho_2 a}^{n}( \tau_a^{-1}(B'))
=
\tau_a^{-1} (\alpha_{\rho_2}^n (B') )
$.
Hence, for $|a|$ large enough,
\be
\omega_{\rho(a)}
\left(
  \tau_a^{-1}(B)A \alpha_{\rho (a)}^{n}(A' \tau_a^{-1}(B'))
\right)
=
\omega_{\rho(a)}
\left(
  A \alpha_{\rho_1}^{n}(A')
  \tau_a^{-1}(B \alpha_{\rho_2}^{n}( B') )
\right)
.
\ee
The representation of the last expectation in terms of classical
expectations is given by finite sums of classical expectations like
$
\la
B(\alpha (a) , \; \beta (a))
\ra_{\rho (a)}
$,
where
$ \alpha (a) = \alpha_1 + \alpha_2 a$ and
$ \beta (a) = \beta_1 + \beta_2 a$, for local cochains
$\alpha_{1, \, 2}$ and $ \beta_{1, \, 2}$, where the  cochains
$\alpha_1$ and $ \beta_1$ are related to the operators
$ A\alpha^{n}_{\rho_1}(A')$
and
where the  cochains
$\alpha_2$ and $ \beta_2$ are related to the operators
$ B\alpha^{n}_{\rho_2}(B')$.

Let us now consider the representation of
$
\la
B(\alpha (a) , \; \beta (a))
\ra_{\rho (a)}
$ in terms of cluster expansions.
It is given by
\be
\begin{array}{lll}
\displaystyle
\la
B(\alpha (a) , \; \beta (a))
\ra_{\rho (a)}
   = & &
 \\
 & &
 \\
\displaystyle
\sum_{(M, \, E)\in\mbox{\footnotesize{Conn}}_1 (\alpha (a) ) \atop
(P, \, D)\in\mbox{\footnotesize{Conn}}_2 (\beta(a) ) }
\left[ D-\beta(a) : \, E-\alpha (a) \right]
\left[ D-\beta(a) : \, - \tilde{ \epsilon(a) } \right]
\left[ - \tilde{\mu }(a) : \, E-\alpha(a) \right]
\left[
\prod_{p\in P} g (D(p))
\right]
 & &
\\
 & &
\\
\displaystyle
\times
\left[
\prod_{b\in M}  h (E(b))
\right]
\exp
\left(
\sum_{\Gamma \in {\cal G}_{clus}}
c_{\Gamma }
\left(
a_{(M, \, E), \, \alpha(a) }^{\Gamma} \,
b_{(P, \, D), \, \beta(a) }^{\Gamma}
-1
\right)
a_\epsilon(a)^\Gamma
b_\mu(a)^\Gamma
\mu^{\Gamma}
\right) ,  & &
\label{euiyrwboewetAurnDjtr}
\end{array}
\ee

Let us assume  $|a|$ so large as to include the set
$\mbox{sup }\alpha_1 \cup \mbox{sup }\beta_1 \cup
\mbox{sup }\alpha_2 \cup \mbox{sup }\beta_2$ in the ball of radius
$|a|/8$ centered at the origin and
let us consider two infinite cylinders $C_1$ and $C_2=C_1 + a$ with
radius $ |a|/4$, parallel to the euclidean time axis and extending from
$+\infty$ to $ -\infty$. The cylinder $ C_1$ contains the set
$ \mbox{sup }\alpha_1 \cup \mbox{sup }\beta_1 $
and $ C_2$ contains the set
$(\mbox{sup }\alpha_2 \cup \mbox{sup }\beta_2) \, + \, a$.
By construction, the sets
$
\mbox{Conn}_1 (\alpha (a) )
$
and
$
\mbox{Conn}_2 (\beta (a) )
$
will contain some elements which are entirely contained in $ C_1 \cup C_2$
and some which are not. These last ones must have a size larger than
$ |a|/4$ and, therefore, by  arguments analogous to those used in the
proof of Proposition \ref{propdecayexpn}, their contribution to
(\ref{euiyrwboewetAurnDjtr}) decay exponentially with $ |a|$.
So, up to  exponentially falling error, we can restrict the sums
over
$
\mbox{Conn}_1 (\alpha (a) )
$
and
$
\mbox{Conn}_2 (\beta (a) )
$
to elements contained only in $ C_1 \cup C_2$.
The next question is, what happens to the sums over clusters,
provided the elements $ M$ and $ P$ are now contained in
$C_1 \cup C_2 $?

Let us denote
$M_a := M \cap C_a$ and $ P_a := M\cap C_a$, for $ a=1, \; 2$,
with
$ M = M_1 \cup M_2$ and $ P = P_1 \cup P_2$, as disjoint
unions and
let $E=E_1 + E_2$ and $ D = D_1 + D_2$ with
$ \mbox{sup }E_a = M_a$ and
$ \mbox{sup }D_a = P_a$ for $a=1, \; 2$.

We claim that the difference
\be
\begin{array}{ll}
\displaystyle
\sum_{\Gamma \in {\cal G}_{clus}}
c_{\Gamma }
\left(
a_{(M, \, E), \, \alpha(a) }^{\Gamma} \,
b_{(P, \, D), \, \beta(a) }^{\Gamma}
-1
\right)
a_\epsilon(a)^\Gamma
b_\mu(a)^\Gamma
\mu^{\Gamma} \; \; -
 &
\\
 &
\\
\displaystyle
\left[
\sum_{\Gamma \in {\cal G}_{clus}}
c_{\Gamma }
\left(
a_{(M_1, \, E_1), \, \alpha_1 }^{\Gamma} \,
b_{(P_1, \, D_1), \, \beta_1 }^{\Gamma}
-1
\right)
a_{\epsilon_1}^\Gamma
b_{\mu_1}^\Gamma
\mu^{\Gamma}
+
\sum_{\Gamma \in {\cal G}_{clus}}
c_{\Gamma }
\left(
a_{(M_2, \, E_2), \, \alpha_2  a }^{\Gamma} \,
b_{(P_2, \, D_2), \, \beta_2  a }^{\Gamma}
-1
\right)
a_{\epsilon_2 a}^\Gamma
b_{\mu_2 a}^\Gamma
\mu^{\Gamma}
\right]
\end{array}
\ee
decays exponentially to zero with $ |a|$.
For, notice that the difference above is given by sums over clusters
connecting $ C_1$ to $ C_2$, having thus a size larger than $ |a|/2$.
Therefore, by (\ref{clusterTRES}), their contribution decay
exponentially with $ |a|$.

Using now the exact factorization
\be
\begin{array}{ll}
\displaystyle
\left[ D-\beta(a) : \, E-\alpha (a) \right]
\left[ D-\beta(a) : \, - \tilde{ \epsilon(a) } \right]
\left[ - \tilde{\mu }(a) : \, E-\alpha(a) \right]
\left[
\prod_{p\in P} g (D(p))
\right]
\left[
\prod_{b\in M}  h (E(b))
\right]
=
 \\
 &
\\
\displaystyle
\left[ D_1-\beta_1 : \, E_1-\alpha_1 \right]
\left[ D_1-\beta_1 : \, - \tilde{ \epsilon }_1 \right]
\left[ - \tilde{\mu }_1 : \, E_1-\alpha_1 \right]
\left[
\prod_{p\in P_1} g (D(p))
\right]
\left[
\prod_{b\in M_1}  h (E(b))
\right]
\times &
\\
 &
\\
\displaystyle
\left[ D_2-\beta_2 a : \, E_2-\alpha_2 a  \right]
\left[ D_2-\beta_2 a : \, - \tilde{ \epsilon }_2 a\right]
\left[ - \tilde{\mu }_2 a : \, E_2-\alpha_2 a \right]
\left[
\prod_{p\in P_2} g (D(p))
\right]
\left[
\prod_{b\in M_2}  h (E(b))
\right]
 &
\end{array}
\ee
valid in $ C_1\cup C_2$, we get
using the translation invariance of the cluster expansions and
taking $|a|\to\infty$,

\be
\lim_{|a|\to\infty}
\la
B(\alpha (a) , \; \beta (a))
\ra_{\rho (a) }
=
\la
B(\alpha_1 , \; \beta_1)
\ra_{\rho_1 }
\la
B(\alpha_2 , \; \beta_2)
\ra_{\rho_2 }
.
\ee
With this, the proof of Theorem \ref{uuiupdsypspYUIYipkjhlkfhg}
is complete. $ \Fullbox$

\subsection{Proof of Proposition \protect\ref{Tratio}}
\label{ProofofPropositionTratio}
\setcounter{theorem}{0}

Here we establish part {\em i} of Proposition \ref{Tratio}.

Let $ \rho = (\epsilon , \; \mu)$ and $ \rho '= (\epsilon '  , \; \mu ')$
and define
$
\rho_0 = (\epsilon_0 , \; \mu_0) =
(\epsilon - \epsilon ' , \; \mu - \mu ')
$.
We need
first an expression in terms of the cluster expansions for the ratio
$
\omega_\rho (X^{(2\nu + 1)}_{\rho, \, \rho '} )/
\omega_\rho (X^{(2\nu )}_{\rho, \, \rho '} )
$.
Using a pictorial notation, this
ratio can be written in terms of classical expectations as
\be
\frac{
\omega_\rho (X^{(2\nu + 1)}_{\rho, \, \rho '} )
}{
\omega_\rho (X^{(2\nu )}_{\rho, \, \rho '} )
}
=
\frac{
\mbox{
\setlength{\unitlength}{0.0035in}%
\begingroup\makeatletter\ifx\SetFigFont\undefined
\def\x#1#2#3#4#5#6#7\relax{\def\x{#1#2#3#4#5#6}}%
\expandafter\x\fmtname xxxxxx\relax \def\y{splain}%
\ifx\x\y
\gdef\SetFigFont#1#2#3{%
  \ifnum #1<17\tiny\else \ifnum #1<20\small\else
  \ifnum #1<24\normalsize\else \ifnum #1<29\large\else
  \ifnum #1<34\Large\else \ifnum #1<41\LARGE\else
     \huge\fi\fi\fi\fi\fi\fi
  \csname #3\endcsname}%
\else
\gdef\SetFigFont#1#2#3{\begingroup
  \count@#1\relax \ifnum 25<\count@\count@25\fi
  \def\x{\endgroup\@setsize\SetFigFont{#2pt}}%
  \expandafter\x
    \csname \romannumeral\the\count@ pt\expandafter\endcsname
    \csname @\romannumeral\the\count@ pt\endcsname
  \csname #3\endcsname}%
\fi
\fi\endgroup
\begin{picture}(920,675)(25,180)
\thinlines
\put(115,855){\line(-1,-4){ 84.118}}
\put(855,185){\line( 1, 4){ 84.118}}
\put(855,855){\line( 1,-4){ 84.118}}
\put(515,800){\makebox(0.1111,0.7778){\SetFigFont{5}{6}{rm}.}}
\multiput(160,220)(8.98649,0.00000){75}{\makebox(0.1111,0.7778){\SetFigFont{5}{6}{rm}.}}
\multiput(160,795)(8.98649,0.00000){75}{\makebox(0.1111,0.7778){\SetFigFont{5}{6}{rm}.}}
\multiput(160,220)(0.00000,8.98438){65}{\makebox(0.1111,0.7778){\SetFigFont{5}{6}{rm}.}}
\multiput(825,220)(0.00000,8.98438){65}{\makebox(0.1111,0.7778){\SetFigFont{5}{6}{rm}.}}
\thicklines
\put(280,795){\line( 0,-1){575}}
\put(280,220){\line( 0, 1){  5}}
\put(350,795){\line( 0,-1){575}}
\thinlines
\multiput(390,795)(0.00000,-8.04196){72}{\line( 0,-1){  4.021}}
\put(115,185){\line(-1, 4){ 84.118}}
\multiput(420,795)(0.00000,-8.04196){72}{\line( 0,-1){  4.021}}
\put(702,528){\makebox(0,0)[lb]{\smash{\SetFigFont{12}{14.4}{rm}$2\nu + 1$}}}
\multiput(240,797)(0.00000,-8.04196){72}{\line( 0,-1){  4.021}}
\put(684,696){\vector( 0, 1){  0}}
\put(684,696){\vector( 0,-1){363}}
\thicklines
\put(501,330){\framebox(145,365){}}
\put(294,180){\makebox(0,0)[lb]{\smash{\SetFigFont{12}{14.4}{rm}$\rho$}}}
\put(561,279){\makebox(0,0)[lb]{\smash{\SetFigFont{12}{14.4}{rm}$\rho
	- \rho'$}}}
\put(540,723){\makebox(0,0)[lb]{\smash{$\ell_e$ or $\ell_m$}}}
\end{picture}
}
}{
\mbox{
\setlength{\unitlength}{0.0035in}%
\begingroup\makeatletter\ifx\SetFigFont\undefined
\def\x#1#2#3#4#5#6#7\relax{\def\x{#1#2#3#4#5#6}}%
\expandafter\x\fmtname xxxxxx\relax \def\y{splain}%
\ifx\x\y
\gdef\SetFigFont#1#2#3{%
  \ifnum #1<17\tiny\else \ifnum #1<20\small\else
  \ifnum #1<24\normalsize\else \ifnum #1<29\large\else
  \ifnum #1<34\Large\else \ifnum #1<41\LARGE\else
     \huge\fi\fi\fi\fi\fi\fi
  \csname #3\endcsname}%
\else
\gdef\SetFigFont#1#2#3{\begingroup
  \count@#1\relax \ifnum 25<\count@\count@25\fi
  \def\x{\endgroup\@setsize\SetFigFont{#2pt}}%
  \expandafter\x
    \csname \romannumeral\the\count@ pt\expandafter\endcsname
    \csname @\romannumeral\the\count@ pt\endcsname
  \csname #3\endcsname}%
\fi
\fi\endgroup
\begin{picture}(920,675)(25,180)
\thinlines
\put(115,855){\line(-1,-4){ 84.118}}
\put(855,185){\line( 1, 4){ 84.118}}
\put(855,855){\line( 1,-4){ 84.118}}
\put(515,800){\makebox(0.1111,0.7778){\SetFigFont{5}{6}{rm}.}}
\multiput(160,220)(8.98649,0.00000){75}{\makebox(0.1111,0.7778){\SetFigFont{5}{6}{rm}.}}
\multiput(160,795)(8.98649,0.00000){75}{\makebox(0.1111,0.7778){\SetFigFont{5}{6}{rm}.}}
\multiput(160,220)(0.00000,8.98438){65}{\makebox(0.1111,0.7778){\SetFigFont{5}{6}{rm}.}}
\multiput(825,220)(0.00000,8.98438){65}{\makebox(0.1111,0.7778){\SetFigFont{5}{6}{rm}.}}
\thicklines
\put(280,795){\line( 0,-1){575}}
\put(280,220){\line( 0, 1){  5}}
\put(350,795){\line( 0,-1){575}}
\thinlines
\multiput(390,795)(0.00000,-8.04196){72}{\line( 0,-1){  4.021}}
\put(115,185){\line(-1, 4){ 84.118}}
\multiput(420,795)(0.00000,-8.04196){72}{\line( 0,-1){  4.021}}
\put(702,528){\makebox(0,0)[lb]{\smash{\SetFigFont{12}{14.4}{rm}$2\nu $}}}
\multiput(240,797)(0.00000,-8.04196){72}{\line( 0,-1){  4.021}}
\put(684,696){\vector( 0, 1){  0}}
\put(684,696){\vector( 0,-1){363}}
\thicklines
\put(501,330){\framebox(145,365){}}
\put(294,180){\makebox(0,0)[lb]{\smash{\SetFigFont{12}{14.4}{rm}$\rho$}}}
\put(561,279){\makebox(0,0)[lb]{\smash{\SetFigFont{12}{14.4}{rm}$\rho
	- \rho'$}}}
\put(540,723){\makebox(0,0)[lb]{\smash{\SetFigFont{12}{14.4}{rm}$\ell_e$
			  or $\ell_m$}}}
\end{picture}
}
} \qquad .
\ee
The infinite vertical lines indicate the background charges $\rho$ and
the finite loops are constructed over the charge distribution
$\rho - \rho'$. Their
horizontal lines represent the strings $\ell_e$ and/or $ \ell_m$ used in the
definition (\ref{Aroro'})
and their vertical lines have length $2\nu +1$ in the numerator and
$2\nu$ in the denominator, respectively
Notice that $ \rho $ and $ \rho -\rho ' $ may have a non-empty overlap,
a circumstance not shown in the figure for reasons of clarity.

The next step is to find an expansion for the last expression in terms
of our cluster expansions. The result is
\be
\frac{
\omega_\rho (X^{(2\nu + 1)}_{\rho, \, \rho '} )
}{
\omega_\rho (X^{(2\nu )}_{\rho, \, \rho '} )
}
=
\exp
\left\{
\sum_{\Gamma}
\;
c_\Gamma
\;
\left(
a_{\epsilon_0 , \; 0, \; 2\nu +1}^\Gamma
b_{\mu_0  , \; 0, \; 2\nu +1}^\Gamma
-
a_{\epsilon_0  , \; 0, \; 2\nu   }^\Gamma
b_{\mu_0  , \; 0, \; 2\nu   }^\Gamma
\right)\;
a_\epsilon^\Gamma \; b_\mu^\Gamma \; \mu^\Gamma
\right\}
,
\label{rtbdrgheaart}
\ee
where, in an almost self-explanatory notation,
$a_{\epsilon_0 , \; \alpha, \; \beta} (\gamma )$, $ \alpha < \beta$,
represents the winding number of the
magnetic part of the polymer $\gamma $ with the electric loop built by
the horizontal strings $ \ell_e$ located at euclidean times $ \alpha $
and $ \beta \in \Z$ and by the vertical electric lines located over the support
of $ \epsilon_0 $ with length $  \beta - \alpha$. The quantity
$b_{\mu_0 , \; \alpha, \; \beta }$ is defined analogously.

The right hand side of (\ref{rtbdrgheaart}) can be written as
\be
\exp
\left\{
\sum_{\Gamma}
\;
c_\Gamma
\;
\left(
a_{\epsilon_0  , \; 0, \; 1}^\Gamma
b_{\mu_0  , \; 0, \; 1}^\Gamma
-
1
\right)\;
a_{\epsilon_0  , \; - 2\nu , \;  0  }^\Gamma
b_{\mu_0  , \; - 2\nu , \; 0   }^\Gamma
a_\epsilon^\Gamma \; b_\mu^\Gamma \; \mu^\Gamma
\right\}
,
\label{Fuetbghoervew}
\ee
where, above, we used the factorization properties
\bear
a_{\epsilon_0 , \; \alpha, \; \beta }(\gamma )
a_{\epsilon_0 , \; \beta, \; \delta }(\gamma ) & = &
a_{\epsilon_0 , \; \alpha, \; \delta }(\gamma )
\\
b_{\mu_0 , \; \alpha, \; \beta }(\gamma )
b_{\mu_0 , \; \beta, \; \delta }(\gamma ) & = &
b_{\mu_0 , \; \alpha, \; \delta }(\gamma )
,
\eear
for $ \alpha<\beta <\delta \in \Z$ and we used translation invariance.
Taking the limit $ \nu\to\infty$ of expression (\ref{Fuetbghoervew})
is easy and gives
\be
\frac{
\| T_\rho (\rho ') \|
}{
\| T_\rho (\rho  ) \|
}
=
\exp
\left\{
\sum_{\Gamma}
\;
c_\Gamma
\;
\left(
a_{\epsilon_0  , \; 0, \; 1}^\Gamma
b_{\mu_0  , \; 0, \; 1}^\Gamma
-
1
\right)\;
a_{\epsilon_0  , \; - \infty , \;  0  }^\Gamma
b_{\mu_0  , \; - \infty , \; 0   }^\Gamma
a_\epsilon^\Gamma \; b_\mu^\Gamma \; \mu^\Gamma
\right\}
,
\label{Toijerhvbppwe}
\ee
where
$
\displaystyle
a_{\epsilon_0  , \; - \infty , \;  0  } (\gamma )
=
\lim_{j \to\infty}
a_{\epsilon_0 , \; - j , \;  0  } (\gamma )
$ etc., which is a well defined limit for each $\gamma$, since the
polymers are finite
(for each $ \gamma $, the limit is reached at finite $ j$).

Next, we are interested in studying the limit
$
\displaystyle
\lim_{|a|\to\infty}
\frac{
\| T_{\rho_1 (a)} (\rho_2 (a)) \|
}{
\| T_{\rho_2 (a)} (\rho_2 (a)  ) \|
}
$
using its representation in terms of cluster expansions.

The main technical problem we have to confront with is the fact that,
if $ \rho_1 - \rho_2$ have a non-zero total charge,
the strings $\ell_e$ and $\ell_m$ have to connect
elements of the support of $\rho_1 - \rho_2$ with elements of
the support of $(\rho_1 - \rho_2)\cdot a $ and have, hence, a length which
increases with $ |a|$. The crucial observation is, however, that
the left hand side of (\ref{Toijerhvbppwe}) does not depend on the
strings $ \ell_e $ and $\ell_m$, although this independence cannot apparently be
seen from the representation in terms of cluster expansions.

Let us consider two cylinders $C_1(r)$ and
$C_2 (r) = C_1(r) + a$, such that $ C_1(r)$ is  centered on the euclidean time
axis, extending from $-\infty$ to $ \infty$ and has a radius $ r$.
Denote by $ r_0 $ the largest distance from the set
$\mbox{sup }(\rho_1)\cup\mbox{sup }( \rho_2)$ to the origin of the lattice
and consider $| a |$ large enough so that
$\mbox{sup }(\rho_1 )\cup \mbox{sup }( \rho_2)$ is contained in $ C_1 (|a|/8)$
(by taking, say, $|a| > 16 r_0$).

We first observe that the sum over clusters contained in
$ \Z^3 \setminus (C_1 (|a|/8) \cup C_2(|a|/8))$ does not contribute to
(\ref{Fuetbghoervew}). This can be seen at best in
(\ref{rtbdrgheaart}) by noticing that: 1) clusters contained in
$ \Z^3 \setminus (C_1 (|a|/8) \cup C_2 (|a|/8))$
crossing the $ t=0$ euclidean
plane  and having a side smaller than $ 2\nu$
have a zero contribution
(for them, one has
$
a_{\epsilon_0 , \; 0, \; 2\nu +1}^\Gamma
b_{\mu_0  , \; 0, \; 2\nu +1}^\Gamma
=
a_{\epsilon_0  , \; 0, \; 2\nu   }^\Gamma
b_{\mu_0  , \; 0, \; 2\nu   }^\Gamma
$);
2) clusters contained in
$ \Z^3 \setminus (C_1 (|a|/8)\cup C_2(|a|/8)) $ crossing the $ t=2\nu$ euclidean
plane, having a side smaller than $ 2\nu $ and having a non-zero
contribution cancel with their translates
by one unit in euclidean time direction; 3) the only surviving clusters
in $ \Z^3 \setminus (C_1(|a|/8) \cup C_2(|a|/8)) $ must cross the planes
$ t=0 $ and $ t=2\nu $, and therefore, their side is larger than $ 2\nu $
and their contribution decays exponentially when the limit
$ \nu\to\infty $ is taken.

It remains to consider two classes of clusters:
{\bf a)} those entirely
contained in $ C_1 (|a|/4) \cup C_2 (|a|/4)$ and
having a non-empty intersection with $ C_1 (d_0) \cup C_2 (d_0)$
for a fixed  $d_0$ with $r_0 < d_0 < a/8$ and
{\bf b)} those having a
non-empty intersection with both  $ C_1 (d_0) \cup C_2 (d_0)$ and
$\Z^3 \setminus (C_1 (|a|/4) \cup C_2 (|a|/4))$.

The contribution to the clusters belonging to the class {\bf b} decays
exponentially with $ |a|$. For, note that the clusters which give a
non-zero contribution to (\ref{Fuetbghoervew}) must either cross  the
$ t=0$ plane or the $ t=1$ plane (or eventually both).
The clusters of this sort having a
non-empty intersection with both  $ C_1 (d_0) \cup C_2 (d_0)$ and
$\Z^3 \setminus (C_1 (|a|/4) \cup C_2 (|a|/4))$ must have a size
larger that $|a|/8$ and , hence, by (\ref{clusterTRES}),
their contribution decays exponentially with $|a|$.

It remains now to consider the clusters belonging to the class {\bf a} above.
They are entirely contained inside one of the cylinders $ C_1 (|a|/4)$
or $ C_2 (|a|/4)$. Since we have freedom to choose the strings
$\ell_e$ and $ \ell_m$ at our will, we choose them depending on $ a $
such that, inside
of $ C_1 (|a|/4) \setminus C_1 (d_0 ) $ and
   $ C_2 (|a|/4) \setminus C_2 (d_0 ) $ they run parallel to a fixed
direction, say to the positive $ x $-axis of $ \Z^2$.
Now, taking the limit $|a|\to\infty$ is straight forward and gives
$ c_{\rho_1, \; \rho_2}$ independent on the way the sequence $ a$ goes
to infinity. The result is that
$
c_{\rho_1, \; \rho_2} =
d_{\rho_1, \; \rho_2} d_{-\rho_1, \; -\rho_2}
$,
where
\be
d_{\rho_1, \; \rho_2} \; := \;
\exp
\left\{
\sum_{\Gamma \atop {\Gamma \cap C_1(d_0) \neq\emptyset }}
\;
c_\Gamma
\;
\left(
a_{\epsilon_{12}  , \; 0, \; 1 ; \; \infty }^\Gamma
b_{\mu_{12}  , \; 0, \; 1 ; \; \infty }^\Gamma
-
1
\right)\;
a_{\epsilon_{12}  , \; - \infty , \;  0  ; \; \infty }^\Gamma
b_{\mu_{12}  , \; - \infty , \; 0  ; \; \infty  }^\Gamma
a_{\epsilon_1}^\Gamma \; b_{\mu_1}^\Gamma \; \mu^\Gamma
\right\}
,
\label{TfkudghLJZglfkueslr}
\ee
for any sufficiently large $ d_0$,
where
$ \rho_1 - \rho_2 = (\epsilon_{12}, \; \mu_{12})$
and where
\be
a_{\epsilon_{12}  , \; \alpha, \; \beta ; \; \infty } (\gamma)
=
\lim_{|a|\to\infty}
a_{  (\epsilon_1 - \epsilon_2) - (\epsilon_1 - \epsilon_2)\cdot a    ,
\; \alpha, \; \beta   } (\gamma ) ,
\ee
etc., where $ \alpha < \beta \in \Z$ and
in
$
a_{  (\epsilon_1 - \epsilon_2) - (\epsilon_1 - \epsilon_2)\cdot a    ,
\; \alpha, \; \beta   } (\gamma )
$
the strings $ \ell_e $  depend on $ a $ in the way
described above, i.e.,
such that, inside
of $ C_1 (|a|/4) \setminus C_1 ( d_0 ) $ and
   $ C_2 (|a|/4) \setminus C_2 ( d_0 ) $ they point parallel
to the positive $ x $-axis of $ \Z^2$. Note that, for each $ \gamma $,
the limit above is reached at finite values of $ |a| $.
The condition $ \Gamma \cap C_1 (d_0) \neq\emptyset $ means
that the geometrical part
of the cluster $ \Gamma $ must have a non-empty intersection with the
cylinder $C_1 (d_0)$.
Note also that
the convergence of the sum over clusters in (\ref{TfkudghLJZglfkueslr}) can be shown
using the fact that the contributing clusters have a non-empty
intersection with $ C_1 (d_0)$ and with the $t=0$ and/or $ t=1$
euclidean time slices together with the exponential decay provided by
(\ref{clusterTRES}). We can say, for instance, that the sum over
clusters in (\ref{TfkudghLJZglfkueslr}) can be bounded by
\be
const. \;
\sum_{t= -\infty }^{\infty}
\sum_{
\Gamma \atop
{\Gamma \cap \left( C_1(d_0) \cap T_t \right) \neq\emptyset
\atop { \| \Gamma \| \geq t}
}
}
\; |c_\Gamma| \; |\mu^\Gamma |
\leq
const. \; \sum_{t= -\infty }^{\infty} \; e^{-c_a |t|}
< \infty
\ee
with some positive constant $ c_a$
where $ T_t$ is the euclidean time-plane at euclidean time $ t$.

Using the representation above in terms of cluster expansions
one can also easily show that
$ d_{-\rho_1, \; -\rho_2}= \overline{ d_{\rho_1, \; \rho_2} }$.

The next problem is to prove the factorization property (\ref{3.64}).
The arguments used are analogous to those leading to
(\ref{TfkudghLJZglfkueslr}). We can namely prove that
$
\displaystyle
\lim_{b\to\infty}
d_{\rho_1 + \rho_1 ' \cdot b, \; \rho_2 + \rho_2 ' \cdot b }=
d_{\rho_1 , \; \rho_2  }d_{\rho_1 ' , \;  \rho_2 ' }
$.
This can be obtained using the representation
(\ref{TfkudghLJZglfkueslr})
with $ d_0$ depending on $ b $ such that
$C_1 (d_0)$ contains
$ \mbox{sup } \rho_1 \cup \mbox{sup }\rho_2 \cup
 \mbox{sup }( \rho_1'\cdot b) \cup \mbox{sup }(\rho_2 ' \cdot b)
$.
We consider again cylinders
$D_1(|b|/4) = C_1(|b|/4)$ and $D_2(|b|/4) = D_1 (|b|/4) + b$, both
contained in $ C_1 (d_0 (b))$, with
$ D_1 (|b|/4)$ containing the set
$ \mbox{sup } \rho_1 \cup \mbox{sup }\rho_2$
and $ D_2(|b|/4)$ containing the set
$ \mbox{sup } (\rho_1 ' \cdot b) \cup \mbox{sup }(\rho_2 ' \cdot b)$
for some $ |b|$ large enough.
Repeating the previous arguments, we can neglect contributions from
clusters contained outside of
$ \Z^3 \setminus (D_1 (|b|/4)\cup D_2(|b|/4)) $
and take the pieces of the strings $ \ell_e $ and $ \ell_m $ which join the
supports of $ \rho_1 $ and $ \rho_2 $ with the supports of
$ \rho_1 ' \cdot b $ and $ \rho_2 ' \cdot  b$  so that they again
run parallel to the $ x$-axis at sufficiently large distances. The
desired relation will follow again from the usual clustering properties of the
cluster expansions established above.
$ \Fullbox $

\subsection{Completing the Proof of Proposition \protect\ref{Phiroro'}}
\label{CompletingtheProofofPropositionPhiroro}
\setcounter{theorem}{0}

We will here complete some missing points in the proof of
Proposition \ref{Phiroro'}. The ideas are actually contained in
\cite{FredMarcu} and therefore we will concentrate only on the more
relevant details.

To show that the sequence of unit vectors $\Phi_{\rho , \; \rho '}^{n} $,
$ n\in \N$ is a Cauchy sequence it is enough to show that, for any
$\epsilon > 0$, one has
$
 |(\Phi_{\rho , \; \rho '}^{n} , \; \Phi_{\rho , \; \rho '}^{m} ) -1|
 < \epsilon
$
provided $ n$ is large enough, with $ m > n $.
The scalar product
$ (\Phi_{\rho , \; \rho '}^{n} , \; \Phi_{\rho , \; \rho '}^{m} ) $
can be expressed as the exponential of a sum over clusters and, hence,
it is enough to show that this sum is small enough
provided $ n$ is large enough, with $ m > n $.

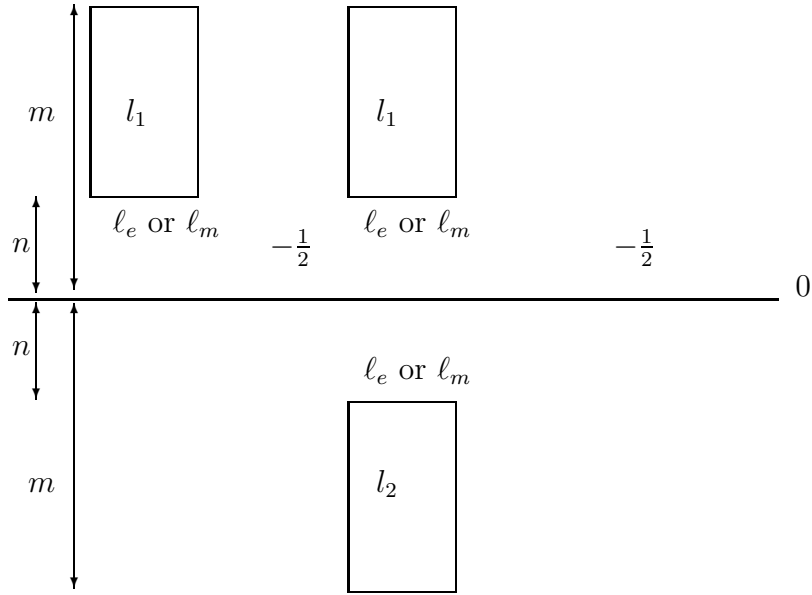
\begin{figure}[hbtp]
\begin{center}
\setlength{\unitlength}{0.006in}%
\begingroup\makeatletter\ifx\SetFigFont\undefined
\def\x#1#2#3#4#5#6#7\relax{\def\x{#1#2#3#4#5#6}}%
\expandafter\x\fmtname xxxxxx\relax \def\y{splain}%
\ifx\x\y
\gdef\SetFigFont#1#2#3{%
  \ifnum #1<17\tiny\else \ifnum #1<20\small\else
  \ifnum #1<24\normalsize\else \ifnum #1<29\large\else
  \ifnum #1<34\Large\else \ifnum #1<41\LARGE\else
     \huge\fi\fi\fi\fi\fi\fi
  \csname #3\endcsname}%
\else
\gdef\SetFigFont#1#2#3{\begingroup
  \count@#1\relax \ifnum 25<\count@\count@25\fi
  \def\x{\endgroup\@setsize\SetFigFont{#2pt}}%
  \expandafter\x
    \csname \romannumeral\the\count@ pt\expandafter\endcsname
    \csname @\romannumeral\the\count@ pt\endcsname
  \csname #3\endcsname}%
\fi
\fi\endgroup
\begin{picture}(687,513)(63,270)
\thinlines
\put(135,615){\framebox(93,165){}}
\put(360,615){\framebox(93,165){}}
\put(360,270){\framebox(93,165){}}
\put( 87,615){\vector( 0, 1){  0}}
\put( 87,615){\vector( 0,-1){ 84}}
\put( 87,522){\vector( 0, 1){  0}}
\put( 87,522){\vector( 0,-1){ 84}}
\put(120,783){\vector( 0, 1){  0}}
\put(120,783){\vector( 0,-1){249}}
\put(120,522){\vector( 0, 1){  0}}
\put(120,522){\vector( 0,-1){249}}
\put(750,528){\makebox(0,0)[lb]{\smash{\SetFigFont{12}{14.4}{rm}$0$}}}
\put( 63,525){\line( 1, 0){672}}
\put(282,564){\makebox(0,0)[lb]{\smash{\SetFigFont{12}{14.4}{rm}
$-\frac{1}{2}$}}}
\put(364,456){\makebox(0,0)[lb]{\smash{\SetFigFont{12}{14.4}{rm}
$\ell_e$ or $\ell_m$}}}
\put(582,564){\makebox(0,0)[lb]{\smash{\SetFigFont{12}{14.4}{rm}
$-\frac{1}{2}$}}}
\put( 66,567){\makebox(0,0)[lb]{\smash{\SetFigFont{12}{14.4}{rm}$n$}}}
\put( 66,477){\makebox(0,0)[lb]{\smash{\SetFigFont{12}{14.4}{rm}$n$}}}
\put(80,681){\makebox(0,0)[lb]{\smash{\SetFigFont{12}{14.4}{rm}$m$}}}
\put(80,357){\makebox(0,0)[lb]{\smash{\SetFigFont{12}{14.4}{rm}$m$}}}
\put(145,584){\makebox(0,0)[lb]{\smash{\SetFigFont{12}{14.4}{rm}
$\ell_e$ or $\ell_m$}}}
\put(364,584){\makebox(0,0)[lb]{\smash{\SetFigFont{12}{14.4}{rm}
$\ell_e$ or $\ell_m$}}}
\put(165,681){\makebox(0,0)[lb]{$l_1$}}
\put(384,681){\makebox(0,0)[lb]{$l_1$}}
\put(384,357){\makebox(0,0)[lb]{$l_2$}}
\end{picture}
\end{center}
\caption{Schematic representation of the expression
$
a^\Gamma_{l_1} b^\Gamma_{l_1}
- \frac{1}{2}
a^\Gamma_{l_1, \; l_2} b^\Gamma_{l_1, \; l_2}
- \frac{1}{2}
$
and the loops $ l_1$ (above) and $ l_2$ (below).
}
\label{FFiGG1}
\end{figure}

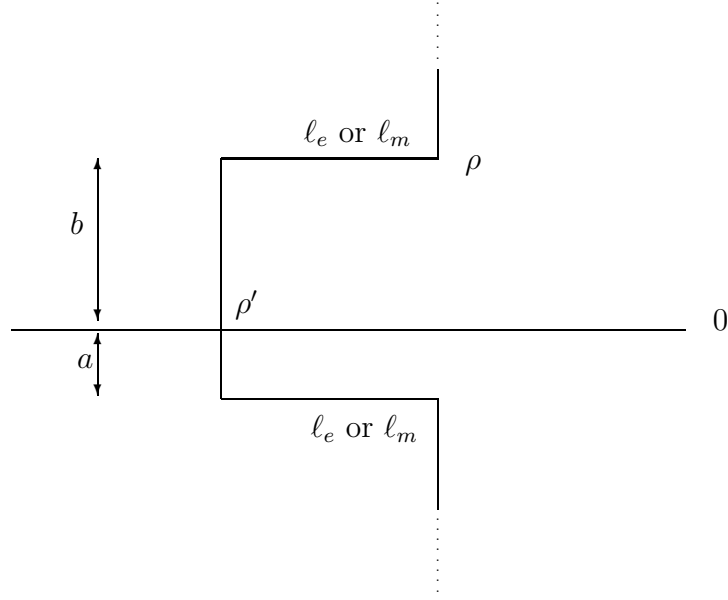
\begin{figure}[hbtp]
\begin{center}
\setlength{\unitlength}{0.006in}%
\begingroup\makeatletter\ifx\SetFigFont\undefined
\def\x#1#2#3#4#5#6#7\relax{\def\x{#1#2#3#4#5#6}}%
\expandafter\x\fmtname xxxxxx\relax \def\y{splain}%
\ifx\x\y
\gdef\SetFigFont#1#2#3{%
  \ifnum #1<17\tiny\else \ifnum #1<20\small\else
  \ifnum #1<24\normalsize\else \ifnum #1<29\large\else
  \ifnum #1<34\Large\else \ifnum #1<41\LARGE\else
     \huge\fi\fi\fi\fi\fi\fi
  \csname #3\endcsname}%
\else
\gdef\SetFigFont#1#2#3{\begingroup
  \count@#1\relax \ifnum 25<\count@\count@25\fi
  \def\x{\endgroup\@setsize\SetFigFont{#2pt}}%
  \expandafter\x
    \csname \romannumeral\the\count@ pt\expandafter\endcsname
    \csname @\romannumeral\the\count@ pt\endcsname
  \csname #3\endcsname}%
\fi
\fi\endgroup
\begin{picture}(612,513)(75,312)
\thinlines
\put(447,768){\line( 0,-1){ 78}}
\put(447,690){\line(-1, 0){189}}
\put(258,690){\line( 0,-1){210}}
\put(258,480){\line( 1, 0){189}}
\put(447,480){\line( 0,-1){ 96}}
\put(447,384){\line( 0, 1){  3}}
\multiput(447,375)(0.00000,-9.00000){8}{\makebox(0.1111,0.7778){\SetFigFont{5}{6}{rm}.}}
\multiput(447,825)(0.00000,-9.60000){6}{\makebox(0.1111,0.7778){\SetFigFont{5}{6}{rm}.}}
\put(150,690){\vector( 0, 1){  0}}
\put(150,690){\vector( 0,-1){141}}
\put(150,537){\vector( 0, 1){  0}}
\put(150,537){\vector( 0,-1){ 54}}
\put( 75,540){\line( 1, 0){588}}
\put(687,540){\makebox(0,0)[lb]{\smash{\SetFigFont{12}{14.4}{rm}$0$}}}
\put(471,681){\makebox(0,0)[lb]{\smash{\SetFigFont{12}{14.4}{rm}$\rho$}}}
\put(126,624){\makebox(0,0)[lb]{\smash{\SetFigFont{12}{14.4}{rm}$b$}}}
\put(132,507){\makebox(0,0)[lb]{\smash{\SetFigFont{12}{14.4}{rm}$a$}}}
\put(321,705){\makebox(0,0)[lb]{\smash{\SetFigFont{12}{14.4}{rm}
$\ell_e$ or $\ell_m$}}}
\put(327,447){\makebox(0,0)[lb]{\smash{\SetFigFont{12}{14.4}{rm}
$\ell_e$ or $\ell_m$}}}
\put(270,555){\makebox(0,0)[lb]{\smash{\SetFigFont{12}{14.4}{rm}$\rho '$}}}
\end{picture}
\end{center}
\caption{Schematic representation of the partial replacement of the
  infinite vertical line representing the background charge distribution
  $\rho$ by $ \rho '$. The connections are performed at euclidean time
  planes $ b$ and $ -a$, $0 \leq a <b$,
through the strings $ \ell_e$ and/or $ \ell_m$.}
\label{FFiGG2}
\end{figure}

This sum can be written as
\be
\sum_{\Gamma} \;
c_{\Gamma}
\left(
a^\Gamma_{l_1} b^\Gamma_{l_1}
- \frac{1}{2}
a^\Gamma_{l_1, \; l_2} b^\Gamma_{l_1, \; l_2}
- \frac{1}{2}
\right)\;
\;
a^\Gamma_{m , \; m} \, b^\Gamma_{m , \; m}
\;
a_{\rho}^{\Gamma} \, b_{\rho}^{\Gamma} \;
\mu^{\Gamma} .
\label{uhgvcbduivbrepw}
\ee
Above $a^\Gamma_{l_1}$, $ b^\Gamma_{l_1} $,
$ a^\Gamma_{l_1, \; l_2} $ and $ b^\Gamma_{l_1, \; l_2}$
are the electric and magnetic winding numbers on the loops
$ l_1$ and $ l_1 \cup \l_2$ schematically represented in Figure \ref{FFiGG1}
(where $l_2 = \theta l_1$, $ \theta $ meaning reflection on the
$ t=0$ euclidean time plane).
Also above
$a^\Gamma_{a , \; b}$ and $b^\Gamma_{a , \; b}$
(with $ 0 \leq a<b$)
are the electric and magnetic winding numbers around the infinite
vertical lines
schematically represented in Figure \ref{FFiGG2}.

By a straight forward inspection we can verify that a cluster $ \Gamma$ with a size
smaller than $ n$ with a non-trivial winding number with, say,
the loop $ l_1 $ are canceled in the sum
(\ref{uhgvcbduivbrepw}) by the contribution of the  reflected
cluster $ \theta \Gamma$.
The contribution of the clusters entirely contained between the
time-slices
$ n$ and $ m$ and the contribution of the clusters entirely contained between the
time-slices
$ -n$ and $ -m$ also cancel mutually.
The only surviving clusters must have
non-trivial winding numbers with both $ l_1$ and $ l_2$
simultaneously and must cross both planes at time $ n$ and $ -n$.
Therefore, they must have a size which increases with
$n$. By estimate (\ref{Xtres}) their contributions disappear when
$ n\to\infty$ uniformly  $ m$,  completing thus the proof.

\subsection{Proof of Theorem \protect\ref{puiDyrbpowuie}}
\label{uyhvbngsSuiry}
\setcounter{theorem}{0}

Let us start proving {\em i)}.
We will first consider the case where $ d^* l_1 \neq  0$ and
$ d^* l_2 \neq 0$. Without loss, we will take $ l_1$ and $ l_2$ as
having support on single lattice links.
According to the definitions we have
\bear
\hat{\calE}_q (l_1 + l_2) \Omega_\rho
& = &
\phi^{el}_{\rho_{1, \, 2}}(l_1 + l_2)
\nonumber \\
 & = &
\lim_{n\to\infty}
\frac{
\pi_{\rho_{1, \, 2}}
\left(
\alpha_{\rho}^n
\left(
A_{\rho_{1, \, 2}, \; \rho } ((l_1 + l_2), \; 0)
\right)
\right)
\Omega_{\rho_{1, \, 2} }
}{
N_1 (n)
} ,
\eear
with
\be
N_1 (n) \; := \;
\left\|
\pi_{\rho_{1, \, 2}   }
\left(
\alpha_{\rho}^n
\left(
A_{\rho_{1, \, 2} , \; \rho } ((l_1 + l_2), \; 0)
\right)
\right)
\Omega_{\rho_{1, \, 2} }
\right\|
\ee
and
\bear
\hat{\calE}_q (l_1) \hat{\calE}_q (l_2)\Omega_\rho
 & = &
\lim_{p\to\infty}
\hat{\calE}_q (l_1)
\frac{
\pi_{\rho_{2} }
\left(
\alpha_{\rho}^p
\left(
A_{\rho_{2} , \; \rho} ( l_2, \; 0)
\right)
\right)
\Omega_{\rho_{2} }
}{
\left\|
\pi_{\rho_{2} }
\left(
\alpha_{\rho}^p
\left(
A_{\rho_{2} , \; \rho} ( l_2, \; 0)
\right)
\right)
\Omega_{\rho_{2} }
\right\|
}
\nonumber \\
 & = &
\lim_{p\to\infty}
\frac{
\pi_{\rho_{1, \, 2}  }
\left(
\alpha_{\rho}^p
\left(
A_{\rho_{2} , \; \rho} ( l_2, \; 0)
\right)
\right)
\phi^{el}_{\rho_{1, \, 2} }(l_1)
}{
\left\|
\pi_{\rho_{2} }
\left(
\alpha_{\rho}^p
\left(
A_{\rho_{2} , \; \rho} ( l_2, \; 0)
\right)
\right)
\Omega_{\rho_{2} }
\right\|
},
\eear
where $ \rho_i = \rho + (d^* l_i , \; 0)$, $ i = 1, \; 2$
and
$ \rho_{1, \, 2} = \rho + (d^* (l_1 + l_2) , \; 0)$,

 The vector in the right hand side can be written as
\be
\lim_{p\to\infty}
\lim_{q\to\infty}
\frac{
\pi_{\rho_{1, \, 2} }
\left(
\alpha_{\rho}^p
\left(
A_{\rho_{2} , \; \rho} ( l_2, \; 0)
\right)
\alpha_{\rho_{2} }^q
\left(
A_{\rho_{1, \, 2} , \; \rho_{2}  } (l_1, \; 0)
\right)
\right)
\Omega_{\rho_{1, \, 2} }
}
{N_2(p) N_3 (q)} ,
\label{Tugoiskfgeg}
\ee
where $N_1(p)$ and $N_2 (q)$ are the normalization factors
\be
N_2 (p) \; := \;
\left\|
\pi_{\rho_{2} }
\left(
\alpha_{\rho}^p
\left(
A_{\rho_{2} , \; \rho} ( l_2, \; 0)
\right)
\right)
\Omega_{\rho_{2} }
\right\|,
\ee
and
\be
N_3(q) \; :=  \;
\left\|
\pi_{\rho_{1, \, 2} }
\left(
\alpha_{\rho_{2}}^q
\left(
A_{\rho_{1, \, 2} , \; \rho_{2}  } (l_1, \; 0)
\right)
\right)
\Omega_{\rho_{1, \, 2} }
\right\|
 .
\ee
The scalar product in (\ref{BTOVCWEIUsjs1}) can now be written as
\be
\lim_{n\to\infty}
\lim_{p\to\infty}
\lim_{q\to\infty}
\frac{
\left(
\Omega_{\rho_{1, \, 2} }
, \; \;
\pi_{\rho_{1, \, 2} }
\left(
A
\right)
\Omega_{\rho_{1, \, 2} }
\right)
}{
N_1 (n) N_2 (p) N_3 (q)
} ,
\label{Tfiusgdoflkgjw}
\ee
with
\be
A \; := \;
\alpha_{\rho}^n
\left(
A_{\rho_{1, \, 2}, \; \rho } ((l_1 + l_2), \; 0)
\right)^*
\alpha_{\rho}^p
\left(
A_{\rho_{2} , \; \rho} ( l_2, \; 0)
\right)
\alpha_{\rho_{2} }^q
\left(
A_{\rho_{1, \, 2} , \; \rho_{2}  } (l_1, \; 0)
\right) .
\ee

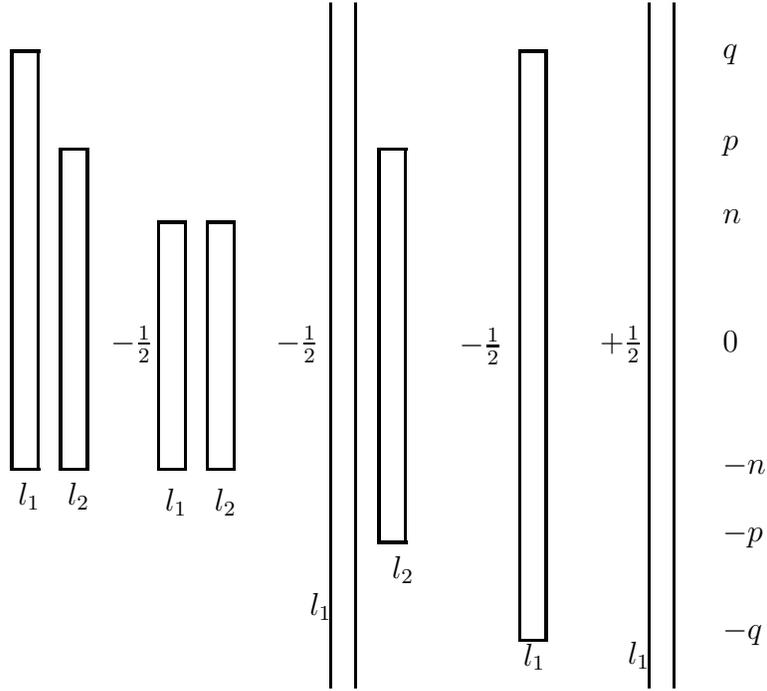
\begin{figure}[hbtp]
\begin{center}
\setlength{\unitlength}{0.00043300in}%
\begingroup\makeatletter\ifx\SetFigFont\undefined%
\gdef\SetFigFont#1#2#3#4#5{%
  \reset@font\fontsize{#1}{#2pt}%
  \fontfamily{#3}\fontseries{#4}\fontshape{#5}%
  \selectfont}%
\fi\endgroup%
\begin{picture}(8712,8424)(889,-8173)
\thicklines
\put(901,-5461){\framebox(300,5100){}}
\put(1501,-5461){\framebox(300,3900){}}
\put(2701,-5461){\framebox(300,3000){}}
\put(3301,-5461){\framebox(300,3000){}}
\put(9001,239){\line( 0,-1){8400}}
\put(8701,239){\line( 0,-1){8400}}
\put(5401,-6361){\framebox(300,4800){}}
\put(5101,239){\line( 0,-1){8400}}
\put(4801,239){\line( 0,-1){8400}}
\put(7126,-7561){\framebox(300,7200){}}
\put(2101,-4036){\makebox(0,0)[lb]{$-\frac{1}{2}$}}
\put(4126,-4036){\makebox(0,0)[lb]{$-\frac{1}{2}$}}
\put(6376,-4061){\makebox(0,0)[lb]{$-\frac{1}{2}$}}
\put(8101,-4036){\makebox(0,0)[lb]{$+\frac{1}{2}$}}
\put(5551,-6811){\makebox(0,0)[lb]{$l_2$}}
\put(970,-5911){\makebox(0,0)[lb]{$l_1$}}
\put(1576,-5911){\makebox(0,0)[lb]{$l_2$}}
\put(2770,-5986){\makebox(0,0)[lb]{$l_1$}}
\put(3376,-5986){\makebox(0,0)[lb]{$l_2$}}
\put(4540,-7261){\makebox(0,0)[lb]{$l_1$}}
\put(7161,-7880){\makebox(0,0)[lb]{$l_1$}}
\put(8440,-7861){\makebox(0,0)[lb]{$l_1$}}
\put(9601,-7561){\makebox(0,0)[lb]{$-q$}}
\put(9601,-6361){\makebox(0,0)[lb]{$-p$}}
\put(9601,-5536){\makebox(0,0)[lb]{$-n$}}
\put(9601,-4036){\makebox(0,0)[lb]{$0$}}
\put(9601,-2461){\makebox(0,0)[lb]{$n$}}
\put(9601,-1561){\makebox(0,0)[lb]{$p$}}
\put(9601,-436){\makebox(0,0)[lb]{$q$}}
\end{picture}
\end{center}
\caption{
Pictorial representation of the expression
$
a_1^{\Gamma}
-\frac{1}{2}
\left(
a_2^{\Gamma} +
a_3^{\Gamma} +
a_4^{\Gamma} -
a_5^{\Gamma}
\right)
$ appearing in (\ref{sudfyvnbgcplse}).
The  $ a_i$'s are winding numbers with respect to the sets of
loops presented in the picture (counted from the left to the right
and separated by the associated factor $\pm 1/2$).
The vertical lines are parallel to the
euclidean time-axis. The open loops cross
$ d^*l_1$ and extend to the euclidean time infinity. At the right we
indicate the different time planes.
}
\label{loops1}
\end{figure}

After expressing the expectation values above in terms of classical
expectations (which involve only closed loops) and these in terms of
cluster expansions, we arrive at the following expression:
\be
\lim_{n\to\infty}
\lim_{p\to\infty}
\lim_{q\to\infty}
\exp
\left(
\sum_{\Gamma}\; c_{\Gamma} \, \mu^{\Gamma}
\;
\left[
a_1^{\Gamma}
-\frac{1}{2}
\left(
a_2^{\Gamma} +
a_3^{\Gamma} +
a_4^{\Gamma}
-
a_5^{\Gamma}
\right)
\right]
a_{\rho_{1, \, 2}}^{\Gamma} \, b_{\rho_{1, \, 2}}^{\Gamma}
\right) ,
\label{sudfyvnbgcplse}
\ee
where $ a_i (\gamma)$ represent winding numbers of $ \gamma$ with
respect to the  loops successively presented in
Figure \ref{loops1}.
The quantities
$a_{\rho_{1, \, 2}} (\gamma)$ and $ b_{\rho_{1, \, 2}} (\gamma) $ are
electric and magnetic winding numbers with respect to the background
charge $ \rho_{1, \, 2}$.

We have to perform a detailed analysis of the sum over clusters
appearing in (\ref{sudfyvnbgcplse}). For the sake of brevity we will
sketch the main arguments.

Define the time planes
$H_a  :=  \{ (\undx,\, x_0 )\in \Z^{3} \mbox{ with } x_0 = a  \}$ and
denote by $ \calG_{B}$ the set of all clusters $ \Gamma $ not crossing
any of the planes $H_{\pm n}$, $ H_{\pm p}$
and $ H_{\pm q}$.
It is easy to verify that for a cluster $ \Gamma \in \calG_B$
one  either has
$
a_1^{\Gamma}
-\frac{1}{2}
\left(
a_2^{\Gamma} +
a_3^{\Gamma} +
a_4^{\Gamma} -
a_5^{\Gamma}
\right) = 0
$
or it happens that its contribution cancels that of other
cluster in $ \calG_B$ obtained by translating  $ \Gamma$ in time direction.
This is, for instance, what happens for clusters located
between $H_p$ and $ H_q$ and translated clusters located between
$H_{-q}$ and $H_{-p} $.

\begin{figure}[hbtp]
\begin{center}
\setlength{\unitlength}{0.00025in}%
\begingroup\makeatletter\ifx\SetFigFont\undefined
\gdef\SetFigFont#1#2#3#4#5{
  \reset@font\fontsize{#1}{#2pt}
  \fontfamily{#3}\fontseries{#4}\fontshape{#5}
  \selectfont}
\fi\endgroup
\begin{picture}(6912,9474)(1201,-8098)
\thicklines
\put(6301,-361){\line( 0,-1){3000}}
\put(6301,-3361){\line( 1, 0){1800}}
\put(8101,-3361){\line( 0, 1){3000}}
\put(2101,-6361){\line( 0, 1){3000}}
\put(2101,-3361){\line( 1, 0){1800}}
\put(3901,-3361){\line( 0,-1){3000}}
\multiput(2001,-6361)(0.00000,-105.00000){16}{.}
\multiput(3821,-6361)(0.00000,-105.00000){16}{.}
\multiput(6241,1364)(0.00000,-105.00000){16}{.}
\multiput(8041,1364)(0.00000,-105.00000){16}{.}
\put(3001,-3286){\makebox(0,0)[lb]{$l$}}
\put(7126,-4136){\makebox(0,0)[lb]{$l$}}
\put(901,-3586){\makebox(0,0)[lb]{$f_l$:}}
\put(5026,-3436){\makebox(0,0)[lb]{$f^t_l$:}}
\end{picture}
\end{center}
\caption{The semi-infinite loops for which $f_l (\gamma)$ and
$ f_l^t  (\gamma)$  are defined. The horizontal lines represent the
link $l$ located at $ H_0$. The vertical lines are parallel to the
euclidean time axis and extend to the negative (left) or positive
(right) euclidean time infinity.
}
\label{loopdledtl}
\end{figure}
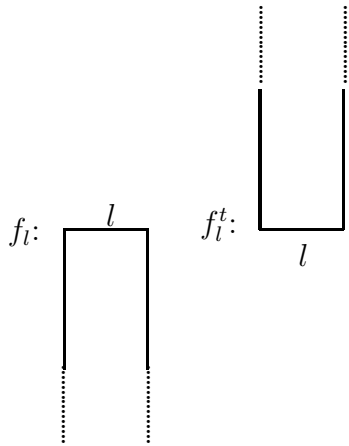

On the other hand, the size of clusters which cross at least two of
the planes $ H_{\pm n}$, $ H_{\pm p}$ or $ H_{\pm q}$
is at least
$ \min \{ 2n, \; p-n, \;q-p\}$ (assuming $q>p>n$).
After the limits
$q\to \infty$, $ p\to\infty$ and $ n\to \infty$ are taken the
total contribution of such clusters is zero, what can be shown using the
exponential decay given in (\ref{clusterTRES}) and noticing
that the size of the loops of Figure \ref{loops1} grows only linearly in $n$, $ p$
or $ q$.
The remaining terms belong to clusters
crossing one and only one of the planes
$ H_{\pm n}$, $ H_{\pm p}$ or $ H_{\pm q}$.
Using the translation invariance of the sum over
clusters, we may express these remaining terms (after the limits are
taken) in the following form:
\be
\begin{array}{l}
\displaystyle
\frac{1}{2}
\sum_{\Gamma \not\sim H_0} \; c_\Gamma \, \mu^\Gamma \;
\left(
  f_{l_1}^{\Gamma}  - (f^t_{l_1})^{\Gamma}
\right)
a_{\rho_{1, \, 2}}^{\Gamma} \, b_{\rho_{1, \, 2}}^{\Gamma}
  +
\frac{1}{2}
\sum_{\Gamma \not\sim H_0} \; c_\Gamma \, \mu^\Gamma \;
\left(
  f_{l_2}^{\Gamma} -  (f_{l_2}^t)^{\Gamma}
\right)
a_{(d^*l_1, \; 0)}^{\Gamma}
a_{\rho_{1, \, 2}}^{\Gamma} \, b_{\rho_{1, \, 2}}^{\Gamma}
+
 \\
 \\
\displaystyle
\frac{1}{2}
\sum_{\Gamma \not\sim H_0} \; c_\Gamma \, \mu^\Gamma \;
\left(
  (f_{l_1}^t)^{\Gamma}   (f_{l_2}^t)^{\Gamma} -
  f_{l_1}^{\Gamma}   f_{l_2}^{\Gamma}
\right)
a_{\rho_{1, \, 2}}^{\Gamma} \, b_{\rho_{1, \, 2}}^{\Gamma}
,
\end{array}
\label{sdeikughvbps}
\ee
where, with some abuse of notation, $\Gamma \not\sim H_0 $ indicates that
the geometric part of at least one polymer composing $\Gamma$ crosses
the plane $ H_0$. Above, $ f_l (\gamma)$ (respectively,  $ f^t_l (\gamma)$)
represents the winding number of the polymer $ \gamma$
with respect to the semi-infinite loops formed by $ l$ and by vertical
lines starting at $ d^* l$ and extending to the negative (positive)
euclidean time infinity. See Figure \ref{loopdledtl}.
Note that the second sum in (\ref{sdeikughvbps}) can be simplified, since
$
a_{(d^*l_1, \; 0)}^{\Gamma}
a_{\rho_{1, \, 2}}^{\Gamma} \, b_{\rho_{1, \, 2}}^{\Gamma}
=
a_{\rho_{ 2}}^{\Gamma} \, b_{\rho_{ 2}}^{\Gamma}
$.

It is easy to show that each of the sums over clusters in
(\ref{sdeikughvbps}) is absolutely convergent. Analogously, sums like
\be
\sum_{\Gamma \not\sim H_0} \; c_\Gamma \, \mu^\Gamma \;
\left(
  f_{l_1}^{\Gamma}  - 1
\right)
a_{\rho_{1, \, 2}}^{\Gamma} \, b_{\rho_{1, \, 2}}^{\Gamma}
\ee
are also absolutely convergent. It can be seen by reflecting polymers
on the plane $ H_0$ that the last expression is the complex
conjugate of
$ \displaystyle
\sum_{\Gamma \not\sim H_0} \; c_\Gamma \, \mu^\Gamma \;
\left(
  (f_{l_1}^t)^{\Gamma}  - 1
\right)
a_{\rho_{1, \, 2}}^{\Gamma} \, b_{\rho_{1, \, 2}}^{\Gamma}
$.
This means that each of the sums over clusters in
(\ref{sdeikughvbps}) is purely imaginary. Defining
\be
z^{el}_{\rho} (l_2) \; := \;
\exp \left\{
\frac{1}{2}
\sum_{\Gamma \not\sim H_0} \; c_\Gamma \, \mu^\Gamma \;
\left(
  f_{l_2}^{\Gamma} -  (f_{l_2}^t)^{\Gamma}
\right)
a_{\rho_{ 2}}^{\Gamma} \, b_{\rho_{ 2}}^{\Gamma}
\right\} ,
\label{Rsekuyvbnpewr}
\ee
which is a pure phase, we conclude from (\ref{sdeikughvbps})
the proof of part {\em i)} of
Theorem \ref{puiDyrbpowuie}. Part {\em ii)} can be proven analogously,
and we do not need to show the details.

The proof of part {\em iii)} is also analogous but with an important
difference. Since in this case
$ l_1$ is a magnetic link and $ l_2$ an electric
one, the closed loops formed by $l_1 $ and
by $ l_2$, appearing in the left Figure \ref{loops1},
can have a nontrivial
winding number, which can contribute to the classical expectations in
the numerator of (\ref{Tfiusgdoflkgjw})
with an additional  $\Z_N $ phase factor, as the phase factor
$ [\beta : \; \alpha ]$ emerging from (\ref{formulamestracomrho2}).
This phase equals $e^{i\la l_1, \; l_2\ra}$.

In order to prove {\em iv)}, consider that the support that $ l_2$ is,
say, an elementary plaquette at $ H_0$. Following the same steps
of the proof of {\em i)} we would arrive at relations like
(\ref{sdeikughvbps}) and (\ref{Rsekuyvbnpewr}) where both
$ f_{l_2}(\gamma) $ and $ f^{t}_{l_2}(\gamma)$
represent the winding number of $ \gamma$ around this plaquette.
Therefore, for any polymer $\gamma $, $ f_{l_2}(\gamma) =
f^t_{l_2}(\gamma)$ and hence $ z^{el}_{\rho}(l_2) = 1$.
The proof of {\em v)} is analogous. $ \Fullbox$

\subsection{Proof of Proposition \protect\ref{Tfifuygorfwiue} }
\label{ProofofPropositionTfifuygorfwiue}
\setcounter{theorem}{0}

We will prove only (\ref{holonomyphase1}) since
(\ref{holonomyphase2}) is analogous. Relation (\ref{holonomyphase1})
means
$ \| 	\phi_\rho^{el} (l_e) - e^{i\bra\mu, \; s\ket}\Omega_\rho \|=0$
and to prove one has only to show that
$
\left(
\Omega_\rho, \; \phi_\rho^{el} (l_e)
\right) = e^{i\bra\mu, \; s\ket}
$.
According to the definitions
\be
\left(
\Omega_\rho, \; \phi_\rho^{el} (l_e)
\right)
\; = \;
\frac{
\left(
\Omega_\rho, \; \;
P_\rho(\rho - (d^* l_e , \; 0))
\pi_{\rho} (A_{\rho , \; \rho - (d^* l_e , \; 0)} (l_e, \; 0))
\Omega_\rho
\right)
}{
\left\|
P_\rho( \rho - (d^* l_e , \; 0))
\pi_{\rho} \left(
A_{\rho , \; \rho - (d^* l_e , \; 0)} (l_e, \; 0)
\right)
\Omega_\rho
\right\|
}.
\label{fdiuvgnpsioef}
\ee
Under the hypothesis $ d^* l_e =0$ and, hence, we can write
the right hand side of (\ref{fdiuvgnpsioef}) as
\be
\lim_{n\to\infty}
\frac{
\left(
\Omega_\rho, \; \;
T_\rho(\rho)^n
\pi_{\rho} \left(
A_{\rho , \; \rho } (l_e, \; 0)
\right)
\Omega_\rho
\right)
}{
\left\|
T_\rho(\rho)^n
\pi_{\rho} \left(
A_{\rho , \; \rho } (l_e, \; 0)
\right) \Omega_\rho
\right\|
}
\; = \;
\lim_{n\to\infty}
\frac{
\left(
\Omega_\rho, \; \;
\pi_{\rho} \left(
\alpha_{\rho }^n (A_{\rho , \; \rho } (l_e, \; 0))
\right)
\Omega_\rho
\right)
}{
\left\|
\pi_{\rho} \left(
\alpha_{\rho }^n(A_{\rho , \; \rho } (l_e, \; 0))
\right)
\Omega_\rho
\right\|
}.
\label{Slkjdtghlsd}
\ee

We now expand the right hand side of (\ref{Slkjdtghlsd})
in terms of our cluster expansions
and treat it  with the same methods used in the proof of
Theorem \ref{puiDyrbpowuie} above. We get
\be
\left(
\Omega_\rho, \; \phi_\rho^{el} (l_e)
\right)
\; = \;
z^{el}_\rho (l_e) \, e^{i\bra\mu, \; s\ket},
\ee
where the $ \Z_N$ phase factor $ e^{i\bra\mu, \; s\ket}$
emerges in this expression as the factor
$ \left[  \tilde{\mu } : \, \alpha \right]$ emerges from
(\ref{formulamestracomrho2}): it represents the winding number of
$ d^* s  $ in the background charge $\rho  $.
Actually
$ e^{i\bra\mu, \; s\ket}= \left[  \tilde{\mu } : \, s \right]$.
Since $ d^* l_e = 0$, one has
$z^{el}_\rho (l_e) =1$ and the proposition is proven.
$ \Fullbox$

\end{appendix}


\newpage

\bibliography{bib}           
\bibliographystyle{unsrt}

\end{document}
